\documentstyle [12pt,epsf] {article}

\newcommand{\nwc}{\newcommand}
\nwc{\cl}  {\clubsuit}

\nwc{\hyp} {\hyphenation}
\nwc{\be}  {\begin{equation}}
\nwc{\ee}  {\end{equation}}
\nwc{\ba}  {\begin{array}}
\nwc{\ea}  {\end{array}}
\nwc{\bdm} {\begin{displaymath}}
\nwc{\edm} {\end{displaymath}}
\nwc{\bea} {\be\ba{rcl}}
\nwc{\eea} {\ea\ee}
\nwc{\ben} {\begin{eqnarray}}
\nwc{\een} {\end{eqnarray}}
\nwc{\bda} {\bdm\ba{lcl}}
\nwc{\eda} {\ea\edm}
\nwc{\bc}  {\begin{center}}
\nwc{\ec}  {\end{center}}
\nwc{\ds}  {\displaystyle}
\nwc{\bmat}{\left(\ba}
\nwc{\emat}{\ea\right)}
\nwc{\non} {\nonumber}
\nwc{\bib} {\bibitem}
\nwc{\lra} {\longrightarrow}
\nwc{\Llra}{\Longleftrightarrow}
\nwc{\ra}  {\rightarrow}
\nwc{\Ra}  {\Rightarrow}
\nwc{\lmt} {\longmapsto}
\nwc{\prl} {\partial}
\nwc{\iy}  {\infty}
\nwc{\ol}  {\overline}
\nwc{\hm}  {\hspace{3mm}}
\nwc{\lf}  {\left}
\nwc{\ri}  {\right}
\nwc{\lm}  {\limits}
\nwc{\lb}  {\lbrack}
\nwc{\rb}  {\rbrack}
\nwc{\ov}  {\over}
\nwc{\pr}  {\prime}
\nwc{\nnn} {\nonumber \vspace{.2cm} \\ }
\nwc{\Sc}  {{\cal S}}
\nwc{\Lc}  {{\cal L}}
\nwc{\Rc}  {{\cal R}}
\nwc{\Dc}  {{\cal D}}
\nwc{\Oc}  {{\cal O}}
\nwc{\Cc}  {{\cal C}}
\nwc{\Pc}  {{\cal P}}
\nwc{\Mc}  {{\cal M}}
\nwc{\Ec}  {{\cal E}}
\nwc{\Fc}  {{\cal F}}
\nwc{\Hc}  {{\cal H}}
\nwc{\Kc}  {{\cal K}}
\nwc{\Xc}  {{\cal X}}
\nwc{\Gc}  {{\cal G}}
\nwc{\Zc}  {{\cal Z}}
\nwc{\Nc}  {{\cal N}}
\nwc{\fca} {{\cal f}}
\nwc{\xc}  {{\cal x}}
\nwc{\Ac}  {{\cal A}}
\nwc{\Bc}  {{\cal B}}
\nwc{\Uc}  {{\cal U}}
\nwc{\Vc}  {{\cal V}}
\nwc{\Th} {\Theta}
\nwc{\th} {\theta}
\nwc{\vth} {\vartheta}
\nwc{\eps}{\epsilon}
\nwc{\si} {\sigma}
\nwc{\Gm} {\Gamma}
\nwc{\gm} {\gamma}
\nwc{\bt} {\beta}
\nwc{\La} {\Lambda}
\nwc{\la} {\lambda}
\nwc{\om} {\omega}
\nwc{\Om} {\Omega}
\nwc{\dt} {\delta}
\nwc{\Si} {\Sigma}
\nwc{\Dt} {\Delta}
\nwc{\al} {\alpha}
\nwc{\vph}{\varphi}
\def\tr{\mathop{\rm tr}}
\def\Tr{\mathop{\rm Tr}}
\def\Det{\mathop{\rm Det}}

\def\VEV#1{\left\langle #1\right\rangle}

\def\abs#1{\left| #1\right|}
\def\pr#1{#1^\prime}

\nwc{\Id}  {{\bf 1}}
\nwc{\diag} {{\rm diag}}
\nwc{\inv}  {{\rm inv}}
\nwc{\mod}  {{\rm mod}}
\nwc{\hal} {\frac{1}{2}}
\nwc{\tpi}  {2\pi i}

\def\slash#1{#1\!\!\!/\!\,\,}

\def\ijmpa#1{Int. J. Mod. Phys. {\bf A#1}}

\def\mpla#1{Mod. Phys. Lett. {\bf A#1}}
\def\npb#1{Nucl. Phys. {\bf B#1}}
\def\nc#1{Nuovo Cim. {\bf #1}}

\def\plb#1{Phys. Lett. {\bf #1B}}
\def\pr#1{Phys. Rev. {\bf #1}}
\def\pra#1{Phys. Rev. {\bf A#1 }}

\def\prc#1{Phys. Rep. {\bf C#1}}
\def\prd#1{Phys. Rev. {\bf D#1 }}
\def\prle#1{Phys. Rev. Lett. {\bf #1}}

\def\zpc#1{Z. Phys. {\bf C#1}}

\def\MeV {\,{\rm  MeV}}
\def\GeV {\,{\rm  GeV}}

\def \lta {\mathrel{\vcenter
     {\hbox{$<$}\nointerlineskip\hbox{$\sim$}}}}
\def \gta {\mathrel{\vcenter
     {\hbox{$>$}\nointerlineskip\hbox{$\sim$}}}}
\newsavebox{\nnin} \sbox{\nnin}{$\hspace{1mm}\in\kern -.8em /
                   \hspace{1mm}$}

\newcommand{\sub}{\subset}
\newsavebox{\nnsub} \sbox{\nnsub}{$\hspace{1mm}\sub\kern -.9em /
            \hspace{1mm}$}

\def\KK{{\rm I\kern -.2em  K}}
\def\NN{{\rm I\kern -.16em N}}
\def\RR{{\rm I\kern -.2em  R}}
\def\ZZ{Z \kern -.43em Z}
\def\QQ{{\rm \kern .25em
             \vrule height1.4ex depth-.12ex width.06em\kern-.31em Q}}
\def\CC{{\rm \kern .25em
             \vrule height1.4ex depth-.12ex width.06em\kern-.31em C}}
\def\ZZZ{Z\kern -0.31em Z}

\nwc{\olnu}  {\ol{\nu}}
\nwc{\olla}  {\ol{\la}}
\nwc{\olm}   {\ol{m}}
\nwc{\olmu}  {\ol{\mu}}
\nwc{\olh}   {\ol{h}}
\nwc{\olpsi} {\ol{\psi}}
\nwc{\olsi}  {\ol{\sigma}}
\nwc{\olgm}  {\ol{\gm}}
\nwc{\vp}    {\varphi}
\nwc{\prlt}  {\frac{\prl}{\prl t}}
\nwc{\ttau}  {\tilde{\tau}}
\nwc{\tP}    {\tilde{P}}
\nwc{\tU}    {\tilde{U}}
\nwc{\teps}  {\tilde{\eps}}
\nwc{\tla}   {\tilde{\la}}
\nwc{\tit}    {\tilde{t}}
\nwc{\iddq}  {\int\frac{d^dq}{(2\pi)^d}}
\nwc{\prpr}  {\prime\prime}
\nwc{\rN}    {\left(\frac{\rho}{N}\right)}
\nwc{\rNt}    {\left(\frac{\rho}{N}\right)^{\frac{N-2}{2}}}
\nwc{\rnN}   {\left(\frac{\rho_0}{N}\right)}
\nwc{\rnNt}    {\left(\frac{\rho_0}{N}\right)^{\frac{N-2}{2}}}
\nwc{\rnNf}    {\left(\frac{\rho_0}{N}\right)^{\frac{N-4}{2}}}
\nwc{\rNs}    {\left(\frac{\rho_0}{N}\right)^{\frac{N-6}{2}}}
\nwc{\kNt}    {\left(\frac{\kappa}{N}\right)^{\frac{N-2}{2}}}
\nwc{\kNf}    {\left(\frac{\kappa}{N}\right)^{\frac{N-4}{2}}}
\nwc{\kNs}    {\left(\frac{\kappa}{N}\right)^{\frac{N-6}{2}}}

\newcounter{app}
\def\app{\par
 \addtocounter{app}{1}
 \def\thesection{\Alph{app}}
 \def\ksection{\Alph{app}}}

\def\appendix#1{\app\sect{#1}}

\newcommand{\sect}[1]{ \section{#1} \setcounter{equation}{0} }

\begin{document}

\begin{titlepage}

\title{Effective Action for the \\
       Chiral Quark--Meson Model\thanks{Supported by the Deutsche
       Forschungsgemeinschaft}}

\author{{\sc D.--U. Jungnickel\thanks{Email:
D.Jungnickel@thphys.uni-heidelberg.de}} \\
 \\ and \\ \\
{\sc C. Wetterich\thanks{Email: C.Wetterich@thphys.uni-heidelberg.de}}
\\ \\ \\
{\em Institut f\"ur Theoretische Physik} \\
{\em Universit\"at Heidelberg} \\
{\em Philosophenweg 16} \\
{\em 69120 Heidelberg, Germany}}

\date{}
\maketitle

\begin{picture}(5,2.5)(-350,-450)
\put(12,-115){HD--THEP--95--7}
\put(12,-148){\today}
\end{picture}

\thispagestyle{empty}

\begin{abstract}

  The scale dependence of an effective average action for mesons and
  quarks is described by a nonperturbative flow equation.
  The running couplings lead to spontaneous chiral symmetry
  breaking. We argue that for
  strong Yukawa coupling between quarks and mesons the low momentum
  physics is essentially determined by infrared fixed points. This
  allows us to establish relations between various parameters related
  to the meson potential. The results for $f_\pi$ and
  $\VEV{\olpsi\psi}$ are not very sensitive to the poorly
  known details of the quark--meson effective action at scales where
  the mesonic bound states form. For realistic constituent quark
  masses we find $f_\pi$ around $100\MeV$.

\end{abstract}

\end{titlepage}

\sect{Introduction}

Quantum chromodynamics as the theory of strong interactions and its
symmetries are well tested both for high and low momenta. For momenta
$q^2\gta(2\GeV)^2$ asymptotic freedom \cite{GW73-1} permits the use of
perturbation theory for a quark--gluon description with small gauge
coupling $g_s$. The long distance behavior for $q^2\lta(300\MeV)^2$
can partially be described by chiral perturbation theory
\cite{Wei79-1,GL82-1}. Here the picture is based on a nonlinear or linear
$\si$--model \cite{GML60-1} for the pseudo--scalar mesons. The latter
can also be extended to describe in addition
scalar, vector and pseudo--vector mesons. Such
effective models
incorporate the chiral symmetries of QCD and use several
free couplings to parameterize the unknown strong interaction
dynamics. The parameters are determined phenomenologically
\cite{GL82-1,RRY85-1}, but
on a more fundamental level the question arises how they can be
related to the parameters of short distance QCD, i.e. the strong fine
structure constant $\alpha_s=\frac{g_s^2}{4\pi}$ and the current quark
masses. For example, one may ask how the most prominent quantity of
the mesonic picture, namely the pion decay constant $f_\pi$ which
measures the strength of spontaneous chiral symmetry breaking, can be
computed from $\alpha_s$ or vice versa.

Important progress in this question has been achieved by numerical
simulations in lattice gauge theories \cite{Lue}. Serious difficulties
in this approach remain, however, related to the treatment of
dynamical quarks and chiral symmetry.
There is also a vast amount of literature on various analytical
attempts to attack this problem. Examples can be found in \cite{Bij95-1}.
In this paper we employ a new
analytical method based on nonperturbative flow equations for scale
dependent effective couplings. These couplings parameterize the
effective average
action $\Gm_k$ \cite{Wet91-1} which is a type of coarse grained free
energy. It includes the effects of all quantum fluctuations with
momenta larger than an infrared cutoff $\sim k$. In the limit where
the average scale $k$ tends to zero $\Gm_{k\ra0}$ becomes therefore
the usual effective action, i.e. the generating functional of $1PI$
Green functions \cite{Wet93-2}. The scale dependence of $\Gm_k$ can be
described by an exact nonperturbative evolution equation
\cite{Wet93-1,Wet93-2}
\be
 \prlt \Gm_k [\vp] =
 \hal\Tr\left\{\left(
 \Gm_k^{(2)}[\vp]+R_k\right)^{-1}
 \frac{\prl R_k}{\prl t}\right\}
 \label{ERGE}
\ee
where $t=\ln(k/\La)$ with $\La$ some suitable high momentum scale. The
trace represents here a momentum integration as well as a summation
over internal indices and we note the appearance on the right hand
side of the {\em exact} inverse propagator $\Gm_k^{(2)}$ as given by
the second
functional variation of $\Gm_k$ with respect to the field variables
$\vp$. The function $R_k(q)$ parameterizes the detailed form of the
infrared cutoff or the averaging procedure. With the
choice\footnote{$Z_{\vp,k}$ is an appropriate wave function
  renormalization constant which will be specified later.}
\be
 R_k(q)=\frac{Z_{\vp,k}q^2 e^{-q^2/k^2}}{1- e^{-q^2/k^2}}
 \label{Rk(q)}
\ee
we observe that the momentum integration in eq. (\ref{ERGE}) is both
infrared and ultraviolet finite. For fluctuations with small momenta
$q^2\ll k^2$ the infrared cutoff $R_k\sim Z_{\vp,k}k^2$ acts like an
additional mass term in the propagator, whereas for $q^2\gg k^2$ it is
ineffective. The only difference between the flow equation
(\ref{ERGE}) and the $k$--derivative of a one--loop expression with
infrared cutoff $R_k$ concerns the appearance of $\Gm_k^{(2)}$ instead
of the second functional derivative of the {\em classical} action. This
turns eq. (\ref{ERGE}) into an exact equation, but also transmutes it
into a complicated functional differential equation which can
only be solved approximately by truncating the most
general form of $\Gm_k$. As it should be, eq. (\ref{ERGE}) can be
shown \cite{BAM93-1} to be equivalent to earlier forms of the exact
renormalization group equation \cite{WH73-1}. It may be interpreted as
a differential form of the Schwinger--Dyson equations
\cite{Dys49-1}. The difficult part is, however, not so much the
establishment of an exact flow equation but rather the finding of a
suitable nonperturbative truncation scheme which allows to solve the
differential equation. Then the flow equation can be integrated from
some short distance scale $\La$, where $\Gm_\La$ can be taken as the
classical action, to $k\ra0$ thus solving the model approximately.

Within the formalism of exact flow equations for the average action it
is possible to change the relevant degrees of freedom
\cite{EW94-1}. The idea is now to start from the exact flow equations
for quarks and gluons for $k>k_\vp$, and to use a similar exact
flow equation for quarks and mesons for $k<k_\vp$. The transition at
the scale $k_\vp$ ($600-700\MeV$) between the two pictures can be
encoded into an exact identity \cite{EW94-1} which replaces
multi--quark interactions in the quark--gluon picture by mesonic
interactions in the quark--meson picture. We emphasize that in the
quark--meson description the quarks remain important degrees of freedom as
long as $k$ is larger than a typical constituent quark mass
$m_q\simeq300\MeV$. We have therefore to deal with an effective
quark--meson model, where the mesons are described by a linear
$\si$--model with Yukawa coupling $h$ to the quarks. A first attempt
to describe the transition to a quark--meson model within a
QCD--inspired model with four--quark interactions \cite{EW94-1} has
been very encouraging. Spontaneous chiral symmetry breaking was
observed for low $k$, with a chiral condensate of the right order of
magnitude. In the following, the formalism has been generalized to QCD
\cite{Wet95-1}, with a method where the gluonic fluctuations with
$q^2>k^2$ are integrated out subsequently as $k$ is lowered.
So far, the treatment of
the quark--meson model for scales $k<k_\vp$ has been very rough,
however, since only quark fluctuations were included in
ref. \cite{EW94-1}. It is the purpose of this paper to present a
systematic study of the scale dependence of the quark--meson effective
action, including both quark and meson fluctuations.

Our main tool are nonperturbative flow equations which describe the
change of shape of the effective meson potential and the running of
the Yukawa coupling $h(k)$. In the perturbative limit these equations
reproduce the running of the couplings \cite{CGS93-1,BHJ94-1} in a
$U_L(N)\times U_R(N)$ model, which has been investigated in the context
of dynamical top quark condensation \cite{BHL91-1}. In our context $N$
stands for the number of quark flavors. The most important
nonperturbative ingredient in the flow equations will turn out to be
the appearance of effective mass threshold functions which account for
the decoupling of modes with mass larger than $k$. The solution of our
approximate flow equations allows us to express typical low momentum
quantities like $f_\pi$ in terms of the ``initial values'' for the
quark--meson model at the scale $k_\vp$, as for example the meson mass
term $\olm^2(k_\vp)$ or the wave function renormalization constant
$Z_\vp(k_\vp)$.

Not too surprisingly, the effective quark--meson Yukawa coupling $h$
will turn out to be rather strong. This can easily be seen by noting
that for $k\ra0$ this coupling is related to the ratio of a
constituent quark mass $m_q$ to $f_\pi=93\MeV$, namely
\be
 h(k=0)=\frac{2m_q}{f_\pi}\simeq6.5\; .
 \label{ValueOfHr}
\ee
For larger $k$ the Yukawa coupling must be even stronger, with a
typical nonperturbative initial value $h^2(k_\vp)/16\pi^2\gta1$.
The presence
of a strong coupling has important consequences for the predictive
power of the quark--meson model. Generically, the system of flow
equations exhibits (partial) infrared fixed points in the absence of a
mass scale. Due to the large Yukawa interaction the couplings are
driven very fast
towards these fixed points and the system ``looses its memory'' on the
detailed form of the initial values at $k_\vp$. Despite the fact that
the running is finally stopped by the formation of the chiral
condensate this infrared stability implies that the low momentum
quantities essentially depend only on one ``relevant'' parameter at
the scale $k_\vp$, i.e. the ratio $\olm^2(k_\vp)/k_\vp^2$. Using the
value (\ref{ValueOfHr}) for $h(0)$ and $h(k_\vp)$ between $12$ and
$100$ we find in a simplified model
\be
 f_\pi\simeq(83-100)\MeV
\ee
in good agreement with the observed value $f_\pi=93\MeV$.
An estimate of the error as well
as a more complete treatment including more accurately
the effects of the chiral
anomaly and the strange quark mass is postponed to future work.

Besides the exciting prospect of computing $f_\pi$ and other
parameters of the low momentum meson interactions from QCD our
approach seems also capable to deal with other issues. Once the
parameters at the transition scale $k_\vp$ are fixed
either by a QCD--computation or by fitting
low momentum observational data,
it is straightforward to study the quark--meson
system at nonvanishing temperature. With methods described in ref.
\cite{TW93-1} the temperature dependence of $f_\pi$ or
$\VEV{\olpsi\psi}$ can be investigated. For $T\lta k_\vp/2\pi$ a study
within the effective quark--meson model with $T$--independent initial
values at $k_\vp$ should be sufficient, whereas for larger temperatures
the $T$--dependence of initial parameters like $\olm^2(k_\vp)$ starts
to become an important effect. One may therefore hope to gain new
insight into the nature of the chiral phase transition in QCD
\cite{PW84-1}. Another interesting issue concerns the use of
quark--meson models to describe hadronization in high energy scattering
experiments involving quarks or gluons \cite{KG-1}. Here our approach
may help to compute the phenomenological parameters used in those
models. Finally, the scalar--fermion models have been extensively
studied in the large--$N_c$ limit
\cite{BHJ94-1,BHL91-1}, and
our nonperturbative flow equations may help to access smaller values
of $N_c$.

Our paper is structured as follows: in section
\ref{TheChiralQuarkMesonModel} we give a brief phenomenological
introduction to the chiral quark--meson model with $N$ flavors. The
scale dependence of the effective meson potential is then described in
section \ref{ScaleDependenceOfTheEffectiveMesonPotential} and the
scalar wave function renormalization can be found in section
\ref{ScalarAnomalousDimension}. In section \ref{EvolutionEquationForH}
we derive the $\beta$--function for the running Yukawa coupling
between quarks and mesons as well as the quark wave function
renormalization. Section \ref{TheO(4)SymmetricSigmaModel} is
devoted to a short discussion of the chiral anomaly and the
presentation of two simplified models with two quark flavors. The
first model is based on the symmetry $U_L(2)\times U_R(2)$ and
neglects the effects of the chiral anomaly, whereas the second one
based on $O(4)$ neglects all scalars whose
masses obtain contributions from the chiral anomaly. In section
\ref{Results} we discuss in detail
the infrared stability properties for models with strong Yukawa
couplings and the consequences for the ``prediction'' of
$f_\pi$. Section \ref{Discussion} finally contains our quantitative
estimates for $f_\pi$ and the chiral condensate
$\VEV{\olpsi\psi}_0$. Conclusions are drawn in section
\ref{Conclusions}.

\sect{The chiral quark--meson model}
\label{TheChiralQuarkMesonModel}

We describe the low--energy degrees of freedom of QCD by an effective
action for quarks and mesons. We concentrate in this paper on
pseudo--scalar and scalar
mesons $\vp$ which transform in the $(\ol{\bf N},{\bf N})$
representation of the flavor symmetry group $SU_L(N)\times SU_R(N)$
for $N$ flavors. We consider the chiral limit where the
current quark masses are neglected. In its simplest form the
effective action $\Gm_k$ for quarks and mesons contains kinetic
terms, a potential for the scalar fields and a Yukawa coupling
between quarks and mesons:
\bea
 \Gm_k &=&\ds{\int d^4 x\left\{
 Z_\vp (k) \prl_\mu \vp^*_{ab} \prl^\mu \vp^{ab}+
 U_k(\vp,\vp^\dagger ) \right.}\nnn
 &+& \ds{\left. iZ_\psi (k)
 \olpsi^a \gm^\mu \prl_\mu \psi_a +
 \olh (k) \olpsi^a \left[
 \frac{1+\ol{\gm}}{2}\vp_a^{\;\;b}-
 \frac{1-\ol{\gm}}{2}{(\vp^\dagger )}_a^{\;\;b} \right]
 \psi_b \right\} }.
 \label{Truncation}
\eea
Our Euclidean conventions ($\olh (k)$ is real) are specified in appendix
\ref{LinearSigmaModel}. The scalar potential is assumed to be a
function of the invariants
\bea
 \rho &=& \ds{\tr\left(\vp^\dagger \vp \right) }\nnn
 \tau_2 &=& \ds{\frac{N}{N-1} \tr\left(\vp^\dagger \vp \right)^2 -
 \frac{1}{N-1} \rho^2 }\nnn
 \xi &=& \ds{
 \det\vp +\det\vp^\dagger }
 \label{BasicInvariants}
\eea
where we neglect the dependence on additional higher order invariants
present for $N\geq 3$ (cf. appendix \ref{ScalarMassSpectrum}).

Spontaneous chiral symmetry breaking with
a residual vector--like $SU(N)$ flavor symmetry occurs if the
potential has a minimum for $\si_0\neq0$
\be
 \vp_0 = \left(
 \ba{cccc}
 \si_0 & & & \\
  & \si_0 & & \\
  & & \ddots & \\
  & & & \si_0
 \ea\right)\; ,\;\;\;
 \rho_0=N\abs{\si_0}^2 \; .
 \label{Minimum}
\ee
In this case we consider a quartic approximation for the potential
\be
 U_k = -\olmu^2 (k) \rho +
 \hal  \olla_1 (k) \rho^2 + \frac{N-1}{4}
 \olla_2 (k) \tau_2 -\hal \olnu(k)\xi
 \label{QuarticPotentialSSB}
\ee
where $\olmu^2(k)$ is related to the $k$--dependent minimum value
$\rho_0 (k)$ by
\be
 \olmu^2(k) =
 \olla_1(k) \rho_0(k) - \frac{|\olnu(k)|}{2}
 \left(\frac{\rho_0 (k)}{N}\right)^{\frac{N-2}{2}} \; .
\ee
Without loss of generality we will restrict ourselves
to positive $\olnu$. Up to an irrelevant constant we can also write
\be
 U_k=\hal \olla_1 (k)
 \left(\rho -\rho_0 (k)\right)^2 +
 \frac{N-1}{4}\olla_2 (k) \tau_2 -
 \hal \olnu(k)\xi
 +\hal\olnu(k)
 \left(\frac{\rho_0(k)}{N}\right)^{\frac{N-2}{2}}
 \rho \; .
\ee
On the other hand, at short distance scales spontaneous chiral
symmetry breaking is not yet visible and $U_k$ is in the symmetric
regime ($\si_0=0$), where
we use the parameterization
\be
 U_k = \olm^2 (k) \rho +
 \hal  \olla_1 (k) \rho^2 + \frac{N-1}{4}
 \olla_2 (k) \tau_2 -\hal \olnu(k)\xi \; .
 \label{PotentialSymRegime}
\ee
With these approximations our model can be described in terms of the
renormalized couplings
\bea
 h(k) &=& \ds{
 Z_\vp^{-1/2}(k) Z_\psi^{-1}(k) \olh (k) }\nnn
 \la_{1,2}(k) &=& \ds{
 Z_\vp^{-2}(k) \olla_{1,2}(k) }
 \label{RenCoupl}
\eea
and either the mass term
\be
 m^2 (k)=Z_\vp^{-1}(k) \olm^2 (k)
 \label{RenMass}
\ee
or the location of the potential minimum
\be
 \rho_R (k)=Z_\vp (k) \rho_0 (k) \; .
 \label{RenRho}
\ee
The dimension of $\olnu$ depends on $N$ and the renormalized
coupling is
\be
 \nu_R (k)=Z_\vp^{-\frac{N}{2}}(k) \olnu(k) \; .
\ee
For $\olnu=0$ the model has an additional axial $U_A (1)$ symmetry
which, however, is broken in QCD through the axial anomaly.

The quark--meson model is supposed to be obtained as an
effective model at some scale $k_\vp$, say $k_\vp\simeq600\MeV$.
It should be derivable from QCD by integrating out the gluonic
degrees of freedom and converting nonlocal four--quark interactions
into an effective quark--meson theory by the change of variables
described in \cite{EW94-1}. This gives a direct relation between
$\vp$ and a suitably defined \cite{EW94-1} composite
quark bilinear operator $\Oc_a^{\;\; b}=\VEV{\olpsi^b \psi_a}$
according to
\be
 \Oc_a^{\;\; b}=\frac{2\olm^2 (k_\vp) Z_\psi (k_\vp)}
 {\olh (k_\vp)}\vp_a^{\;\; b} \; .
\ee
The aim of this paper is to follow the evolution of $\Gm_k$ from the
``initial value'' at $k=k_\vp$ to $k=0$. The effective action
$\Gm =\Gm_{k=0}$ then describes the 1PI Green functions for the
collective meson fields or quark bilinears. In
particular, the chiral condensate is related to the vacuum
expectation value of $\vp$ corresponding to the minimum of the
effective potential $U=U_{k=0}$ through\footnote{We note that
  (\ref{Codensate}) defines the quark condensate at the scale $k_\vp$
  which may be different from the scale of usual chiral perturbation
  theory estimates.}
\be
 \VEV{\olpsi\psi}_0 =
 \frac{2\olm^2 (k_\vp) Z_\psi (k_\vp)}
 {\olh (k_\vp)} \si_0
 \label{Codensate}
\ee
with\footnote{Neglecting fermion masses the phase of $\si_0$ is
  arbitrary and we employ here a real and positive
  $\si_0$. Correspondingly, $\VEV{\olpsi\psi}_0$ stands only for the
  magnitude of the chiral condensate.}
\be
 \si_0 = \left(
 \frac{\rho_0 (k=0)}{N}\right)^{1/2} \; .
\ee

For $\si_0$ different from zero the chiral symmetry is spontaneously
broken and the spectrum of scalars contains $N^2 -1$ Goldstone bosons
corresponding to the pions. Their interactions are described by the
nonlinear $\si$--model. Neglecting the explicit $SU_V(N)$
breaking through quark masses the pion decay constant
$f_\pi$ is given in our conventions by
\be
 f_\pi = 2\si_R
\ee
with renormalized expectation value
\be
 \si_R = Z_\vp^{1/2} (k=0) \si_0 =
 \left(\frac{Z_\vp (0)\rho_0 (0)}{N}\right)^{1/2} \; .
\ee
In our normalization the
experimental value reads $f_\pi =93\MeV$ or
\be
 \si_R=46.5\MeV\; .
\ee
Another interesting quantity in our picture is the renormalized quark
mass ($h\equiv h(0)$)
\be
 m_q = h \si_R
\ee
This corresponds to a constituent mass generated by chiral symmetry
breaking. (We remind that we consider the approximation of vanishing
current quark masses here.) A typical value should be around
$300\MeV$ and the renormalized Yukawa coupling therefore be relatively
large, $h\approx 6.5$.

The scalar spectrum is discussed in detail in appendix
\ref{ScalarMassSpectrum}.
For $\si_R\neq 0$ the meson sector contains besides the $N^2-1$
massless Goldstone bosons the $\si$--field (radial mode). With
$\la_1 =\la_1 (0)$ and $\nu_R=\nu_R(0)$ its mass is given by
\be
 m_\si^2 = 2N\la_1 \si_R^2 -\hal\nu_R(N-2)\si_R^{N-2} \; .
\ee
One meson acquires a mass through the chiral anomaly. For the
realistic case of $N=3$ this can be identified with the
$\eta^\prime$ meson whereas the $\eta$ meson remains massless in the
chiral limit as one of the Goldstone bosons. For $N=2$
we are left
with the three pions as massless degrees of freedom
whereas the $K$--mesons and the $\eta$--meson are absent from the
spectrum. We will also for $N=2$ associate
the anomalously massive meson with the
$\eta^\prime$ meson. In our model its mass is given by
\be
 m_{\eta^\prime}^2 = \frac{N}{2}\nu_R \si_R^{N-2} \; .
\ee
The remaining $N^2 -1$ massive scalar fields in the adjoint
representation of the diagonal flavor symmetry group $SU(N)$ have
mass (for $\la_2 =\la_2 (0)$)
\be
 m_a^2 = N\la_2 \si_R^2 + \nu_R\si_R^{N-2} \; .
\ee
The neutral component can be associated with the $a_0$
meson with mass $983\MeV$. For realistic meson masses
($m_{\eta^\prime} = 958\MeV$) the couplings would
be
\bea
 \nu_R &\simeq& \left( 958\MeV\right)^2\;\;\;
 {\rm for}\;\;\; N=2 \nnn
 \nu_R &\simeq& 13158\MeV\;\;\;
 {\rm for}\;\;\; N=3
 \label{ValuesForNyR}
\eea
\bea
 \la_2 &\simeq& 11 \;\;\;
 {\rm for}\;\;\; N=2 \nnn
 \la_2 &\simeq& 55 \;\;\;
 {\rm for}\;\;\; N=3 \; .
\eea
One should, however, notice that the values for $\nu_R$ and $\la_2$
are rather sensitive to the precise association of
$m_a$ or $m_{\eta^\prime}$ with known particle masses
and should therefore only be taken as a rough estimate. In particular,
for $N=3$ the effects of a nonzero strange quark mass have to be
incorporated for a more realistic estimate.

\sect{Scale dependence of the effective meson potential}
\label{ScaleDependenceOfTheEffectiveMesonPotential}

The meson degrees of freedom can be
introduced \cite{EW94-1} at
some short distance scale $k_\vp$ by
inserting the identity
\bea
 1 &=& \ds{{\rm const} \int\Dc\si_A \Dc\si_H
 \exp\left\{ -\hal \left[\left(\si_A^\dagger -
 K_A^\dagger \tilde{G}-\Oc^\dagger [\psi]\tilde{G}\right)
 \tilde{G}^{-1}\left(\si_A -\tilde{G}K_A -
 \tilde{G}\Oc [\psi]\right)\right.\right. }\nnn
 &+& \ds{\left.\left.\left(\si_H^\dagger -
 K_H^\dagger \tilde{G}-\Oc^{(5)\dagger} [\psi]\tilde{G}\right)
 \tilde{G}^{-1}\left(\si_H -\tilde{G}K_H -
 \tilde{G}\Oc^{(5)} [\psi]\right)\right]\right\} }
 \label{identity}
\eea
into the functional integral for the effective
average action for quarks. Here $K_{A,H}$ are sources for the
collective fields and correspond to the antihermitian and hermitian
parts\footnote{The fields $\ol{\si}_{A,H}$ associated to $K_{A,H}$ by
a Legendre transformation obey $\ol{\si}_A =-\frac{i}{2}(\vp
-\vp^\dagger )$, $\ol{\si}_H =\hal (\vp
+\vp^\dagger )$.} of
$\vp$. They are associated to the fermion bilinear operators $\Oc
[\psi]$,
$\Oc^{(5)}[\psi ]$ whose Fourier components read
\bea
 \Oc_{\;\; b}^a (q) $=$ \ds{
 -i\int\frac{d^4 p}{(2\pi)^4} g(-p,p+q)
 \olpsi^a (p)\psi_b (p+q) }\nnn
 \Oc_{\;\;\;\;\;\; b}^{(5)a} (q) $=$ \ds{
 -\int\frac{d^4 p}{(2\pi)^4} g(-p,p+q)
 \olpsi^a (p)\ol{\gm}\psi_b (p+q) }\; .
\eea
The wave function renormalization $g(-p,p+q)$ and the propagator
$\tilde{G}(q)$ are chosen such that the four--quark interaction
contained in (\ref{identity}) cancels the dominant part of the
QCD--induced nonlocal four--quark interaction in the
effective average action
formulated only for quarks. As a result, the introduction of
collective fields by (\ref{identity}) replaces the dominant part of
the
four--quark interaction by terms quadratic and linear
in the meson field. The resulting effective quark--meson
interactions are more general than those of the model described
in the last section.
The momentum dependence of the kinetic terms and the Yukawa couplings
can be described by an extended truncation of the effective average
action which, for general space--time dimensions $d$, is given
by
\bea
 \Gm_k &=& \ds{ \int d^d x\;
 U_k (\vp ,\vp^\dagger )}\nnn
 &+& \ds{
 \iddq\left\{
 Z_{\vp ,k}(q) q^2 \tr\left(
 \vp^\dagger (q)\vp (q)\right) +
 Z_{\psi ,k}(q)\olpsi(q)\gm^\mu q_\mu \psi (q) \right. }\nnn
 &+& \ds{ \left.
 \iddq
 \olh _k (-q,q-p) \olpsi(q) \left(
 \frac{1+\ol{\gm}}{2}\vp (p)-
 \frac{1-\ol{\gm}}{2}\vp^\dagger (-p) \right)
 \psi (q-p) \right\} \; .}
 \label{EffActAnsatz}
\eea
At the scale $k_\vp$ the average potential is
then purely quadratic
\be
 U_{k_\vp} = \olm^2\tr\left(\vp^\dagger\vp\right)
 \label{InitialU}
\ee
and the inverse scalar propagator is related to $\tilde{G}(q)$ in eq.
(\ref{identity}) by
\bea
 \tilde{G}^{-1}(q) &=& 2\olm^2 +2\ol{Z}_\vp (q) q^2 \nnn
 \ol{Z}_\vp (q) &\equiv& Z_{\vp ,k_\vp}(q) \; .
\eea
The initial value of the Yukawa coupling corresponds to the ``quark
wave function in the meson'' in eq. (\ref{identity}), i.e.
\be
 \olh _{k_\vp}(-q,q-p) = g(-q,q-p)
\ee
which can be normalized with $\olh_{k_\vp}(0,0)=g(0,0)=1$.
The propagator $\tilde{G}$ and the wave function $g(-q,q-p)$ should be
optimized for a most complete elimination of terms quartic in the
quark fields. Neglecting the remaining $\psi^4$ terms and terms of
higher order in $\psi$ (e.g, $\psi^6$) one arrives at the initial
value for $\Gm_{k_\vp}$. In the present paper we often do not want to
keep the complete momentum dependence of $Z_{\psi ,k}$, $Z_{\vp ,k}$
and $\olh _k$. Useful definitions of the initial values of the
parameters of the model in section
\ref{TheChiralQuarkMesonModel} are then
\bea
 Z_\psi (k_\vp ) &=& \ds{\left.
 Z_{\psi ,k_\vp}(q)\right|_{q^2 =0} }\nnn
 Z_\vp (k_\vp ) &=& \ds{ \left.\frac{1}{2q^2}\left(
 \tilde{G}^{-1}(q)-\tilde{G}^{-1}(0)\right)
 \right|_{q^2 =k_\vp^2} }\nnn
 \olm^2 (k_\vp ) &=& \ds{
 \hal \tilde{G}^{-1}(0) }\nnn
 \olh (k_\vp ) &=& \ds{\left.
 \olh _{k_\vp}(-q,q)\right|_{q^2 =0}\equiv 1 }\; .
 \label{InitialValues}
\eea

Although the results of \cite{EW94-1} should only be considered as
rough estimates it seems convenient to use them as a guide for the
choice of initial values of the various couplings. The values found
for the transition scale $k_\vp$ and the scalar mass at this scale are
\cite{EW94-1} $k_\vp=630\MeV$, $\olm(k_\vp)=120\MeV$. The
$q^2$--dependence of $\tilde{G}$ was not computed very reliably in ref
\cite{EW94-1}. Large--$N_c$ estimates use a rather weak
$q^2$--dependence \cite{BHJ94-1}.
As a typical guess we consider here, somewhat
arbitrarily, that $\tilde{G}^{-1}(q^2=k_\vp^2)$ exceeds
$\tilde{G}^{-1}(q^2=0)$ by $15\%$. This leads to
$Z_\vp(k_\vp)=0.15\frac{\olm^2(k_\vp)}{k_\vp^2}\simeq\frac{1}{180}$. With
$Z_\psi(k_\vp)=1$, $\olh(k_\vp)=1$ this corresponds to a large
renormalized Yukawa coupling of $h^2(k_\vp)=180$. We will see later
(section \ref{Results}) that for strong initial Yukawa couplings the
decisive parameter is the ratio
\be
 \teps_0=Z_\psi^2(k_\vp)\frac{\olm^2(k_\vp)}{k_\vp^2}\; .
\ee
The values of \cite{EW94-1} correspond to $\teps_0=0.036$.
Both, $Z_\psi$ and
$\olm$, may be somewhat lower than the values from \cite{EW94-1}
and we will often use typical values $\olm Z_\psi(k_\vp)=89\MeV
(63\MeV)$ for which
$\teps_0\equiv\olm^2Z_\psi^2(k_\vp)/k_\vp^2\simeq0.02(0.01)$ and
$h^2(k_\vp)\simeq330(660)$ if the same assumption on the momentum
dependence of $\tilde{G}^{-1}(q)$ is made as above. The dependence of our
results on the choice of initial values will be discussed in detail in
section \ref{Results}.

We note that the use of the identity (\ref{identity})
does not lead to anomalous
$U_A(1)$ violating meson interactions
and (\ref{EffActAnsatz}) conserves the axial $U_A(1)$ symmetry.
Consequently the solution of the flow equations for the
$k$--dependence of the average potential also conserves this
symmetry. The formalism has therefore to be extended to incorporate
anomalous fermion interactions of the type
$\Det\left(\olpsi^a\psi_b\right)$ into the mesonic picture.
This issue may be addressed for the time being by introducing a term
$-\hal\olnu\xi$ into $U_k$ of (\ref{InitialU}) as a phenomenologically
determined coupling. We leave this for future work and
concentrate here on the $U_A(1)$ conserving case $\olnu=0$ and later
(section \ref{TheO(4)SymmetricSigmaModel}) on the opposite extreme
$\olnu\ra\infty$ for $N=2$.
We also observe that a generalization of the
formalism of \cite{EW94-1} may lead to nonvanishing meson
self--interactions $\olla_1$, $\olla_2$ at the scale $k_\vp$
as predicted by large--$N_c$ results \cite{BHJ94-1}.

\begin{figure}
\unitlength1.0cm
\begin{picture}(13.,9.)
\put(.7,5.){\bf $\ds{\frac{m,\si_R}{\MeV}}$}
\put(8.,0.5){\bf $k/\MeV$}
\put(6.,3.5){\bf $\si_R$}
\put(11.5,5.){\bf $m$}
\put(-0.8,-11.5){
\epsfysize=22.cm
\epsffile{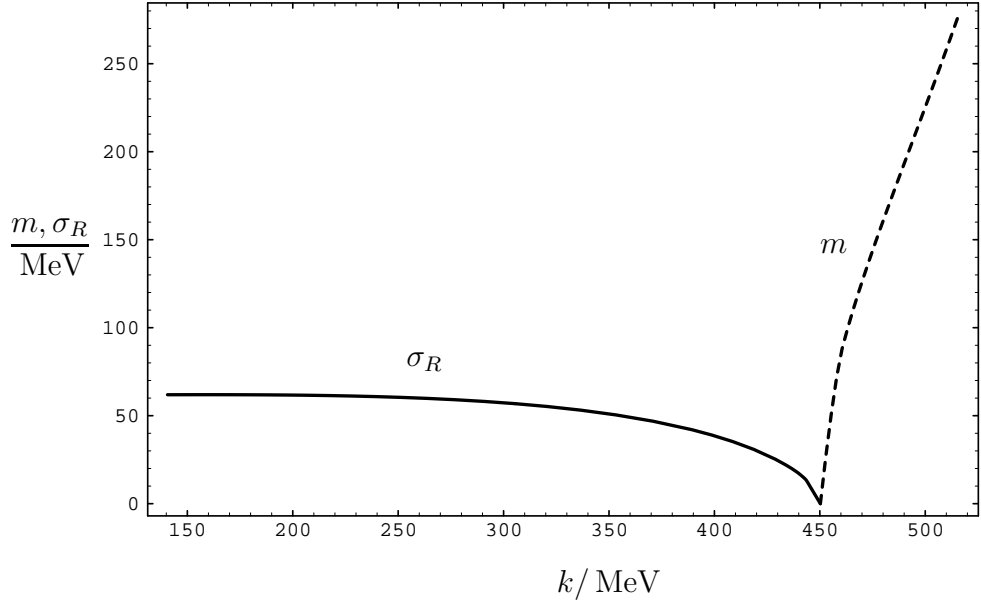}
}
\end{picture}
\caption{\footnotesize Evolution of the
  renormalized mass $m$ in the symmetric regime (dashed line) and the
  vacuum expectation value $\si_R$ of the scalar field in the SSB regime
  (solid line) as functions of $k$ for the $U_L(2)\times U_R(2)$
  model. Initial values are
  $\la_1(k_\vp)=\la_2(k_\vp)=0$ for
  $k_\vp=630\MeV$ with $h^2(k_\vp)=300$ and
  $\teps_0=0.01$.}
\label{Fig1}
\end{figure}

At the scale $k_\vp$ the effective potential (\ref{InitialU})
has its minimum at the origin. As a result of quantum
fluctuations with momenta $q^2<k_\vp^2$ one expects that the potential
changes its shape and ends up at $k=0$ with a minimum for $\rho>0$,
resulting in a spontaneous breaking of chiral symmetry.
The aim of this paper is to derive flow equations for the
$k$--dependence of $Z_\psi$, $Z_\vp$, $U_k$ and $\olh$ and
to compute the observable quantities at $k=0$ described in the
last section  from the initial values (\ref{InitialValues}).
Solving the flow equations numerically we find that chiral symmetry
breaking indeed occurs as demonstrated in figure \ref{Fig1}.

We begin with the evolution equation for the effective average
potential $U_k$. Except for the $k$--dependence of $\olnu$ the evolution
equation for the
potential can be obtained (c.f. appendix
\ref{LinearSigmaModel})
by studying a constant scalar field
configuration which is real and
diagonal
\be
 \vp_{ab}=\vp_a \delta_{ab}=
 \widehat{m}_a \delta_{ab}
\ee
We evaluate the exact evolution equation (\ref{ERGE}) for this
configuration and insert the truncation (\ref{Truncation}) into the
right hand side. The flow equation has a fermionic and a bosonic
contribution. The bosonic part follows by neglecting for a moment
the quarks \cite{Wet93-1,Wet93-2}:
\bea
 \ds{\prlt U_k} &=& \ds{
 \hal\iddq\prlt R_k(q)
 \left\{\sum_a \left[ \frac{1}{Z_\vp P+M_{Ra}^2}+
 \frac{1}{Z_\vp P+M_{Ia}^2}\right]\right. }\nnn
 &+& \ds{ \left.\sum_{a\neq b}\left[
 \frac{1}{Z_\vp P+(M_{Rab}^+)^2}+
 \frac{1}{Z_\vp P+(M_{Rab}^-)^2}\right.\right.}\nnn
 &&\;\;\;\;\, \ds{\left.\left. +
 \frac{1}{Z_\vp P+(M_{Iab}^+)^2}+
 \frac{1}{Z_\vp P+(M_{Iab}^-)^2}\right]\right\} } \; .
 \label{UkEvol}
\eea
We observe the appearance of the (massless) inverse average propagator
\be
 P(q)=q^2+Z_\vp^{-1}R_k(q)=\frac{q^2}{1-e^{-\frac{q^2}{k^2}}}
\ee
which incorporates the infrared cutoff function $R_k$ (\ref{Rk(q)}).
The dependence of the various mass eigenvalues on $\vp_a$ can be found
in appendix \ref{ScalarMassSpectrum}. We restrict the discussion here
to the approximation $\olnu=0$ which
corresponds to neglecting the mass difference between the pseudo--scalar
pion triplet and the $\eta^\prime$ singlet. The mass eigenvalues on
the right hand side of (\ref{UkEvol}) are then
given by (\ref{MIa})---(\ref{MRac}).
In order to express the eigenvalues
$\widehat{m}_a^2$ in terms of $\rho$ and $\tau_2$ we consider a
particular configuration $\vp$ where $N-1$ eigenvalues are equal
to $\widehat{m}_1^2$ such that
\be\ba{rcl}
 \rho &=& \ds{
 (N-1)\widehat{m}_1^2 +\hat{m}_N^2 } \nnn
 \tr\phi^2 &=& \ds{\frac{N-1}{N}\left(
 \widehat{m}_1^2-\widehat{m}_N^2 \right)^2\; ,\;\;\;
 \phi=\vp^\dagger\vp-\frac{1}{N}\rho }
\ea\ee
or
\be\ba{rcccl}
 \vp_1^2 &=& \widehat{m}_1^2 &=&
 \ds{\frac{1}{N}\left(\rho +\sqrt{\tau_2}\right)}\nnn
 \vp_N^2 &=& \widehat{m}_N^2 &=&
 \ds{\frac{1}{N}\left(\rho -(N-1)\sqrt{\tau_2}\right)}\; .
 \label{configuration}
\ea\ee
This defines the right hand side of the evolution equation
(\ref{UkEvol}) as a function of $\rho$ and $\tau_2$. In the symmetric
regime the evolution of the couplings $\olla_1$, $\olla_2$ and
the mass term $\olm^2$ can now be extracted by suitable
differentiation of eq. (\ref{UkEvol}) with respect to
$\rho$ and $\tau_2$, evaluated for $\rho=\tau_2=0$. One finds for the
bosonic contributions
\begin{eqnarray}
 \ds{\prlt\olm^2} &=& \ds{
 -\hal\iddq\frac{\prl R_k}{\prl t}
 \frac{2(N^2+1)\olla_1 +(N^2-1)\olla_2}
 {(Z_\vp P+\olm^2)^2} }
 \label{BosEvolMSymReg}\\[2mm]
 \ds{\prlt\olla_1} &=& \ds{
 \iddq\frac{\prl R_k}{\prl t}
 \frac{2(N^2+4)\olla_1^2 +2(N^2-1)\olla_1 \olla_2+
 (N^2-1)\olla_2^2}
 {(Z_\vp P+\olm^2)^3} }
 \label{BosEvolLa1SymReg}\\[2mm]
 \ds{\prlt\olla_2} &=& \ds{
 \iddq\frac{\prl R_k}{\prl t}
 \frac{12\olla_1 \olla_2 +2(N^2-3)\olla_2^2}
 {(Z_\vp P+\olm^2)^3} }\; .
 \label{BosEvolLa2SymReg}
\end{eqnarray}
If in the course of the evolution towards smaller values of $k$ the
mass term $\olm^2$ becomes negative we should switch to the
couplings appropriate to the regime with spontaneous symmetry
breaking (SSB regime). There we define
\bea
 \olla_1 &=& U_k^{\prpr}(\rho_0,\tau_2 =0) \nnn
 \olla_2 &=& \ds{\frac{4}{N-1}\frac{\prl U_k}{\prl\tau_2}
 (\rho_0,\tau_2 =0)}
 \label{DefLambdas}
\eea
where $\rho_0$ corresponds to the $k$--dependent minimum of the
potential. We use that
\be
 U_k^\prime (\rho_0)=0
\ee
is valid for all $k$ and therefore obtain
\be
 \prlt\rho_0 =
 -\frac{1}{\olla_1}\prlt
 U_k^\prime (\rho_0) \; .
 \label{DefRhoEvenN}
\ee
The evolution equations for $\rho_0$, $\olla_1$ and $\olla_2$ follow
directly from the definitions (\ref{DefLambdas}),
(\ref{DefRhoEvenN}). For $\olnu =0$ they read
\be
 \prlt\rho_0 =
 \hal\iddq\frac{\prl R_k}{\prl t}\left\{
 \frac{N^2}{(Z_\vp P)^2}+\frac{3}{(Z_\vp P+2\olla_1
 \rho_0)^2}+
 \frac{(N^2-1)\left(1+\frac{\olla_2}{\olla_1}\right)}
 {(Z_\vp P+\olla_2 \rho_0)^2}\right\}
 \label{BosEvolRhoSSB}
\ee
\be
 \prlt\olla_1 =
 \iddq\frac{\prl R_k}{\prl t}\left\{
 \frac{N^2 \olla_1^2}{(Z_\vp P)^3}+
 \frac{9\olla_1^2}{(Z_\vp P+2\olla_1 \rho_0)^3}+
 \frac{(N^2-1)\left(\olla_1+\olla_2\right)^2}
 {(Z_\vp P+\olla_2 \rho_0)^3}\right\}
 \label{BosEvolLa1SSB}
\ee
\bea
 \ds{\prlt\olla_2} &=& \ds{
 \iddq\frac{\prl R_k}{\prl t}\left\{
 \frac{N^2}{4}\frac{\olla_2^2}{(Z_\vp P)^3} \right. }\nnn
 &+& \ds{
 \frac{9(N^2-4)}{4}\frac{\olla_2^2}
 {(Z_\vp P+\olla_2 \rho_0)^3}
 + \frac{N^2\olla_2}{4\rho_0}\left[
 \frac{1}{(Z_\vp P+\olla_2 \rho_0)^2}-
 \frac{1}{(Z_\vp P)^2}\right] }\nnn
 &+& \ds{\left.
 \frac{3\olla_2 (\frac{1}{4}\olla_2+\olla_1)}
      {\rho_0 (\hal\olla_2 -\olla_1)}\left[
 \frac{1}{(Z_\vp P+2\olla_1 \rho_0)^2}-
 \frac{1}{(Z_\vp P+\olla_2 \rho_0)^2}\right]\right\} }\; .
 \label{BosEvolLambda2SSB}
\eea
It is straightforward to check that in the limits $\olm^2 \ra 0$,
$\rho_0 \ra 0$ the flow equations for $\olla_1$, $\olla_2$ coincide
in the symmetric and SSB regime. We also note that the evolution
equation for $\olla_2$ depends on the precise definition of this
coupling. This issue is shortly addressed in
appendix \ref{EvolEquLa2} where we
also give an alternative formulation of
eq. (\ref{BosEvolLambda2SSB}).

Next we turn to the fermionic contribution to the evolution equation
for the effective average potential which we denote by $\prl
U_{kF}/\prl t$. Using the general formulae of \cite{Wet90-1} it can
be computed without additional effort for the extended ansatz
(\ref{EffActAnsatz})
where we keep the momentum dependence of
$Z_{\psi,k}$ and part of the momentum dependence of the Yukawa
coupling with $\olh_k(q)\equiv\olh(-q,q)$. With $P_F$ given in
appendix \ref{InfraresCutoffFermions} and setting for a
moment $Z_{\psi,k} (q)=1$ one obtains:
\be
 \prlt U_{kF} =
 -2^{\frac{d}{2}-1}N_c \iddq\sum_{a=1}^N
 \frac{\prl P_F(q)}{\prl t}\left( P_F (q)+m_a^2 (q)\right)^{-1} \; .
\ee
Here $m_a^2 (q)$ are the $N$ real nonnegative eigenvalues of the
$N\times N$ matrix $\olh_k (q)\olh_k^* (q)\vp^\dagger\vp$ with momentum
dependent Yukawa couplings defined by (\ref{SYuk}). Here we have taken
into account the $N_c$ colors of the quarks. For a given value of
$q$ we may use the identity (with $\prl /\prl t$ acting only on $P_F$,
$m^2 ={\rm diag}(m_a^2)$ and $\tr$ taken in flavor space)
\bea
 \ds{\sum_{a=1}^N \frac{\prl P_F}{\prl t}
 \left( P_F +m_a^2\right)^{-1}} &=& \ds{
 \prlt\tr\ln\left( P_F+m^2\right) =
 \prlt\ln\det\left( P_F+m^2\right)}\nnn
 &=& \ds{\prlt
 \ln\det\left( P_F+\abs{\olh_k (q)}^2\vp^\dagger\vp\right) }\nnn
 &=& \ds{
 \frac{\prl P_F}{\prl t}
 \tr\left( P_F+\abs{\olh_k (q)}^2\vp^\dagger\vp\right)^{-1} }\; .
\eea
This gives an expression for general $\vp$
\be
 \prlt U_{kF} =
 -2^{\frac{d}{2}-1}N_c
 \iddq\frac{\prl P_F}{\prl t}
 \tr\left( P_F+\abs{\olh_k (q)}^2\vp^\dagger\vp\right)^{-1} \; .
 \label{UFEvol}
\ee
Using the particular configuration with $N-1$ equal eigenvalues and
the relation (\ref{configuration}) one finally obtains
\bea
 \ds{\prlt U_{kF} } &=& \ds{
 -2^{\frac{d}{2}-1}N_c
 \iddq\frac{\prl P_F}{\prl t}
 \left\{ (N-1)\left( P_F +
 \frac{1}{N}\abs{\olh_k}^2(\rho +\sqrt{\tau_2})\right)^{-1}
 \right. }\nnn
 &+& \ds{\left.
 \left( P_F +
 \frac{1}{N}\abs{\olh_k}^2(\rho -(N-1)\sqrt{\tau_2})\right)^{-1}
 \right\} }\; .
 \label{UFEvolN=2}
\eea
We emphasize that the Yukawa couplings in (\ref{EffActAnsatz})
conserve the axial $U_A(1)$ symmetry. The fermionic contribution to
$\prl U_{kF}/\prl t$ is therefore independent of $\xi$. The fermionic
wave function renormalization $Z_{\psi,k} (q)$ is easily restored if
we replace the function $P_F (q)$ in (\ref{UFEvolN=2}) by
$Z_{\psi,k}^2 (q)P_F(q)$ and note that the partial derivative
$\widehat{\prlt}(Z_{\psi,k}^2 P_F)$ only acts on the
pieces related to the infrared cutoff $R_k$. Within the truncation
(\ref{EffActAnsatz}) the fermionic contribution to the evolution
equation (\ref{UFEvol}) is then exact. Eq. (\ref{UFEvolN=2}) gives
the exact result for $N=2$ whereas for
$N>2$ one has an additional dependence on invariants $\tau_i$, $i\geq
3$, defined in appendix \ref{ScalarMassSpectrum}, which can be
extracted from (\ref{UFEvol}). The derivatives of
$\prlt U_{kF}$ with respect to $\rho$ and $\tau_2$ can be written in
a suggestive form as
\begin{eqnarray}
 \ds{\prlt U_{kF}^\prime} &=& \ds{
 -2^{\frac{d}{2}-1} \frac{N_c}{N}
 \iddq\abs{\olh_k}^2 }
 \label{UFprime} \\[2mm]
 &\times& \ds{
 \widehat{\prlt} \left\{
 \frac{N-1}{Z_{\psi,k}^2 P_F +
 \frac{1}{N}\abs{\olh_k}^2(\rho +\sqrt{\tau_2})} +
 \frac{1}{Z_{\psi,k}^2 P_F +
 \frac{1}{N}\abs{\olh_k}^2(\rho -(N-1)\sqrt{\tau_2})}
 \right\} } \nnn
 \ds{\prlt\frac{\prl U_{kF}}{\prl\tau_2}} &=& \ds{
 -2^{\frac{d}{2}-2} N_c\frac{N-1}{N} \frac{1}{\sqrt{\tau_2}}
 \iddq\abs{\olh_k}^2 }
 \label{UFtau2}\\[2mm]
 &\times& \ds{
 \widehat{\prlt} \left\{
 \frac{1}{Z_{\psi,k}^2 P_F +
 \frac{1}{N}\abs{\olh_k}^2(\rho +\sqrt{\tau_2})} -
 \frac{1}{Z_{\psi,k}^2 P_F +
 \frac{1}{N}\abs{\olh_k}^2(\rho -(N-1)\sqrt{\tau_2})}
 \right\} } \nonumber
\end{eqnarray}
with the formal definition
(cf. appendix \ref{InfraresCutoffFermions})
\be
 \ds{\widehat{\prlt}}
 \equiv \ds{
 \frac{1}{Z_{\vp,k}}\frac{\prl R_k}{\prl t}
 \frac{\prl}{\prl P} }
 + \ds{
 \frac{2}{Z_{\psi,k}} \frac{P_F}{1+r_F}
 \frac{\prl \left[Z_{\psi,k} r_F\right]}{\prl t}
 \frac{\prl}{\prl P_F} }\; .
\ee

We may now combine the bosonic and fermionic contributions to the
running of the renormalized couplings. Here we restrict the
discussion again to momentum independent $Z_\vp$, $Z_\psi$ and (real)
$\olh$, i.e. we replace similarly to (\ref{InitialValues})
$Z_{\vp,k}(q)\ra Z_\vp (k)$, $Z_{\psi,k}(q)\ra Z_\psi (k)$ and
$\olh_k(q)\ra\olh (k)=\olh^*(k)$. For arbitrary $d$ it is
convenient to introduce dimensionless couplings in analogy to
(\ref{RenCoupl})---(\ref{RenRho}):
\bea
 h^2 &=& \ds{
 Z_\vp ^{-1} Z_\psi^{-2} k^{d-4} \olh^2 } \nnn
 \la_{1,2} &=& \ds{
 Z_\vp^{-2} k^{d-4} \olla_{1,2} } \nnn
 \kappa &=& \ds{
 k^{2-d} \rho_R\; =\; Z_\vp k^{2-d} \rho_0 }\nnn
 \eps &=& \ds{k^{-2} m^2\; =\; Z_\vp^{-1} k^{-2} \olm^2 }
 \label{DimensionlessCouplings}
\eea
and to define the anomalous dimensions for the scalar field,
$\eta_\vp$, and the fermion field, $\eta_\psi$, by
\be
 \eta_\vp=-\prlt\ln Z_{\vp,k} \; ,\;\;\;
 \eta_\psi=-\prlt\ln Z_{\psi,k}\; .
 \label{AnoDimensions}
\ee
We also use dimensionless integrals
\ben
 l_n^d (w;\eta_\vp) &=& \ds{
 l_n^d (w) - \eta_\vp\hat{l}_n^d (w)
 }\nnn
 &=& \ds{
 \frac{n}{4} v_d^{-1}k^{2n-d} \iddq
 \left(\frac{1}{Z_\vp} \frac{\prl R_k(q)}{\prl t}\right)
 \left( P+w k^2\right)^{-(n+1)} }\nnn
 &=& \ds{
 -\hal k^{2n-d} \int_0^{\infty} dx x^{\frac{d}{2}-1}
 \widehat{\prlt} \left( P+w k^2\right)^{-n} }\\[2mm]
 \ds{ l_{n_1,n_2}^d(w_1,w_2;\eta_\vp) }
 &=& \ds{
 \l_{n_1,n_2}^d(w_1,w_2)
 -\eta_\vp \hat{l}_{n_1,n_2}^d(w_1,w_2)}\nnn
 &=& \ds{
 -\hal k^{2(n_1+n_2)-d}
 \int_0^\infty dx\, x^{\frac{d}{2}-1}
 \widehat{\prlt} \left\{\left(
 P+w_1k^2\right)^{-n_1}
 \left( P+w_2k^2\right)^{-n_2} \right\} \nonumber
 }
\een
where the part $\sim\eta_\vp\hat{l}_n^d(w)$ arises from the
$t$--derivative acting on $Z_\vp$ within $R_k$ (cf. (\ref{Rk(q)})) and
\be
 v_d^{-1} = 2^{d+1} \pi^\frac{d}{2} \Gm\left(\frac{d}{2}\right) \; .
\ee
The ``fermionic integrals''
$l_n^{(F)d} (w;\eta_\psi)=l_n^{(F)d} (w)-
\eta_\psi\check{l}_n^{(F)d} (w)$
are defined analogously,
with $P$ replaced by $P_F$. Combining
(\ref{BosEvolMSymReg})---(\ref{BosEvolLa2SymReg}) with
(\ref{UFprime}), (\ref{UFtau2}) we obtain the evolution equations for
the symmetric regime:
\begin{eqnarray}
 \ds{ \frac{\prl\eps}{\prl t} } &=& \ds{
 -(2-\eta_\vp )\eps -2v_d\left\{
 [2(N^2+1)\la_1+(N^2-1)\la_2] l_1^d (\eps;\eta_\vp)
 \right. }\nnn
 &-& \ds{\left.
 2^\frac{d}{2}N_c h^2
 l_1^{(F)d}(\eta_\psi) \right\}
 \label{FlowOfEpsilon} }\\[2mm]
 \ds{\frac{\prl\la_1}{\prl t} } &=& \ds{
 (d-4+2\eta_\vp )\la_1
 + 2v_d \left\{
 [2(N^2+4)\la_1^2+(N^2-1)\la_2(2\la_1+\la_2)]l_2^d(\eps;\eta_\vp)
 \right. }\nnn
 &-& \ds{ \left.
 2^\frac{d}{2} \frac{N_c}{N} h^4
 l_{2}^{(F)d} (\eta_\psi)\right\} }\\[2mm]
 \ds{\frac{\prl\la_2}{\prl t} } &=& \ds{
 (d-4+2\eta_\vp )\la_2
 + 2v_d \left\{\left[
 12\la_1\la_2 +2(N^2-3)\la_2^2\right]
 l_2^d (\eps;\eta_\vp) \right. }\nnn
 &-& \ds{ \left.
 2^{\frac{d}{2}+1} \frac{N_c}{N} h^4
 l_{2}^{(F)d}(\eta_\psi)\right\}\; . }
\een
Similarly, the evolution equations for the SSB regime read
\ben
 \ds{ \frac{\prl\kappa}{\prl t} } &=& \ds{
 (2-d-\eta_\vp )\kappa + 2v_d \left\{
 N^2l_1^d(\eta_\vp)
 +3l_1^d (2\la_1 \kappa;\eta_\vp) \right. }\nnn
 &+& \ds{ \left.
 (N^2-1)\left[ 1+\frac{\la_2}{\la_1}\right]
 l_1^d (\la_2\kappa;\eta_\vp)-2^\frac{d}{2}N_c \frac{h^2}{\la_1}
 l_{1}^{(F)d} (\frac{1}{N}h^2 \kappa;\eta_\psi)
 \right\}
 \label{FlowOfKappa} }\\[2mm]
 \ds{\frac{\prl\la_1}{\prl t} } &=& \ds{
 (d-4+2\eta_\vp )\la_1 +2v_d \left\{
 N^2\la_1^2 l_2^d(\eta_\vp)
 +9\la_1^2 l_2^d (2\la_1\kappa;\eta_\vp)
 \right.}\nnn
 &+& \ds{\left.
 (N^2-1)\left[\la_1 +\la_2\right]^2
 l_2^d (\la_2\kappa;\eta_\vp)-2^\frac{d}{2}\frac{N_c}{N}h^4
 l_{2}^{(F)d} (\frac{1}{N}h^2\kappa;\eta_\psi)
 \right\} }\\[2mm]
 \ds{\frac{\prl\la_2}{\prl t} } &=& \ds{
 (d-4+2\eta_\vp )\la_2 +2v_d \left\{
 \frac{N^2}{4}\la_2^2 l_2^d(\eta_\vp) +
 \frac{9}{4}(N^2-4)\la_2^2 l_2^d (\la_2\kappa;\eta_\vp)
 \right. }\nnn
 &-& \ds{
 \hal N^2 \la_2^2
 l_{1,1}^d(0,\la_2\kappa;\eta_\vp) +
 3[\la_2+4\la_1]\la_2
 l_{1,1}^d(2\la_1\kappa,\la_2\kappa;\eta_\vp)
 \label{SSBFlowOfLambda2}
 }\\[2mm]
 &-& \ds{\left.
 2^{\frac{d}{2}+1}\frac{N_c}{N}h^4
 l_{2}^{(F)d} (\frac{1}{N}h^2\kappa;\eta_\psi)\right\}
 }\nonumber
\een
where we have defined
\be
 l_n^d(\eta_\vp)\equiv l_n^d(0;\eta_\vp)\; ,\;\;\;
 l_{n}^{(F)d}(\eta_\psi)\equiv l_{n}^{(F)d}(0;\eta_\psi) \; .
\ee
No explicit dependence on $k$ appears in this scaling form of the flow
equations. In the limit $\eps,\kappa\ra 0$ one recovers for both
regimes to leading order in the coupling constants the known
\cite{CGS93-1,BHJ94-1} perturbative one--loop beta functions for
$\la_1$ and $\la_2$.

The system of flow equations
(\ref{FlowOfEpsilon})--(\ref{SSBFlowOfLambda2}) is the central piece
of this work. For the quark--meson model ($d=4$) we fix the initial
conditions at $k=k_\vp$ ($t=0$) with $\la_1(k_\vp)=\la_2(k_\vp)=0$,
$\eps(k_\vp)=Z_\vp^{-1}(k_\vp)k_\vp^{-2}\olm^2$. The solution of the
evolution equations should then reveal the phenomenon of spontaneous
chiral symmetry breaking for $k\ra 0$ ($t\ra -\infty$). More precisely,
we expect for some scale $k_s>0$ that the mass term vanishes, i.e.
$\eps(k_s)=0$. Subsequently, for $k<k_s$ we follow the evolution of
the minimum of the potential using the system
(\ref{FlowOfKappa})--(\ref{SSBFlowOfLambda2}), with initial condition
$\kappa(k_s)=0$. (The couplings $\la_1$ and $\la_2$ are continuous at
$k_s$.) For sufficiently small values of $k$ the minimum of the
potential will not move anymore, i.e.
\bea
 \ds{\lim_{k\ra 0} Z_\vp(k) }
 &=& \ds{Z_\vp}\nnn
 \ds{ \lim_{k\ra 0}\rho_R(k) }
 &=& \ds{ \lim_{k\ra 0}Z_\vp(k)\rho_0(k)
 =\rho_R=Z_\vp \rho_0 }\nnn
 &=& \ds{ N\si_R^2 = NZ_\vp \si_0^2
 =\frac{N}{4}f_\pi^2 }\nnn
 \ds{ \lim_{k\ra 0}\kappa(k)}
 &\ra& \ds{
 \rho_R k^{-2} \; .}
\eea

Supplementing the flow equations for $\kappa$, $\la_1$ and $\la_2$ by
the one for the Yukawa coupling $h^2$ and inserting the anomalous
dimensions --- these quantities will be computed in the next two
sections --- we can now study how the shape of the average potential
changes as $k$ is lowered. We have integrated the flow equations
numerically for $d=4$ and $N_c=3$ from
$t_i=0$ corresponding to $k=k_\vp$
to $t_f=\ln(m_\pi/k_\vp)$. This endpoint of the numerical integration
simulates the pion mass threshold which is neglected in our
approximation of vanishing quark masses. We ignore here all dependence
of the threshold functions on the anomalous dimensions $\eta_\vp$ and
$\eta_\psi$. This approximations will be justified in section
\ref{Results}. We use first $k_\vp=630\MeV$, $\teps_0=0.01$ and
$Z_\psi(k_\vp)=1$, $h^2(k_\vp)=Z_\vp^{-1}(k_\vp)=300$. For the initial
values of $\la_1$ and $\la_2$ we employ two different sets of boundary
conditions. One corresponds to the approximations used in
\cite{EW94-1}
\be
 \la_1(k_\vp) = \la_2(k_\vp)=0 \; .
 \label{EWBoundaryConditions}
\ee
The other set is obtained in the large--$N_c$ limit of the
$U_L(N)\times U_R(N)$ model \cite{BHJ94-1} and reads
\bea
 \la_1(k_\vp) &=& \ds{\frac{2}{N}h^2(k_\vp)} \nnn
 \la_2(k_\vp) &=& \ds{\frac{4}{N}h^2(k_\vp)} \; .
 \label{LNBoundaryConditions}
\eea

\begin{figure}
\unitlength1.0cm
\begin{picture}(13.,9.)
\put(1.,5.){\bf $\la_{1/2}$}
\put(8.,0.5){\bf $k/\MeV$}
\put(-0.8,-11.5){
\epsfysize=22.cm
\epsffile{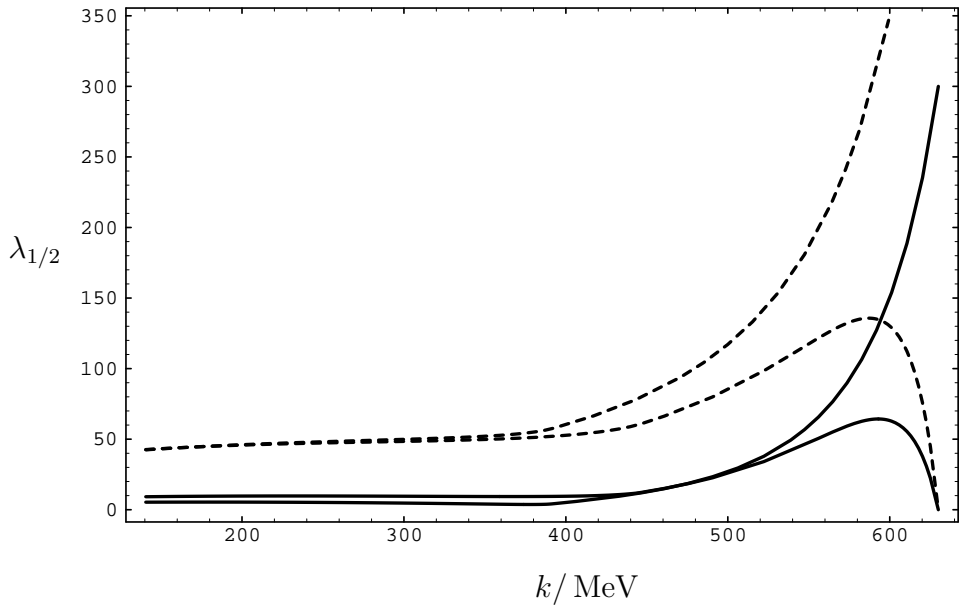}
}
\end{picture}
\caption{\footnotesize Flow of $\la_1$ (solid
  lines) and  $\la_2$ (dashed lines) with $k$,
  for the $U_L(2)\times U_R(2)$ model with two sets
  of initial conditions at $k_\vp=630\MeV$ with $h^2(k_\vp)=300$ and
  $\teps_0=0.01$.}
\label{Fig2}
\end{figure}
For the numerical investigations we will concentrate in the present
work on $N=2$.
In fig. \ref{Fig1} we have plotted the renormalized mass
$m$ as a function of $k$.
Starting from a very large value
$m(k_\vp)=1091\MeV$ the mass rapidly decreases and reaches zero for
$k_s\simeq 450\MeV$. For $k<k_s$ the minimum of the potential occurs
for $\rho_0>0$ and we have to use the flow equations for the SSB
regime. In fig. \ref{Fig1} we also show the $k$--dependence of the
location of the minimum, $\si_R(k)=\left(\rho_R(k)/N\right)^\hal$,
and see how it stabilizes
for $k\lta250\MeV$. The final result for $\si_R$ is rather
insensitive to the exact choice of the endpoint $k_f=m_\pi$.
In fig. \ref{Fig2} we display the $k$--dependence of the quartic
couplings $\la_1$ and $\la_2$ for the two sets of boundary
conditions. We observe that the result for $k=k_f$ does not depend
strongly on the initial values. This is a first manifestation of the
infrared stability mentioned in the introduction. We will explain the
origin of this behavior in more detail later (section \ref{Results}),
since for an understanding we first need to discuss the anomalous
dimensions and the running of $h^2$.

\sect{Scalar anomalous dimension}
\label{ScalarAnomalousDimension}

A computation of $f_\pi$ in terms of the four--quark interaction at
the scale $k_\vp$ requires the ratio $Z_\vp(0)/Z_\vp(k_\vp)$.
An evaluation of this ratio is the subject of this section. We will
begin with the determination of the flow equation for the
momentum dependent scalar wave function
renormalization $Z_{\vp ,k} (q)$ or, equivalently, the scalar
anomalous dimension $\eta_{\vp ,k} (q)\equiv -Z_{\vp ,k}^{-1}(q)\prl_t
Z_{\vp ,k}(q)$.
For this purpose we have to consider
a spatially varying scalar field configuration. We choose a nondiagonal
distortion of the constant vacuum configuration (\ref{Minimum}):
\be
 \vp_{ab}(x) = \vp\dt_{ab} +
 \left[\dt\vp e^{-iQx} + \dt\vp^* e^{iQx}\right] \Si_{ab}
 =\vp_{ab}^*(x)
 \label{ConfAnDi}
\ee
with
\be
 \Si_{ab}=\dt_{a1}\dt_{b2}-\dt_{a2}\dt_{b1}
\ee
or, in momentum space (with $\dt(q_1,q_2)=(2\pi)^d\dt(q_1-q_2)$),
\bea
 \ds{\vp_{ab}(q) }
 &=& \ds{ \vp\dt(q,0)\dt_{ab}+
 \left[\dt\vp\,\dt(q,Q) + \dt\vp^*\,\dt(q,-Q)\right]\Si_{ab} }\nnn
 &\equiv& \ds{\vp\dt(q,0)\dt_{ab} + \Dt(q,Q)\Si_{ab} }
\eea
Expanding around $\vp$ at the potential minimum, we observe that
$\dt\vp$ corresponds to an excitation of the
massless charged pion $\pi^\pm$.
We keep the discussion here for general $\vp$.
If we supplement the scalar configuration by a fermionic background
\be
 \psi_\al = \olpsi_\al =0
\ee
$\Gm^{(2)}_k$ is easily seen to be
block--diagonal. It decays into matrices acting in scalar and fermion
subspaces, $\Gm_{Sk}^{(2)}$ and $\Gm_{Fk}^{(2)}$,
respectively. Hence, we can read off from (\ref{ERGE})
\bea
 \ds{
 \frac{\prl}{\prl t} Z_{\vp ,k}(Q) }
 &=& \ds{
 \frac{1}{4Q^2}\left(
 \lim_{\dt\vp ,\dt\vp^* \ra 0}
 \frac{\prl}{\prl (\dt\vp \dt\vp^*)}
 \left\{\hal \Tr \left[\left(\Gm^{(2)}_{Sk}+R_k\right)^{-1}
 \frac{\prl R_k}{\prl t}\right] \right.\right. } \nnn
 &-& \ds{ \left.\left.
 \Tr\left[\left(\Gm^{(2)}_{Fk}+R_{Fk}\right)^{-1}
 \frac{\prl R_{Fk}}{\prl t}\right]\right\}
 - \left( Q\ra 0\right) \right) }
 \label{Zvpfull}
\eea
The right hand side may be
evaluated by expanding the traces in powers of $\dt\vp$ and
$\dt\vp^*$ up to order $\dt\vp\dt\vp^*$
and subtracting all $Q$--independent
terms. The flow equations for the renormalization constant $Z_\vp (k)$
or equivalently $\eta_\vp (k)$ are now determined as
\be
 \frac{\prl}{\prl t} Z_\vp(k) =
 \lim_{Q^2 \ra 0}
 \frac{\prl}{\prl t} Z_{\vp ,k}(Q) \; ,\;\;\;
 \eta_\vp (k)=\lim_{Q^2 \ra 0}\eta_{\vp ,k}(Q)
 =-\prlt\ln Z_\vp(k) \;.
\ee
In order to evaluate the right hand side we split
\be
 \Gm_{k}^{(2)} = \Gm_{k,0}^{(2)} +
 \Dt \Gm_{k}^{(2)}
\ee
such that all $\dt\vp$, $\dt\vp^*$ dependence is contained in $\Dt
\Gm_{k}^{(2)}$. We may then expand
\bea
 \ds{ \Tr\left[\left(\Gm_{k}^{(2)}+R_k\right)^{-1}
 \frac{\prl R_k}{\prl t}\right] }
 &=& \ds{
 \Tr\left[\left(\Gm_{k,0}^{(2)}+R_k\right)^{-1}
 \frac{\prl R_k}{\prl t}\right] }\nnn
 &+& \ds{
 \Tr\left[\widehat{\frac{\prl}{\prl t}}
 \left\{
 \left(\Gm_{k,0}^{(2)}+R_k\right)^{-1}
 \Dt \Gm_{k}^{(2)} \right\}\right]} \\[2mm]
 &-& \ds{
 \hal\Tr\left[\widehat{\frac{\prl}{\prl t}}
 \left\{
 \left(\Gm_{k,0}^{(2)}+R_k\right)^{-1}
 \Dt \Gm_{k}^{(2)}
 \left(\Gm_{k,0}^{(2)}+R_k\right)^{-1}
 \Dt \Gm_{k}^{(2)} \right\}\right] }\nnn
 &+& \ds{ {\cal O}(\Dt^3) }
 \label{TraceExp}
\eea
and compute the right hand side of (\ref{Zvpfull}) from the first terms.
Details of the calculation can be found in appendix
\ref{ScalarWaveFunctionRenormalization}. Taking the limit $Q^2\ra 0$
and neglecting all momentum
dependence of the Yukawa coupling and wave function
renormalizations we find for the SSB regime
\bea
 \ds{ \eta_\vp} &=& \ds{
 8\frac{v_d}{d}\kappa\left\{
 2\la_1^2 m_{2,2}^d(0,2\kappa\la_1;\eta_\vp)+
 \frac{N^2-2}{4}\la_2^2 m_{2,2}^d(0,\kappa\la_2;\eta_\vp)
 \right\} }\nnn
 &+& \ds{
 2^{\frac{d}{2}+2} \frac{v_d}{d} N_c h^2
 m_4^{(F)d} (\frac{1}{N}\kappa h^2; \eta_\psi ) }
 \label{EtaVpBR}
\eea
and for the symmetric regime
\be
 \eta_\vp =2^{\frac{d}{2}+2} \frac{v_d}{d} N_c h^2
 m_4^{(F)d} (0; \eta_\psi )\; .
 \label{EtaVpSR}
\ee
Here we have defined the threshold functions
\bea
 \ds{m_{n_1,n_2}^d (w_1,w_2;\eta_\vp) } &\equiv& \ds{
 m_{n_1,n_2}^d (w_1,w_2) - \eta_\vp
 \hat{m}_{n_1,n_2}^d (w_1,w_2) }\nnn
 && \ds{ \hspace{-1cm}
 = -\hal k^{2(n_1+n_2-1)-d}
 \int_0^\infty dx\, x^{\frac{d}{2}}
 \widehat{\frac{\prl}{\prl t}} \left\{
 \frac{\dot{P} (x)}
 {[P(x)+k^2 w_1]^{n_1} }
 \frac{\dot{P} (x)}
 {[P(x)+k^2 w_2]^{n_2}} \right\} }
 \label{mn1n2d}
\eea
and
\bea
 \ds{m_4^{(F)d} (w; \eta_\psi )}
 &=& \ds{
 m_4^{(F)d} (w)-\eta_\psi \check{m}_4^{(F)d} (w) }\nnn
 &=& \ds{
 -\hal k^{4-d}
 \int_0^\infty dx\, x^{\frac{d}{2}+1}
 \widehat{\prlt} \left(
 \frac{\prl}{\prl x}
 \frac{1+r_F(x)}{P_F(x)+k^2w}\right)^2 }
 \label{m4Fd}
\eea
where
\be
  x= q^2\; ,\;\; P(x)\equiv P(q)\; ,\;\;
 \dot{P}(x)\equiv\frac{\prl}{\prl x}P(x)\; ,\;\;
 \widehat{\prlt}\dot{P}\equiv\frac{\prl}{\prl x}\widehat{\prlt}P\; .
\ee

{}From the anomalous dimension $\eta_\vp(k)$ the wave function
renormalization can be obtained by numerical integration of
(\ref{AnoDimensions}). We have plotted the result in fig.
\ref{Fig3} for $k_\vp=630\MeV$, $\teps_0=0.01$,
$\la_1(k_\vp)=\la_2(k_\vp)=0$ and two values of $h^2(k_\vp)=300$ and
$10^4$.
\begin{figure}
\unitlength1.0cm
\begin{picture}(13.,9.)
\put(1.5,5.){\bf $Z_\vp$}
\put(8.,0.5){\bf $k/\MeV$}
\put(-0.8,-11.5){
\epsfysize=22.cm
\epsffile{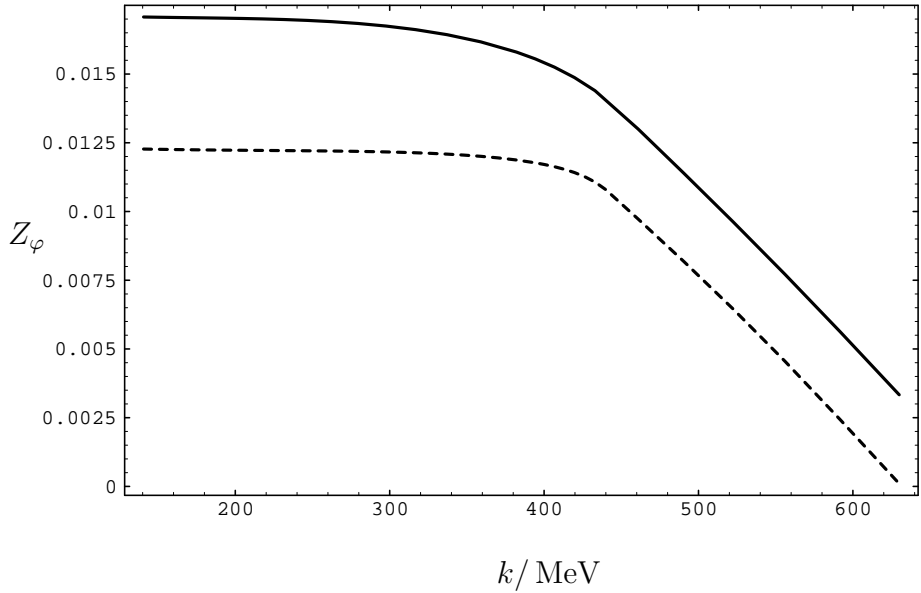}
}
\end{picture}
\caption{\footnotesize Evolution of $Z_\vp$ with $k$, for the
  $U_L(2)\times U_R(2)$ model with
  $k_\vp=630\MeV$, $\teps_0=0.01$,
  $\la_1(k_\vp)=\la_2(k_\vp)=0$ and
  two values $h^2(k_\vp)=300$ (solid line) and $h^2(k_\vp)=10^4$ (dashed
  line).}
\label{Fig3}
\end{figure}
We note that $Z_\vp$ increases strongly in the symmetric regime for
$k>450\MeV$ and stabilizes for low values of $k$. Again, the final
value at $k=k_f$ does not depend very much on the initial conditions
for $h^2$.

\sect{Evolution equation for the Yukawa coupling and fermion anomalous
dimension}
\label{EvolutionEquationForH}

To determine the evolution equation for the Yukawa coupling and the
fermionic wave function renormalization constant we will turn to a
field configuration ($\vp=\vp^*$)
\bea
 \ds{\vp_{ab}(x)} &=& \ds{\vp\dt_{ab}}\nnn
 \ds{\psi_a^{\hat{\al}}(x)} &=& \ds{\psi_a^{\hat{\al}}e^{-iQx}}\nnn
 \ds{\olpsi^a_{\hat{\al}}(x)} &=& \ds{\olpsi^a_{\hat{\al}}e^{iQx}} \; .
\eea
Furthermore, we will approximate the momentum dependence of the Yukawa
coupling by $\olh_k(-q,q^\prime)\simeq\olh_k(\frac{q+q^\prime}{2})$.
This amounts to neglecting its dependence on the external scalar
momentum $\frac{q-q^\prime}{2}$ in the Yukawa vertex in
(\ref{EffActAnsatz}). Accordingly, the renormalized Yukawa coupling is
defined via
\be
 h_k(q)=k^{\frac{d}{2}-2}Z_{\psi,k}^{-1}(q)
 Z_{\vp,k}^{-\hal}(0)\olh_k(q)\; .
 \label{YukRen}
\ee
The matrix of second functional derivatives of $\Gm_k$ simplifies
considerably for the above configuration. Omitting spinor indices
one finds
\be
 \frac{\dt^2\Gm_k}
 {\dt\olpsi^a (q)\dt\psi_b (q^\prime )}
 =\left( Z_{\psi ,k}(q)\slash{q}
 +\olh_k(q)\vp\olgm \right)
 (2\pi)^d\dt(q-q^\prime)\dt_a^b \; .
 \label{FFF}
\ee
The derivation of the flow equations for $\olh$ and $Z_\psi$
follows similar lines as for the
scalar anomalous dimension discussed in section
\ref{ScalarAnomalousDimension}. For details of the calculation we
refer to appendix \ref{FermionWaveFunctionRenormalization}.
Neglecting the effects of the chiral anomaly ($\olnu=0$) as well as
the momentum dependence of the wave function renormalizations and the
Yukawa coupling we find in the limit $Q\ra0$
\bea
 \ds{ \prlt h^2}
 &=& \ds{ \left[ d-4+2\eta_\psi +\eta_\vp\right] h^2
 -\frac{4}{N}v_d h^4 \left\{
 N^2l_{1,1}^{(FB)d} (\frac{1}{N}\kappa h^2,\eps;
 \eta_\psi,\eta_\vp) \right. }\nnn
 &-& \ds{\left.
 (N^2-1)l_{1,1}^{(FB)d} (\frac{1}{N}\kappa h^2,\eps+\kappa\la_2;
 \eta_\psi,\eta_\vp)
 -l_{1,1}^{(FB)d} (\frac{1}{N}\kappa h^2,\eps+2\kappa\la_1;
 \eta_\psi,\eta_\vp)
 \right\} }
 \label{RunningOfh2}
\eea
where
\bea
 \ds{ l_{n_1,n_2}^{(FB)d}(w_1,w_2;\eta_\psi,\eta_\vp) }
 &=& \ds{
 l_{n_1,n_2}^{(FB)d}(w_1,w_2)
 -\eta_\psi \check{l}_{n_1,n_2}^{(FB)d}(w_1,w_2)
 -\eta_\vp \hat{l}_{n_1,n_2}^{(FB)d}(w_1,w_2) }\nnn
 && \ds{ \hspace{-2cm}
 = -\hal k^{2(n_1+n_2)-d}
 \int_0^\infty dx\, x^{\frac{d}{2}-1}
 \widehat{\prlt}\left\{
 \frac{1}{[P_F(x)+k^2w_1]^{n_1} [P(x)+k^2w_2]^{n_2} } \right\} }\; .
 \label{ln1n2FBd}
\eea
Similarly, the fermionic anomalous dimension reads
\bea
 \ds{\eta_\psi }
 &=& \ds{
 \frac{4}{N}\frac{v_d}{d} h^2 \left\{
 N^2m_{1,2}^{(FB)d}(\frac{1}{N}h^2\kappa,\eps;\eta_\psi,\eta_\vp)
 +m_{1,2}^{(FB)d}(\frac{1}{N}h^2\kappa,\eps+2\kappa\la_1;
 \eta_\psi,\eta_\vp) \right. }\nnn
 &+& \ds{ \left.
 (N^2-1)m_{1,2}^{(FB)d}(\frac{1}{N}h^2\kappa,\eps+\kappa\la_2;
 \eta_\psi,\eta_\vp) \right\} }
 \label{EtaPsi}
\eea
with
\bea
 \ds{m_{n_1,n_2}^{(FB)d}(w_1,w_2; \eta_\psi,\eta_\vp)}
 &=& \ds{
 m_{n_1,n_2}^{(FB)d}(w_1,w_2)
 -\eta_\psi \check{m}_{n_1,n_2}^{(FB)d}(w_1,w_2)
 -\eta_\vp \hat{m}_{n_1,n_2}^{(FB)d}(w_1,w_2) }\nnn
 && \ds{ \hspace{-2.5cm}
 = -\hal k^{2(n_1+n_2-1)-d}
 \int_0^\infty dx\, x^{\frac{d}{2}}
 \widehat{\prlt}\left\{
 \frac{1+r_F(x)}{[P_F(x)+k^2w_1]^{n_1}}
 \frac{\dot{P}(x)}{[P(x)+k^2w_2]^{n_2}} \right\} \; . }
 \label{mn1n2FBd}
\eea
In summary, the equations (\ref{EtaVpBR}) and
(\ref{EtaPsi}) constitute a linear system for $\eta_\vp$ and
$\eta_\psi$ with solution
\bea
 \ds{\eta_\vp} &=& \ds{
 \frac{A_1(1+A_6)-A_3A_4}{(1+A_2)(1+A_6)-A_3A_5} }\nnn
 \ds{\eta_\psi} &=& \ds{
 \frac{A_4(1+A_2)-A_1A_5}{(1+A_2)(1+A_6)-A_3A_5} }
 \label{EtaSolution}
\eea
where $A_1,\ldots,A_6$ are defined by writing (\ref{EtaVpBR})
and (\ref{EtaPsi}) in an obvious notation as
\bea
 \ds{\eta_\vp} &=& \ds{ A_1-A_2\eta_\vp -A_3\eta_\psi }\nnn
 \ds{\eta_\psi} &=& \ds{
 A_4-A_5\eta_\vp-A_6\eta_\psi} \; .
 \label{DefinitionOfAs}
\eea

We note that in four dimensions the integrals
\be
 l_{1,1}^{(FB)4}(0,0)=m_4^{(F)4}(0)=m_{1,2}^{(FB)4}(0,0)=1
\ee
are independent of the particular choice of the infrared cutoff. We
therefore find in the limit of small masses $\kappa$, $\eps$
in both
regimes to leading order in the coupling constants the known
\cite{CGS93-1,BHJ94-1} perturbative one--loop results for both
anomalous dimensions:
\bea
 \ds{\eta_\vp} &=& \ds{
 \frac{N_c}{8\pi^2}h^2 }\nnn
 \ds{\eta_\psi} &=& \ds{
 \frac{N}{16\pi^2}h^2 } \; .
\eea
This in turn yields the correct perturbative one--loop result
\be
 \prlt h^2 =(2\eta_\psi+\eta_\vp)h^2=
 \frac{N+N_c}{8\pi^2}h^4 \; .
\ee
We observe that for large $h^2$ the running of the Yukawa coupling is
very fast due to the term $\sim h^4$ in the flow equation
(\ref{RunningOfh2}). This explains why different large values of
$h^2(k_\vp)$ lead to very similar results for $h_R^2=h^2(0)$. We
demonstrate this in fig. \ref{Fig4} where two different initial values
$h^2(k_\vp)=300$ and $h^2(k_\vp)=10^4$ are compared.
\begin{figure}
\unitlength1.0cm
\begin{picture}(13.,9.)
\put(1.5,5.){\bf $h^2$}
\put(8.,0.5){\bf $k/\MeV$}
\put(-0.8,-11.5){
\epsfysize=22.cm
\epsffile{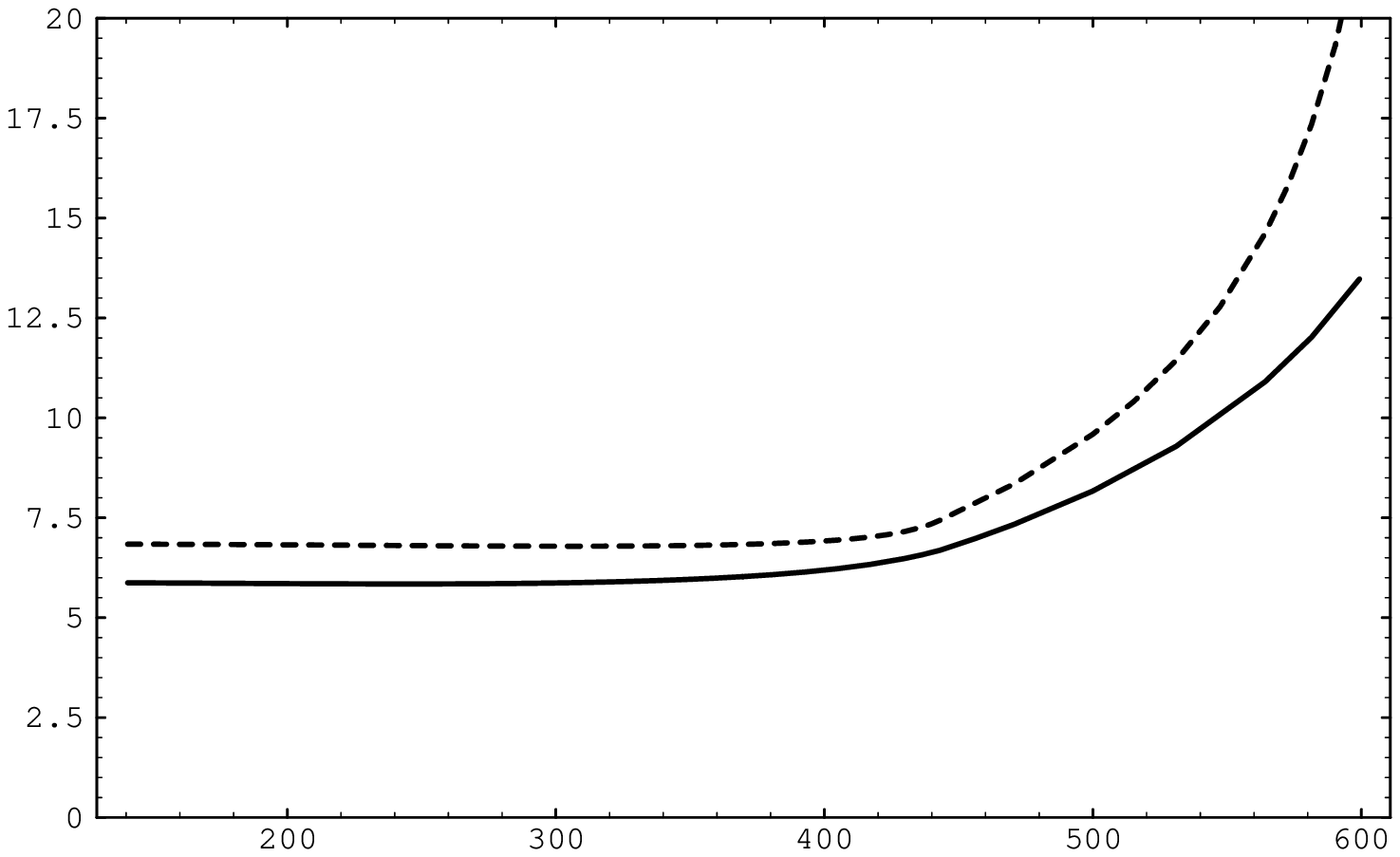}
}
\end{picture}
\caption{\footnotesize Dependence of $h^2$ on $k$,
  for the $U_L(2)\times U_R(2)$ model with two different initial
  values, $h^2(k_\vp)=300$ (solid line)
  and $h^2(k_\vp)=10^4$ (dashed line). We use $k_\vp=630\MeV$,
  $\teps_0=0.01$ and $\la_1(k_\vp)=\la_2(k_\vp)=0$.}
\label{Fig4}
\end{figure}
Fig. \ref{Fig5} shows the corresponding evolution of $Z_\psi(k)$.
\begin{figure}
\unitlength1.0cm
\begin{picture}(13.,9.)
\put(1.5,5.){\bf $Z_\psi$}
\put(8.,0.5){\bf $k/\MeV$}
\put(-0.8,-11.5){
\epsfysize=22.cm
\epsffile{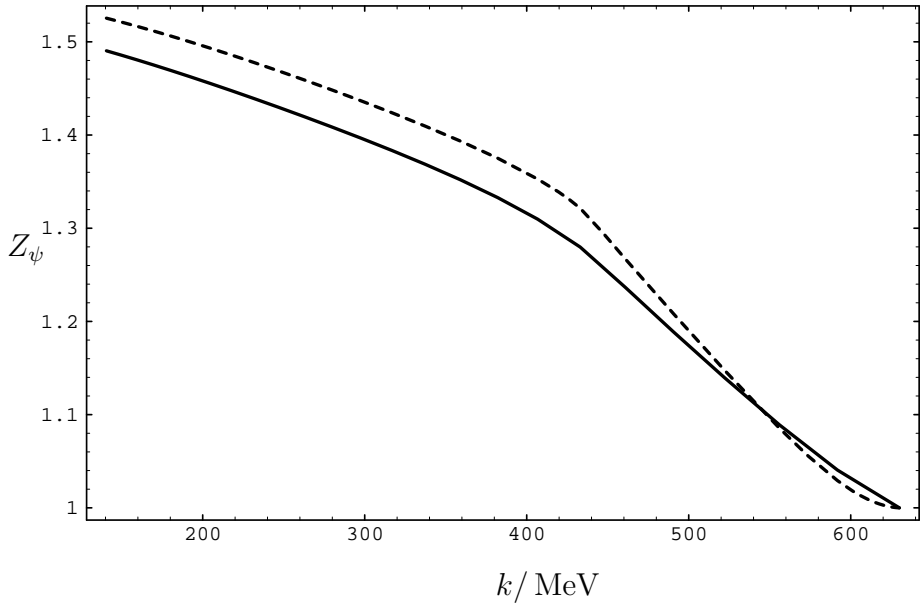}
}
\end{picture}
\caption{\footnotesize Running of $Z_\psi$ with
  $k$, for the $U_L(2)\times U_R(2)$ model with
 two different initial values $h^2(k_\vp)=300$ (solid line),
  $h^2(k_\vp)=10^4$ (dashed line), and $k_\vp=630\MeV$,
  $\teps_0=0.01$, $\la_1(k_\vp)=\la_2(k_\vp)=0$.}
\label{Fig5}
\end{figure}

\sect{The chiral anomaly}
\label{TheO(4)SymmetricSigmaModel}

So far we have considered the somewhat unrealistic limit $\olnu\ra 0$
where the effects of the chiral anomaly are neglected.
In view of the large value of $\nu_R$
(\ref{ValuesForNyR}) as compared
to $k_\vp\simeq630\MeV$, however, it appears that the opposite
limit, $\olnu\ra\infty$, should be closer to reality.
For $N=2$ it is straightforward to take the effects of the chiral
anomaly in this limit into account. To see this we
notice that the complex $({\bf 2},{\bf 2})$ representation $\vp$ of
the global symmetry group $SU_L(2)\times SU_R(2)\simeq O(4)$
decomposes into two irreducible real vector representations of $O(4)$
(cf. the discussion at the end of appendix \ref{ScalarMassSpectrum}):
\be
 \vp=\hal\left( \si-i\eta^\prime\right) +\hal\left(
 a^k+i\pi^k\right)\tau_k \; .
\ee
By taking $\olnu\ra\infty$ while keeping
$\olm^2-\hal\olnu$ (or
$\olmu^2+\hal\olnu$) fixed,
$m_a$ and $m_{\eta^\prime}$
diverge and the four corresponding mesons decouple.
Hence, we are left with the real
vector representation $\vec{\phi}=(\si,\pi_1,\pi_2,\pi_3)$ of
$O(4)$. Its potential reads in the symmetric regime
\be
 U_k=\hal(\olm^2-\hal\olnu)\phi_a\phi^a
 +\frac{1}{8}\la_1\left(\phi_a\phi^a\right)^2
\ee
and similarly in the SSB regime.
We therefore end up with the $O(4)$ symmetric linear sigma model
coupled to fermions. The flow equations in the
symmetric regime are given by \cite{Wet93-1,BW93-1}
\bea
 \ds{\frac{\prl\eps}{\prl t} }
 &=& \ds{
 -(2-\eta_\vp)\eps-2v_d\left\{6\la_1
 l_1^d(\eps;\eta_\vp)
 -2^{\frac{d}{2}}N_c h^2 l_1^{(F)d}(\eta_\psi)\right\} } \nnn
 \ds{\frac{\prl\la_1}{\prl t} }
 &=& \ds{
 (d-4+2\eta_\vp)\la_1+2v_d\left\{
 12\la_1^2 l_2^d(\eps;\eta_\vp)
 -2^{\frac{d}{2}-1}N_c h^4l_2^{(F)d}(\eta_\psi) \right\} }\nnn
 \ds{\frac{\prl h^2}{\prl t}}
 &=& \ds{
 (d-4+2\eta_\psi+\eta_\vp)h^2
 -4v_dh^4l_{1,1}^{(FB)d}(0,\eps;\eta_\psi,\eta_\vp) }\nnn
 \eta_\vp &=& \ds{
 2^{\frac{d}{2}+2}\frac{v_d}{d}N_c h^2
 m_4^{(F)d}(0;\eta_\psi) }\nnn
 \eta_\psi &=& \ds{
 8\frac{v_d}{d}h^2
 m_{1,2}^{(FB)d}(0,\eps;\eta_\psi,\eta_\vp) }
\eea
where $\eps$ is defined here by $\eps=Z_\vp^{-1}k^{-2}(\olm^2-\hal\olnu)$.
For the SSB regime we find
\ben
 \ds{\frac{\prl\kappa}{\prl t} }
 &=& \ds{
 (2-d-\eta_\vp)\kappa+2v_d\left\{
 3l_1^d(\eta_\vp)+3l_1^d(2\la_1\kappa;\eta_\vp)
 -2^{\frac{d}{2}}N_c \frac{h^2}{\la_1}
 l_1^{(F)d}(\hal h^2\kappa;\eta_\psi)\right\} }\nnn
 \ds{\frac{\prl\la_1}{\prl t} }
 &=& \ds{
 (d-4+2\eta_\vp)\la_1 }\nnn
 &+& \ds{
 2v_d\left\{
 3\la_1^2 l_2^d(\eta_\vp)
 +9\la_1^2 l_2^d(2\la_1\kappa;\eta_\vp)
 -2^{\frac{d}{2}-1}N_c h^4
 l_2^{(F)d}(\hal h^2\kappa;\eta_\psi) \right\} }\nnn
 \ds{\frac{\prl h^2}{\prl t}}
 &=& \ds{
 (d-4+2\eta_\psi+\eta_\vp)h^2 }\\[2mm]
 &-& \ds{
 2v_d h^4 \left\{
 3l_{1,1}^{(FB)d}(\hal h^2\kappa,0;\eta_\psi,\eta_\vp)
 -l_{1,1}^{(FB)d}(\hal h^2\kappa,2\la_1\kappa;
 \eta_\psi,\eta_\vp)\right\} }\nnn
 \eta_\vp &=& \ds{
 4\frac{v_d}{d}\left\{
 4\kappa\la_1^2 m_{2,2}^d(0,2\la_1\kappa;\eta_\vp)
 +2^{\frac{d}{2}}N_c h^2
 m_4^{(F)d}(\hal h^2\kappa;\eta_\psi) \right\} }\nnn
 \eta_\psi &=& \ds{
 2\frac{v_d}{d}h^2 \left\{
 3m_{1,2}^{(FB)d}(\hal h^2\kappa,0;\eta_\psi,\eta_\vp)
 +m_{1,2}^{(FB)d}(\hal h^2\kappa,2\la_1\kappa;
 \eta_\psi,\eta_\vp)\right\} }\; . \nonumber
\een

The difference between the results of the $O(4)$ model
($\olnu\ra\infty$) and the $U_L(2)\times U_R(2)$ model ($\olnu=0$)
can be taken as a measure for the uncertainty due to the rough
treatment of the anomaly in the present work. Here the $U_L(2)\times
U_R(2)$ model exhibits the effects of additional scalar degrees of
freedom beyond the pions (and the $\si$--mode). Since the additional
modes are relatively heavy, the $O(4)$ model should be closer to
reality. The best model with $N=3$, where the strange quark mass and
the chiral anomaly are properly taken into account, is expected to
deviate from the $O(4)$ model in the same direction as the
$U_L(2)\times U_R(2)$ model.
Since there are three light
quark flavors in nature with masses
smaller than $k_\vp$ one might naively expect the case $N=3$ to
correspond to the
most realistic description of the real world.
However, we have
neglected quark masses and in particular the strange quark mass in this
work. In the $SU_L(3)\times SU_R(3)\times U_V(1)$ model the four
$K$--mesons will therefore appear as massless
Goldstone degrees of freedom which will unnaturally
drive the evolution of all parameters even at scales much lower than
their physical masses of approximately $500\MeV$. The same holds, of
course, for the three pions. The effects of their physical masses
$m_\pi\simeq140\MeV$ can, however, be mimicked by stopping the
running for $k_f=m_\pi$. We therefore expect the case $N=2$ to yield
more realistic results than $N=3$ as long as the strange quark mass is
neglected. In addition, we note that for $N=3$ the scalar self coupling
$\la_1$ turns negative for positive but small values of $\olm^2$. This
happens despite the fact that $\la_1$ has acquired first a large
positive value due to the strong initial Yukawa
coupling. The  cause is a large value of $\la_2$ which
can drive $\la_1$ negative when
the scalar loop contributions to the running of $\la_1$ become
numerically important around the scale $k_s$. We interpret this
quartic instability of the
truncation (\ref{PotentialSymRegime}) of the effective potential as a
signal for a first order phase transition in the mass parameter
even within the $U_L(3)\times U_R(3)$ model without chiral anomaly. A
proper treatment of the case $N=3$ therefore requires a more general
truncation of the effective potential \cite{BTW95-1}. We will leave
this problem for future
work. However, we would like to point out that the
inclusion of the cubic (for $N=3$)
$\xi$--term into the potential will change the phase transition to
first order anyway. One expects that the vacuum
expectation value of $\vp$ jumps discontinuously to a finite value
already for a large value of $\eps$ such that the scalar
contributions to
the evolution of $\la_1$ never become strong enough to turn it
negative.

The difference between the $O(4)$ model and the $U_L(2)\times U_R(2)$
model is exemplified in fig. \ref{Fig6}.
\begin{figure}
\unitlength1.0cm
\begin{picture}(13.,9.)
\put(1.,5.){\bf $\ds{\frac{f_\pi}{\MeV}}$}
\put(8.,0.5){\bf $k_\vp$}
\put(8.,4.){\bf $O(4)$}
\put(7.,6.5){\bf $U_L(2)\times U_R(2)$}
\put(-0.8,-11.5){
\epsfysize=22.cm
\epsffile{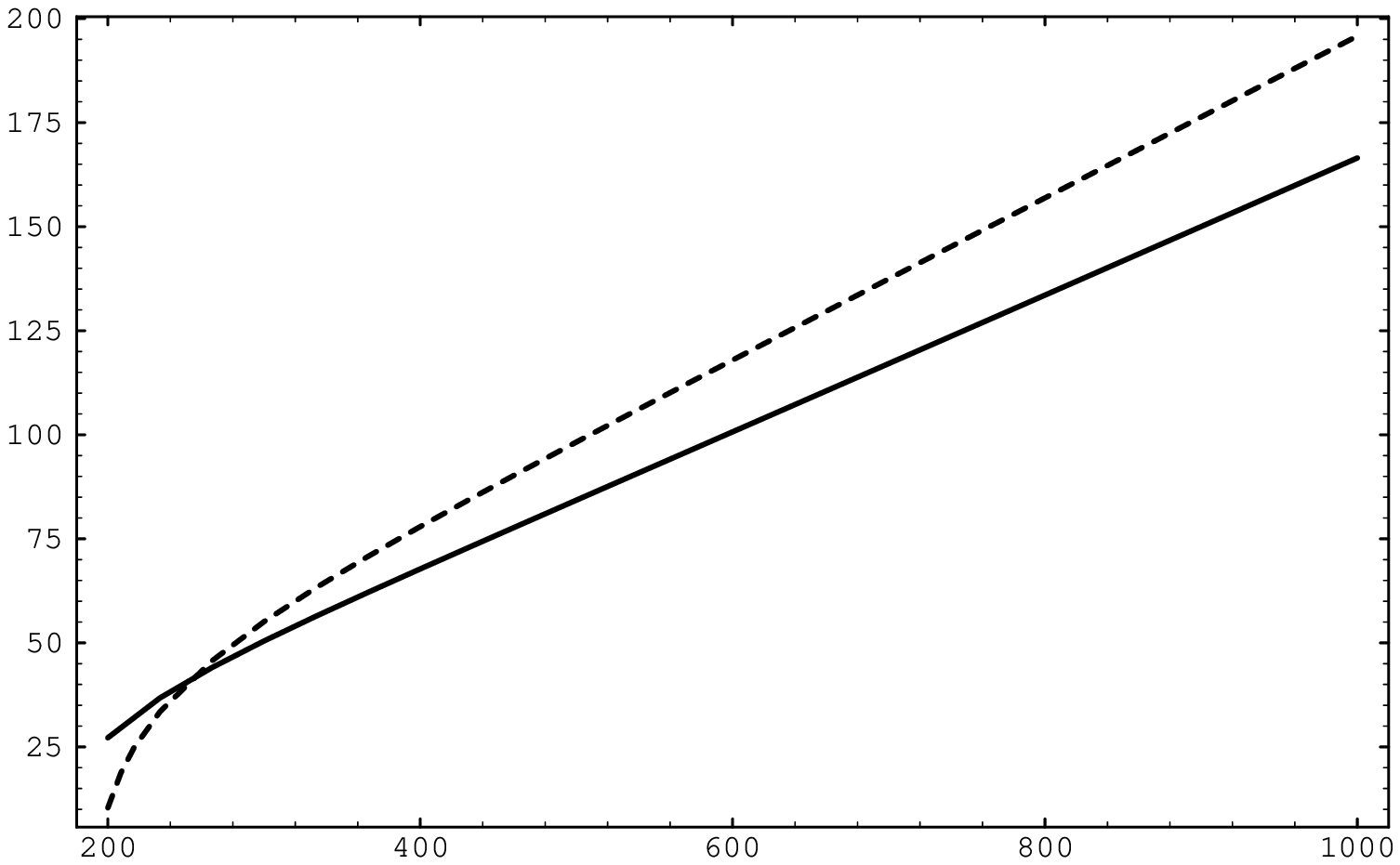}
}
\end{picture}
\caption{\footnotesize The pion decay constant
  $f_\pi$ as a function
  of $k_\vp$ for $\la_1(k_\vp)=\la_2(k_\vp)=0$,
  $h^2(k_\vp)=300$ and $\teps_0=0.01$.}
\label{Fig6}
\end{figure}
There we show the ``prediction'' of $f_\pi$ as a function of the
initial scale $k_\vp$. The difference between the two models is not
very large. We observe for not too small values of $k_\vp$ a linear
behavior $f_\pi\sim k_\vp$. For fixed initial values of the
dimensionless
parameters like $\teps=\olm^2Z_\psi^2/k_\vp^2$, $h^2$ etc., this
proportionality follows on pure dimensional grounds if no other mass
scale is present. The bending of the curves for small $k_\vp$ is
therefore purely a consequence of the additional infrared cutoff
$k_f=m_\pi$ which obviously plays a negligible role for a realistic
size of $k_\vp$. Looking at fig. \ref{Fig6} the reader may prematurely
conclude that the predictive power for $f_\pi$ is severely limited by
the arbitrariness of the choice of the scale $k_\vp$ from where on the
meson description is used. We should emphasize that for a full
treatment along the lines of \cite{EW94-1,Wet95-1} this is
actually not the case: If one lowers the transition scale $k_\vp$ more
fluctuations are included in the momentum range $q^2>k_\vp^2$ where a
quark--gluon picture is used. As a result, the pole like structure
in the effective four--quark interaction becomes stronger and $\olm^2$
therefore decreases. This is the same effect as found in the
quark--meson picture used for the fluctuations with $q^2<k_\vp^2$: the
mass term $\olm^2$ decreases with smaller $k$ as a result of the
Yukawa coupling to the quarks. In the limit where the pole like
structure dominates the evolution of $\tilde{G}(0)=\frac{1}{2\olm^2}$
in the quark gluon picture the running of $\olm^2$ is identical in
both pictures for $k$ larger or smaller than $k_\vp$. The initial
value $\olm^2(k_\vp)$ as a function of
$k_\vp$ follows therefore the same renormalization group trajectory as
given by the flow equation (\ref{BosEvolMSymReg}). In this ideal case
the choice of the transition scale $k_\vp$ does not affect the
results, since the initial conditions move on trajectories of constant
physics. In practice, this ideal scenario will often not be fully
realized, since different types of fluctuations are included in the
quark--gluon and the quark--meson description. The dependence of the
results on $k_\vp$ can then be used as a quantitative check of the
reliability of the employed truncations for the effective action.

\sect{Infrared stability and predictive power for $f_\pi$}
\label{Results}

Comparing the results for $f_\pi$ from fig. \ref{Fig6} with the
experimental value $f_\pi=93\MeV$ we find a surprisingly good
agreement for $k_\vp=630\MeV$ as infered from ref. \cite{EW94-1}. The
question arises to what extent this result depends on the particular
choice of initial values at the scale $k_\vp$. In principle, the
values of the parameters of the quark--meson system at the scale
$k_\vp$ can be computed from QCD \cite{EW94-1,Wet95-1}. In practice,
however, many quantities will not be available with high accuracy,
since one has to deal with a problem involving strong interactions. If
$f_\pi$ would depend very sensitively on such quantities, a
computation of $f_\pi$ with satisfactory precision
would be extremely difficult. We will argue in this section that
for small enough $Z_\vp$ the opposite situation is realized. In this
event the prediction for $f_\pi$ turns out to be almost independent
of the initial values of many couplings. The reason is that a small
$Z_\vp$ corresponds to a strong Yukawa interaction.
The large value of $h$
induces then a very fast running of almost all couplings towards
values which are determined by an infrared attractive behavior. More
precisely, the ratios $\la_1/h^2$ and $\la_2/h^2$ are determined by
infrared fixed points of the type first found in the electroweak
standard model \cite{Wet81-1}. This explains the insensitivity with respect
to the initial values as demonstrated in fig. \ref{Fig2}. The Yukawa
coupling itself is also strongly renormalized and predicted to be in
the vicinity of the upper bound of the relevant infrared
interval\footnote{The infrared fixed point for $h$ is $h_*=0$ if no
  infrared cutoff is present. Due to a finite amount of running from
  $k_\vp$ to $k_s$ this translates into an infrared interval.}
\cite{Hill81-1}. Here the upper bound of the infrared interval is
essentially determined by the scale $k_s$ where spontaneous symmetry
breaking sets in and the effective quark masses constitute an
infrared cutoff. This can clearly be observed in fig. \ref{Fig4} where
also the insensitivity with respect to the initial value $h(k_\vp)$
becomes apparent. The only relevant parameter will turn out to be the
ratio
\be
 \teps_0=\frac{\eps(k_\vp)}{h^2(k_\vp)}=
 \frac{\olm^2(k_\vp)Z_\psi^2(k_\vp)}{k_\vp^2} \; .
\ee
This value determines $k_s$ and $f_\pi$ as well as all other
couplings at the scale $k_f$.

For small enough $Z_\vp$ one starts in a regime where
$\eps=m^2/k^2=\olm^2Z_\vp^{-1}k^{-2}$ is large.
For large $\eps$ the scalar fluctuations
are suppressed by inverse powers of $\eps$ appearing in the threshold
functions. Then the scalar fluctuations can be neglected and only
quark fluctuations drive the flow of the couplings. This is the
approximation used in ref. \cite{EW94-1} which remains valid as long
as $\eps\gg1$. The Yukawa coupling $\olh$ is normalized according to
(\ref{InitialValues}) as
$\olh(0)\equiv\olh_{k_\vp}(0,0)=1$. Consequently, the initial value
for the renormalized Yukawa coupling is
\be
 h_0^2\equiv
 h^2(k_\vp)=\frac{1}{Z_\vp(k_\vp)Z_\psi^2(k_\vp)}\; .
\ee
We use here a normalization of the fermion kinetic term such that
$Z_\psi(k_\vp)=1$. For small $Z_\vp(k_\vp)$ we therefore start with a
strong Yukawa coupling. In the limit $\eps\gg1$ the flow equations
simplify considerably.
If we define $\teps\equiv\eps/h^2$ and
$\tla_i\equiv\la_i/h^2$ we find
\bea
 \ds{\frac{\prl\teps}{\prl t} } &=& \ds{
 -(d-2)\teps+2^{\frac{d}{2}+1}v_dN_c }\nnn
 \ds{\frac{\prl\tla_1}{\prl t} } &=& \ds{
 2^{\frac{d}{2}+1}v_dN_ch^2
 \left[\frac{2}{d}\tla_1-\frac{1}{N}\right] }\nnn
 \ds{\frac{\prl\tla_2}{\prl t} } &=& \ds{
 2^{\frac{d}{2}+2}v_dN_ch^2
 \left[\frac{1}{d}\tla_2-\frac{1}{N}\right] }\nnn
 \ds{\frac{\prl h^2}{\prl t} } &=& \ds{
 (d-4)h^2+2^{\frac{d}{2}+2}\frac{v_d}{d}N_ch^4 }\; .
 \label{ReducedSystem}
\eea
In the following we will specialize to the case $d=4$.\footnote{The
  system (\ref{ReducedSystem}) remains solvable for general $d$.}

As a first observation we notice that the $\tla_1$--$\tla_2$ system
exhibits an infrared fixed point given by
\be
 \tla_{1*}=\hal\tla_{2*}=\frac{2}{N}\; .
 \label{FixedPoint}
\ee
This fixed point corresponds exactly to the large--$N_c$ estimate of
\cite{BHJ94-1}. The explicit solution of the differential equations
(\ref{ReducedSystem}) reads
\bea
 \ds{\teps(t) } &=& \ds{
 4v_4N_c\left[ 1-e^{-2t}\right]+\teps_0e^{-2t} }\nnn
 \ds{\tla_1(t)} &=& \ds{
 \frac{\frac{\la_{10}}{h_0^2}-8\frac{N_c}{N}v_4h_0^2 t}
 {1-4N_cv_4h_0^2t} }\nnn
 \ds{\tla_2(t)} &=& \ds{
 \frac{\frac{\la_{20}}{h_0^2}-16\frac{N_c}{N}v_4h_0^2 t}
 {1-4N_cv_4h_0^2t} }\nnn
 h^2(t) &=& \ds{
 \frac{h_0^2}{1-4N_cv_4h_0^2t} }
\eea
with $\teps_0\equiv\teps(t=0)$, etc.
The system crosses into the SSB regime when $\teps(t_s)=0$,
corresponding to a scale
\be
 t_s=\hal\ln\left[1-\frac{\teps_0}{4N_cv_4}\right]\; ,\;\;\;
 k_s^2=\left(1-\frac{8\pi^2}{3}\teps_0\right)k_\vp^2 \; .
 \label{ttilde}
\ee
Here we have used $v_4=(32\pi^2)^{-1}$ and $N_c=3$ in the last
expression. Around the scale $k_s$
our approximation (large $\eps$) breaks down.
Nevertheless, it becomes apparent already at this stage that values of
$\teps_0$ substantially larger than $0.04$ are incompatible with
chiral symmetry breaking, since $\teps$ would remain positive for all
$k$ in this case.
For large $\teps_0$ the effect of the quark fluctuations is
simply not strong enough to turn the scalar mass term negative.
We notice that $\teps_0\ll1$
for $\olm(k_\vp)\sim\Oc(100\MeV)$ and $Z_\psi(k_\vp)\lta 1$.
If we furthermore assume $-4N_cv_4h_0^2t_s\gg 1$ or
\be
 \teps_0\gg 4N_cv_4\left[
 1-\exp\{-\frac{1}{2N_cv_4h_0^2}\}\right] \; ,
 \label{tepsCondition}
\ee
we find for not too large $\la_{10}/h_0^2$, $\la_{20}/h_0^2$
\be
 \tla_1(t_s)\simeq\tla_{1*}\; ,\;\;\;
 \tla_2(t_s)\simeq\tla_{2*}\; ,\;\;\;
 h^2(t_s)\simeq\left[2N_cv_4\ln\frac{4N_cv_4}
 {4N_cv_4-\teps_0}\right]^{-1} \; .
 \label{FixedPointH2}
\ee
This result may be interpreted as follows:
Even though (\ref{ttilde}) should only give an approximate
estimate for the
entering point into the SSB regime this is sufficient to imply
that $\tla_1$ and $\tla_2$ approximately reach their fixed
point long before $\eps$ goes through zero provided
(\ref{tepsCondition}) is
fulfilled. Furthermore $h^2$ becomes
asymptotically independent of $h_0^2$ and is only a function of
$\teps_0$. We therefore conclude that for small $Z_\vp(k_\vp)$
the system is governed by an
infrared fixed point of $\tla_1$ and $\tla_2$ in the symmetric regime
and looses almost all its
information on the initial values at $k=k_\vp$. Infrared quantities
of the SSB regime like $f_\pi$ or meson masses will therefore
merely depend on $\teps_0$ for a given $k_\vp$.
It is straightforward to see that the same analysis holds for the
$O(4)$ model discussed in section \ref{TheO(4)SymmetricSigmaModel}.
The approximate flow equations for $\teps$, $\tla_1$ and $h^2$ are the
same as (\ref{ReducedSystem}). The difference between the
$U_L(2)\times U_R(2)$ and the $O(4)$ model arises principally from
the behavior of the running couplings around the scale $k_s$.

For larger values of $Z_\vp(k_\vp)$ or smaller values of $h(k_\vp)$
the attraction of the infrared fixed points becomes weaker. As a
consequence, the dependence of $f_\pi$ on the initial values of the
couplings becomes more important as demonstrated in fig. \ref{Fig7}.
\begin{figure}
\unitlength1.0cm
\begin{picture}(13.,9.)
\put(1.,5.){\bf $\ds{\frac{f_\pi}{\MeV}}$}
\put(8.,0.5){\bf $h_0$}
\put(-0.8,-11.5){
\epsfysize=22.cm
\epsffile{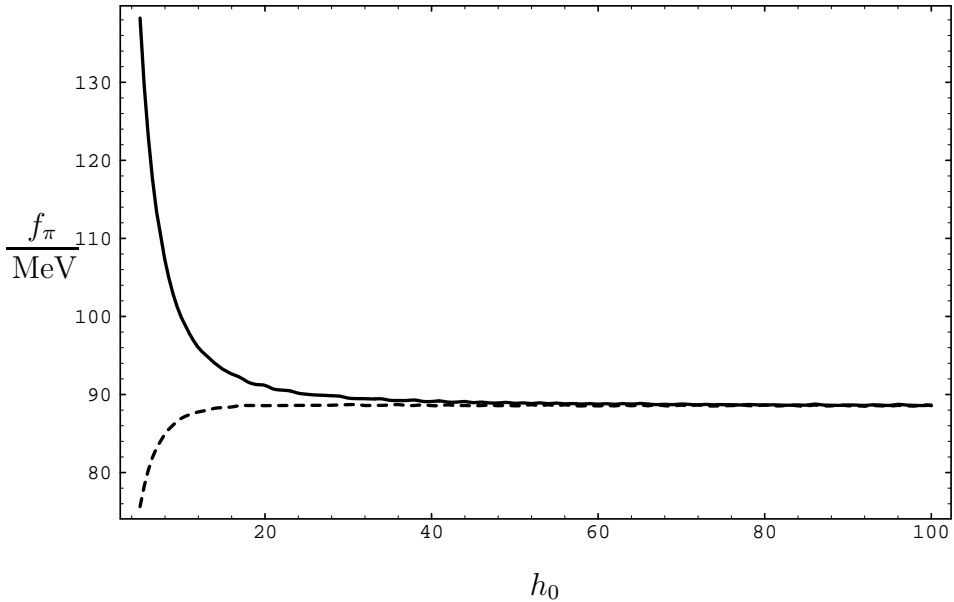}
}
\end{picture}
\caption{\footnotesize The
  $O(4)$ model pion decay constant $f_\pi$
  as a function of $h_0\equiv h(k_\vp)$ for $k_\vp=630\MeV$,
  $\teps_0=0.02$ and
  $\la_1(k_\vp)=0$ (solid line) as well as $\la_1(k_\vp)=h_0^2$
  (dashed line).}
\label{Fig7}
\end{figure}
We conclude that for $h(k_\vp)$ substantially smaller than ten it will
become more and more difficult to obtain an accurate prediction for
$f_\pi$. On the other hand, fig. \ref{Fig7} clearly shows the
approximate independence of $f_\pi$ on $h^2(k_\vp)$ or $\la_1(k_\vp)$
if $h^2(k_\vp)$ exceeds 300.

An additional aspect of strong Yukawa couplings concerns the error in
$f_\pi$ due to the truncations of the quark--meson effective
action. The effects of truncations in the scalar sector are diminished
by the fact that scalar fluctuations are subdominant in the region of
very large Yukawa couplings. A similar argument justifies the
approximation of
neglecting the terms proportional to the anomalous dimensions in
the threshold functions. For the $m$--type functions this
approximation is valid, since
$A_2,A_3,A_5,A_6\ll 1$ in (\ref{EtaSolution}),
(\ref{DefinitionOfAs}). For $l$-type functions one might be
worried that $\eta_\vp$ and $\eta_\psi$ are large for
the initial part of the running in the symmetric regime due to large
values of $h^2$. However, because of the
large values of $\eps$ in this range of scales the contributions from
the $l$--type functions can be neglected altogether. For smaller
values of $\eps$, i.e. for scales closer to $k_s$, the anomalous
dimensions are expected to be already small (as indicated by
figs. (\ref{Fig3}) and (\ref{Fig4})).

\sect{Computation of $f_\pi$ for strong Yukawa coupling}
\label{Discussion}

In the last section we have shown that for large enough Yukawa
couplings, say $h^2(k_\vp)>200$, the value of $f_\pi$ only depends on
the parameter $\teps(k_\vp)\equiv\teps_0$ for given $k_\vp$. We
demonstrate this quantitatively in fig. \ref{Fig8}
\begin{figure}
\unitlength1.0cm
\begin{picture}(13.,9.)
\put(1.,5.){\bf $\ds{\frac{f_\pi}{\MeV}}$}
\put(8.,0.5){\bf $\tilde{\epsilon}_0$}
\put(10.5,5.5){\bf $O(4)$}
\put(5.5,2.5){\bf $U_L(2)\times U_R(2)$}
\put(-0.8,-11.5){
\epsfysize=22.cm
\epsffile{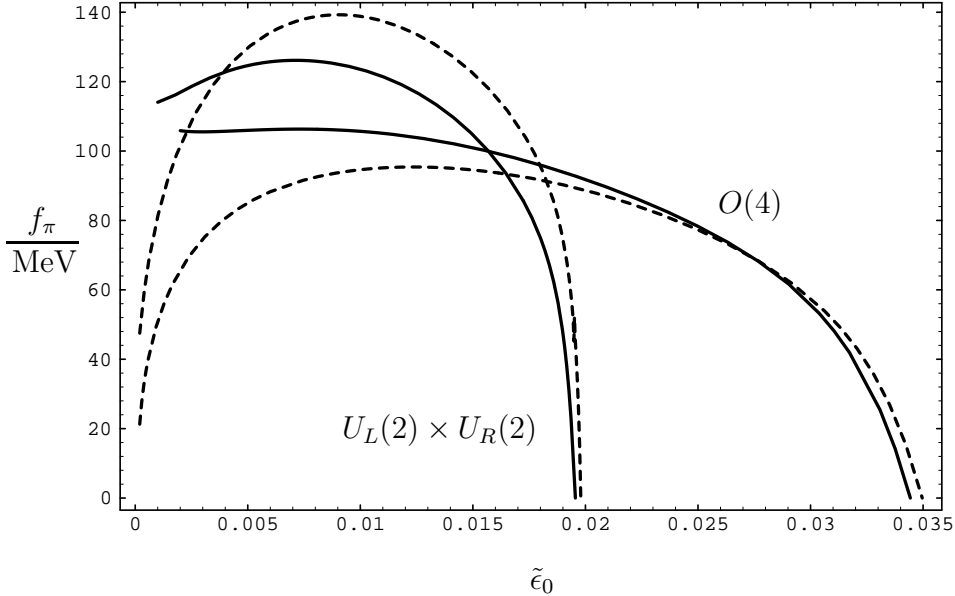}
}
\end{picture}
\caption{\footnotesize The pion decay constant
  $f_\pi$ as a function
  of $\teps_0$ for $k_\vp=630\MeV$, $\la_1(k_\vp)=\la_2(k_\vp)=0$ and
  $h^2(k_\vp)=300$ (solid line) as well as $h^2(k_\vp)=10^4$ (dashed line).}
\label{Fig8}
\end{figure}
where we plot $f_\pi$ as a function of $\teps_0$, for both the
$O(4)$ and the $U_L(2)\times U_R(2)$ model as well as for two
different large initial values of $h^2$. The ``prediction'' for
$f_\pi$ is rather insensitive to $h^2(k_\vp)$ for moderate values of
$\teps_0$.\footnote{The decrease of $f_\pi$ for $h^2(k_\vp)=10^4$
  observed for unnaturally small values of $\teps_0$ results
  simply from the fact that the system enters almost immediately into
  the SSB regime, having no ``time'' for $Z_\vp$ to grow much beyond
  $Z_\vp(k_\vp)$.}
For the $O(4)$ model we also observe an extended plateau where
$f_\pi$ is not very sensitive to $\teps_0$ either.
For this plateau the value
of $f_\pi$ comes out between $80$ and $100\MeV$ which fits very well
with the experimental value of $93\MeV$. Also the renormalized Yukawa
coupling $h_R=h(k_f)$, or, equivalently, the constituent quark mass
$m_q=\hal h_Rf_\pi$ depends essentially only on $\teps_0$. We show
this in fig. \ref{Fig9},
\begin{figure}
\unitlength1.0cm
\begin{picture}(13.,9.)
\put(1.,5.){\bf $h_R$}
\put(8.,0.5){\bf $\tilde{\epsilon}_0$}
\put(8.,3.8){\bf $O(4)$}
\put(2.7,2.){\bf $U_L(2)\times U_R(2)$}
\put(-0.8,-11.5){
\epsfysize=22.cm
\epsffile{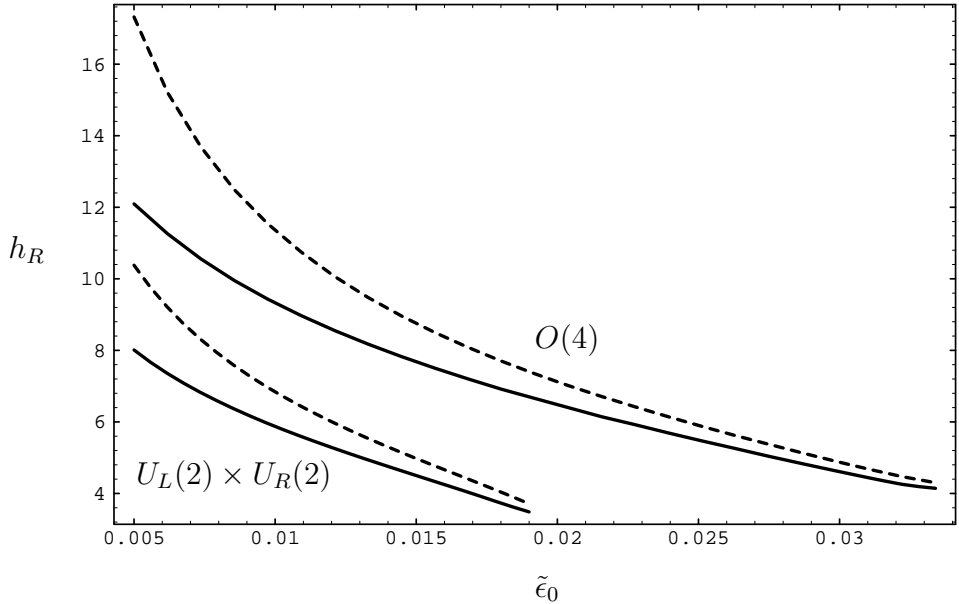}
}
\end{picture}
\caption{\footnotesize The renormalized Yukawa
  coupling $h_R$ as a function of $\teps_0$ for $k_\vp=630\MeV$,
  $\la_1(k_\vp)=\la_2(k_\vp)=0$ and $h^2(k_\vp)=300$ (solid lines),
  $h^2(k_\vp)=10^4$ (dashed lines).}
\label{Fig9}
\end{figure}
again for two different large values of $h^2(k_\vp)$. Since both $h_R$
and $f_\pi$ are functions of only one parameter $\teps_0$,
there arises a
quantitative relation between those two quantities. Consider first the
$O(4)$ model: For a constituent quark mass of $300\MeV$ or
$h_R\simeq6.5$ we read off from fig. \ref{Fig9} that
$\teps_0\simeq0.02$. Inserting this into the plot of fig. \ref{Fig8}
one obtains
\be
 f_\pi\simeq92\MeV \; .
 \label{ResultFpi}
\ee
This value was obtained for $h^2(k_\vp)=300$ and
$\la_1(k_\vp)=\la_2(k_\vp)=0$, but it turns out to be not very
different for $h^2(k_\vp)=10^4$ or different initial values of
$\la_1$ and $\la_2$. A similar procedure gives for the $U_L(2)\times
U_R(2)$ model a value $\teps_0\simeq0.008$ and in turn
$f_\pi\simeq126\MeV$. Repeating this procedure for $k_\vp=700\MeV$ we
can infer from table \ref{tab1}
\begin{table}
\begin{center}
\begin{tabular}{|c|c|c||c|c|c||c|c|c|} \hline
 \multicolumn{2}{|c|}{ } & $\frac{m_q}{\MeV}$\rule[-3mm]{0mm}{8mm} &
 \multicolumn{3}{c||}{300} &
 \multicolumn{3}{c|}{350} \\ \cline{3-9}
 \multicolumn{2}{|c|}{ } &
 $h^2(k_\vp)$ & 64 & 300 & $10^4$ &
 106 & 300 & $10^4$ \\ \hline
 & $\frac{k_\vp}{\MeV}\rule[-3mm]{0mm}{8mm} $
 & & & & & & & \\ \hline\hline
 $O(4)$ & $630$ & & 143.9 & 91.7  & 83.5 & 124.6 & 99.9  & 91.0 \\ \cline{2-9}
        & $700$ & & 159.7 & 101.5 & 92.5
        & 138.2 & 110.7 & 100.7\\ \hline\hline
 $U_L(2)\times U_R(2)$
        & $630$ & & - & 125.7 & 138.3 & - & - & 138.9 \\ \cline{2-9}
        & $700$ & & - & 139.6 & 153.5 & - & - & 154.2 \\ \hline
\end{tabular}
\caption{\footnotesize $f_\pi$ in $\MeV$ for various initial
  conditions at
  $k_\vp$ and two values of the constituent quark mass.}
\label{tab1}
\end{center}
\end{table}
a guess of the uncertainty in $f_\pi$ that can be expected within the
quark--meson model for large Yukawa couplings, as
represented in the table by the values $h^2(k_\vp)=300$ and $10^4$. On
the other hand, a given
value of $m_q$ also implies a minimal value $h_{\rm min}(k_\vp)$
such that the evolution of $h(k)$ can reach the value
$h_R=2m_q/f_\pi$ at all.
We assume here that the result for $\olm^2(k_\vp)$ of ref. \cite{EW94-1}
should not be off by more than a factor of four. We can therefore
conclude that for $k_\vp$ in the range $(630-700)\MeV$ there exists a
lower bound on $\teps_0$, i.e. $\teps_0\gta0.01$.
This in turn amounts for the $O(4)$ model to $h_{\rm min}(k_\vp)\simeq
6.2,8.0,10.3$ for $m_q=250,300,350\MeV$,
respectively. The corresponding values in
table \ref{tab1} give an estimate for
the maximal deviation of $f_\pi$ from its value for strong Yukawa
coupling. For the $U_L(2)\times
U_R(2)$ model a value $h^2(k_\vp)=300$ is already near the lower limit
of what is compatible with realistic values for $m_q$ and $\teps_0$
(cf. figure \ref{Fig9}).

We can also invert these relations and look for the optimum value of
$\teps_0$ and $h^2(k_\vp)$ for fixed $f_\pi=93\MeV$ and $m_q=300\MeV$.
One obtains for the $O(4)$ model and $k_\vp=630\MeV$
\bea
 \ds{\teps_0} &\simeq& \ds{0.02}\nnn
 \ds{h^2(k_\vp)} &\simeq& \ds{280} \; .
 \label{OptimalValues}
\eea
Within the simple QCD inspired model of ref. \cite{EW94-1} the
transition scale was found as $k_\vp=630\MeV$ and the mass term at
$k_\vp$ gave $\teps_0=0.036$. Comparing with fig. \ref{Fig8} we find
that this value of $\teps_0$ is actually too large to induce
spontaneous symmetry breaking if the meson fluctuations are taken into
account. Given the simplified character of the model considered in
\cite{EW94-1}, however, we find the agreement with the order of
magnitude of (\ref{OptimalValues}) very encouraging. On the other
hand, the estimate of $Z_\vp(k_\vp)\simeq0.85$
appears to be very inaccurate for
the model of \cite{EW94-1} and far away from the small values of
$Z_\vp(k_\vp)$ for which the values (\ref{ResultFpi}),
(\ref{OptimalValues}) were obtained.

We finally compute the chiral condensate from (\ref{Codensate}) with
$\olh(k_\vp)=1$ as
\be
 \VEV{\olpsi\psi}_0=\teps_0 Z_\vp^{-\hal}(m_\pi)
 Z_\psi^{-1}(k_\vp)f_\pi k_\vp^2 \; .
\ee
Extracting $Z_\vp(m_\pi)$ from section \ref{ScalarAnomalousDimension}
(fig. \ref{Fig3}) we can use this value for a check of the
self--consistency of our scenario. We observe that three different
small quantities, $\teps_0$, $Z_\vp(m_\pi)$ and $f_\pi/k_\vp$ enter
here, and it is by far not trivial that a reasonable value of the
chiral condensate can be obtained. We normalize the condensate at
$k_\vp$ with $Z_\psi(k_\vp)=1$. The result for $\VEV{\olpsi\psi}_0$ as
a function of $\teps_0$ is plotted in fig. \ref{Fig10}.
\begin{figure}
\unitlength1.0cm
\begin{picture}(13.,9.)
\put(0.3,5.){\bf $\ds{\frac{\abs{\VEV{\olpsi\psi}_0}^{\frac{1}{3}}}{\MeV}}$}
\put(8.,0.5){\bf $\teps_0$}
\put(5.7,3.){\bf $U_L(2)\times U_R(2)$}
\put(12.,3.){\bf $O(4)$}
\put(-0.8,-11.5){
\epsfysize=22.cm
\epsffile{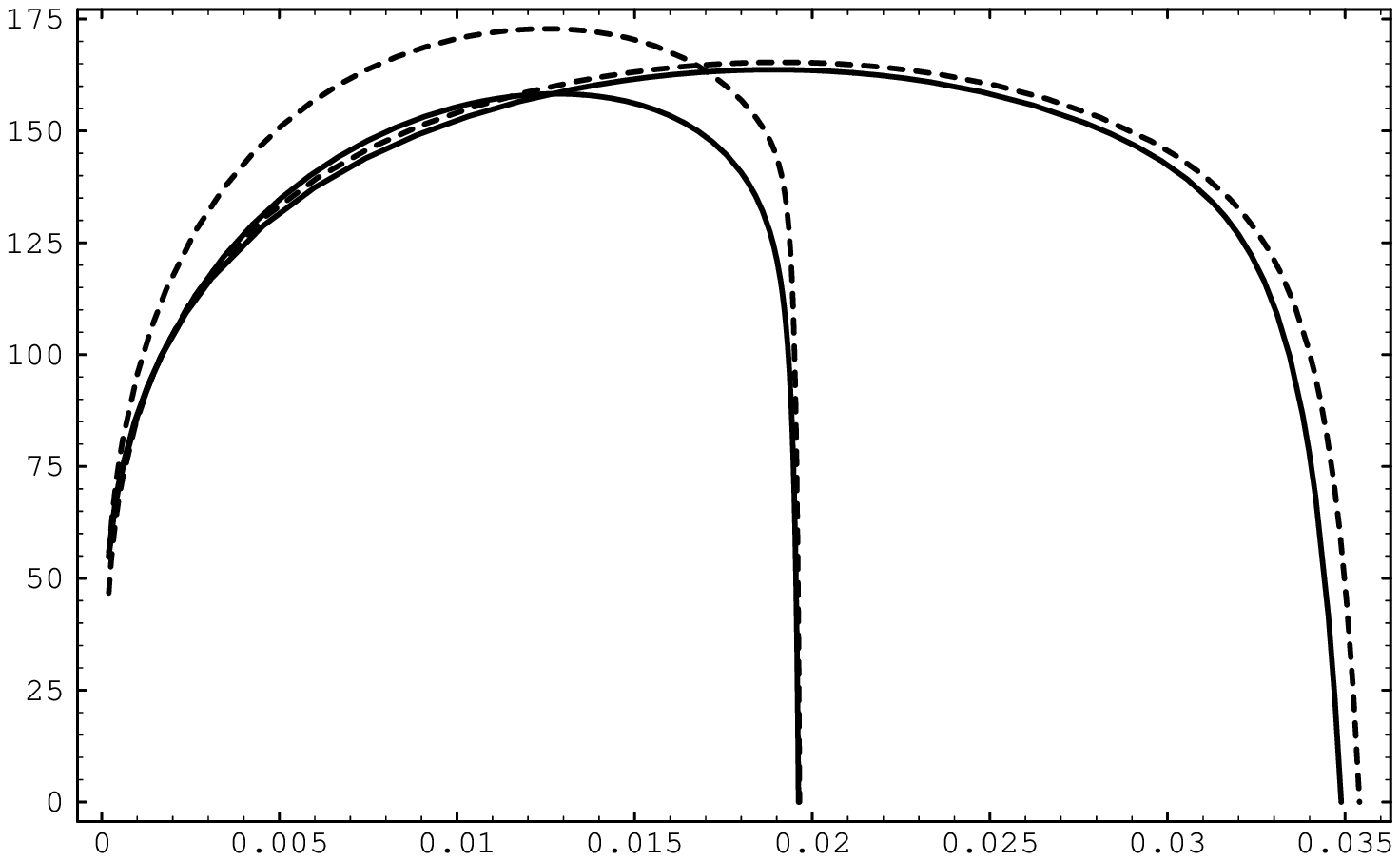}
}
\end{picture}
\caption{\footnotesize The quark condensate as
  a function of $\teps_0$ for $k_\vp=630\MeV$,
  $\la_1(k_\vp)=\la_2(k_\vp)=0$ and $h^2(k_\vp)=300$ (solid line) as
  well as $h^2(k_\vp)=10^4$ (dashed line).}
\label{Fig10}
\end{figure}
For $k_\vp=630\MeV$, $\teps_0=0.02$ and $h^2(k_\vp)=300$
we obtain in the $O(4)$ model
\be
 \abs{\VEV{\olpsi\psi}_0}^{\frac{1}{3}} \simeq 163\MeV \; .
\ee
We may compare this value with a typical value infered from chiral
perturbation theory \cite{GL82-1}
\be
 \abs{\VEV{\olpsi\psi}_{\rm CPT}}^{\frac{1}{3}}(1\GeV)
 \simeq (225\pm25)\MeV \; .
\ee
This value can be scaled down to $k_\vp=630\MeV$ by
exploiting $k\frac{\prl}{\prl k}\VEV{\olpsi\psi}_{\rm
CPT}(k)m_q(k)=0$.
We use here the
three--loop $\beta$--function of QCD ($\ol{\rm MS}$ scheme) for the
running quark mass \cite{JM95-1}
\be
 \frac{k}{m_q(k)}\frac{\prl}{\prl k}m_q(k)=
 -\left[\gm_1\frac{\alpha_s}{\pi}
 +\gm_2\left(\frac{\alpha_s}{\pi}\right)^2
 +\gm_3\left(\frac{\alpha_s}{\pi}\right)^3\right]
\ee
with
\bea
 \ds{\gm_1} &=& \ds{\frac{3}{2}C_F\; ;\;\;\;
 \gm_2=\frac{C_F}{48}\left[
 97N_c+9C_F-10N\right]}\nnn
 \ds{\gm_3} &=& \ds{
 \frac{C_F}{32}\left[
 \frac{11413}{108}N_c^2-\frac{129}{4}N_cC_F
 -\left(\frac{278}{27}+24\zeta(3)\right)NN_c
 +\frac{129}{2}C_F^2
 \right. }\nnn
 &-& \ds{\left.
 \left(23-24\zeta(3)\right)NC_F-\frac{35}{27}N^2\right] }
\eea
and $C_F=\frac{N_c^2-1}{2N_c}$.
This yields
\be
 \frac{m_q(k_\vp)}{m_q(1\GeV)}\simeq 1.72
 \label{BetaFunctionForMq}
\ee
for $k_\vp=630\MeV$ and therefore
\be
\abs{\VEV{\olpsi\psi}_{\rm CPT}}^{\frac{1}{3}}(k_\vp)\simeq
 (188\pm21)\MeV \; .
\ee
This is in satisfactory agreement with our estimate.

\sect{Conclusions}
\label{Conclusions}

We have presented here an effective quark--meson model which is
supposed to describe the strong interaction dynamics between
(constituent) quarks, scalar and pseudo--scalar mesons at momentum
scales smaller than $k_\vp\simeq630\MeV$. The effective average action
$\Gm_k$ for this model depends on a scale $k$ which plays the role of
an infrared cutoff. The scale dependence of the average action obeys
an exact nonperturbative evolution equation. Using a truncation for
the general form of $\Gm_k$ this results in approximate flow equations
for the meson potential and kinetic term as well as the quark kinetic
term and the quark--meson Yukawa coupling $h$. The initial values of
these parameters at the scale $k_\vp$ can, in principle, be computed
\cite{EW94-1,Wet95-1} from evolution equations for QCD which are valid
for scales larger than $k_\vp$. Following the flow equations from
$k_\vp$ to lower scales $k$ one recovers for $k=0$ the effective
action, i.e. the generating functional for the $1PI$ Green functions
for the mesons. In particular, the standard nonlinear $\si$--model
framework of chiral perturbation theory obtains if the expectation
value of the meson field is kept fixed at a nonvanishing vacuum
expectation value.

We have solved the flow equations numerically and observe how the
minimum of the meson potential turns from $\si_0=0$ at high scales to
a nonzero value $\abs{\si_0}>0$ for small $k$. The nonvanishing
expectation value $\si_0$ indicates spontaneous chiral symmetry
breaking. Our numerical solution allows us to compute the mass scales
characteristic for chiral symmetry breaking, i.e. the pion decay
constant $f_\pi$, the constituent quark mass $m_q$ and the chiral
condensate $\VEV{\olpsi\psi}_0$. These quantities are computed as
functions of the initial values for the scalar mass term
$\olm^2(k_\vp)$ and wave function renormalization $Z_\vp(k_\vp)$. The
latter is related to the value of the renormalized Yukawa coupling at
$k_\vp$ by $h^2(k_\vp)=Z_\vp^{-1}(k_\vp)$ once we normalize the
quark wave function with
$Z_\psi(k_\vp)=1$.

The large ratio between the constituent quark mass $m_q$ and $f_\pi$
necessitates a large value of the renormalized Yukawa coupling
$h_R^2=h^2(k=0)$ according to
\be
 h_R^2=\frac{4m_q^2}{f_\pi^2}\simeq 50 \; .
\ee
Since $h^2(k)$ decreases rapidly with decreasing $k$ we conclude that
at the scale $k_\vp$ we have to deal with a strong Yukawa
coupling. Our investigation therefore concentrates on large initial
values $h^2(k_\vp)\gta200$. The most crucial observation of the
present work is that strong Yukawa couplings imply a very fast running
of almost all couplings towards values determined by infrared fixed
points or corresponding infrared intervals. The quark--meson model for
small scales $k$ therefore looses its memory of the exact initial
values of most of the couplings at the scale $k_\vp$. In consequence,
strong Yukawa couplings greatly enhance the chances for a reliable
estimate of $f_\pi$! We find that $f_\pi$ as well as $m_q$ and
$\VEV{\olpsi\psi}_0$ depend only on one ``relevant'' initial value,
namely $\teps_0\sim\olm^2(k_\vp)/k_\vp^2$. The value of $h_R$ may then
be used to fix the required initial value of $\teps_0$ and therefore
to determine $f_\pi$ and $\VEV{\olpsi\psi}_0$. For the simplified
$O(4)$ model discussed in the present paper we obtain
$f_\pi\simeq92\MeV$ and
$\abs{\VEV{\olpsi\psi}_0}^{\frac{1}{3}}\simeq163\MeV$. The comparison
with the experimental result $f_\pi=93\MeV$ and the estimate from
chiral perturbation theory
$\VEV{\olpsi\psi}_{\rm CPT}(k_\vp)=(188\pm21)\MeV$ is very encouraging!

Despite the success of the simplified computation of the present work
our results have partly the character of a feasibility study. Several
steps should considerably improve the accuracy of our computation of
$f_\pi$. First, one should consider the case $N=3$ with a realistic
value of $\olnu$ for the chiral anomaly. This investigation is already
prepared in the appendices of the present paper where the central
results are exhibited for arbitrary $N$ and $\olnu$. Second, the
explicit chiral symmetry breaking due to current quark masses, in
particular the strange quark mass, should be included. Third, the
dependence of the Yukawa coupling and the quark kinetic term on the
quark momentum can be incorporated. This will provide information
on the quark wave function inside the mesons \cite{EW94-1}.

Furthermore, we have not included residual gluon effects for $k<k_\vp$
in the present
work. This would not be necessary if the gluons could completely be
integrated out for the determination of the effective quark--meson
action at the scale $k_\vp$. Since the latter seems to be a quite
difficult task, one may rather use the proposal of ref. \cite{Wet95-1}
and integrate out only the gluons with momenta $q^2>k_\vp^2$. The
remaining gluon fluctuations with momenta $q^2<k_\vp^2$ give then
additional contributions to the flow equations in the quark--meson
model. In the context of a computation of $f_\pi$
the most important effect seems to be a residual gluonic contribution to
$\eta_\psi$ and the effective quark meson vertex and therefore
to the running of the Yukawa coupling. This can be taken into account
by generalizing the average action to include gauge bosons
\cite{RW93-1} and using the formalism of \cite{Wet95-1} to integrate
them out consecutively. A great part of the gluon fluctuation effects
is already included in the contributions from the effective
four--fermion interactions for $k>k_\vp$ or from the quark--meson
interactions considered in the present paper for $k<k_\vp$. The
corrections from residual gluon fluctuations can be found from the
explicit formulae in \cite{Wet95-1}. Since
in addition the confinement scale is
below the constituent quark mass one may
hope that the complicated effects of
gluon condensation do not have a very important influence on
the determination of $f_\pi$.

Finally, a computation of $f_\pi$ within QCD, i.e. as a function of
$\alpha_s(1\GeV)$ and the quark masses, necessitates a
reliable computation of $\teps_0$ and $h^2(k_\vp)$
within the QCD framework for the effective average action
proposed in \cite{Wet95-1}. Thus, the road to an analytical computation
of $f_\pi(\alpha_s(1\GeV))$ is still long. Our results should
encourage to go it.

\newpage

\noindent
{\LARGE Appendices}

\appendix{Linear $\si$--model with fermions}
\label{LinearSigmaModel}

In this appendix we describe the Yukawa couplings of quarks to mesons
in the language of the linear $\si$--model in a Euclidean
formulation. In Euclidean space a Weyl fermion is described by two
independent Grassmann variables $\psi_\al$ and
$\olpsi_{\dot{\al}}$. (We employ the notation of ref. \cite{Wet90-1}.)
We therefore describe the left--handed quarks and their
antiparticles, the right--handed antiquarks, by spinor fields
$\psi_{L\al}$ and $\olpsi_{L\dot{\al}}$, the whereas right--handed
quarks and left--handed antiquarks are contained in
$\psi_R^{\dot{\al}}$ and $\ol{\psi}_R^\al$, respectively. In this
language a Dirac spinor
is composed from left-- and right--handed Weyl spinors:
\be
 \psi =\left(
 \ba{c} \psi_{L\al} \\ \psi_R^{\dot{\al}} \ea
 \right)\; ,\;\;\;
 \olpsi=\left(
 \olpsi_R^\al , \olpsi_{L\dot{\al}}
 \right) \; .
\ee
The tensors contained in the fermion bilinears are described in the
usual way. For example a vector reads
\bea
 \olpsi\gm^m\psi &=& \ds{
 \left(\olpsi_R,\olpsi_L\right)
 \left(\ba{cc}
 0       & \si^m \\
 \olsi^m & 0     \ea\right)
 \left(\ba{c}
 \psi_L \\ \psi_R
 \ea\right) }\nnn
 &=&\ds{ \olpsi_L\olsi^m\psi_L+\olpsi_R\si^m\psi_R }\; .
 \label{Vector}
\eea
The chiral $SU_L(N)\times SU_R(N)$ transformations (with $N$
being the number of quark flavors) act independently on the
left-- and right--handed quarks:
\bea
 \psi &\ra& \ds{\left(
 \ba{cc}
 U_L & 0 \\
 0   & U_R \ea\right)\psi }\nnn
 \olpsi &\ra& \ds{\olpsi\left(
 \ba{cc}
 U_R^\dagger & 0 \\
 0           & U_L^\dagger \ea\right) }
 \label{SU(N)xSU(N)-SymFerm}
\eea
and the $U_B(1)$ symmetry corresponding to conserved baryon number
acts as
\bea
 \psi &\ra& \ds{
 \exp\left(+\frac{i\vth_B}{3}\right)\psi }\nnn
 \olpsi &\ra& \ds{
 \exp\left( -\frac{i\vth_B}{3}\right)\olpsi }\; .
 \label{BarNumFerm}
\eea
Here $U_L$ and $U_R$ are $SU(N)$ matrices with $U_L^\dagger U_L
=U_R^\dagger U_R =1$. The vector (\ref{Vector}) and therefore also
the kinetic term for the quarks is invariant with respect to these
transformations. On the other hand, no mass term is allowed by the
symmetry (\ref{SU(N)xSU(N)-SymFerm}).

Let us now introduce a complex scalar field $\vp$ in the
$({\bf\ol{N}},{\bf N})$ representation of $SU_L(N)\times SU_R(N)$
carrying no baryon number. The appropriate flavor transformations are
\bea
 \vp &\ra& U_R^{ }\vp U_L^\dagger \nnn
 \vp^\dagger &\ra& U_L^{ }\vp^\dagger U_R^\dagger
\eea
where $\vp$ is represented by an $N\times N$ matrix. Its $N^2$
complex components should describe mesons or quark--antiquark
bilinears. The most general Yukawa coupling between mesons and quarks
which is consistent with the $SU_L(N)\times SU_R(N)\times U_B(1)$
flavor symmetry and Osterwalder--Schrader positivity
reads\footnote{In the conventions of ref. \cite{Wet90-1} we take
$\eps =1$ and use $\eta_L =\psi_L$, $\tilde{\eta}_R=\olpsi_R$,
$h_{LR}=h_{RL}=\olh$, $\olh_{LR}^\prime =\olh_{RL}^\prime =-\olh^*$,
$\olh_{RL}=\olh_{LR}=h_{LR}^\prime =h_{RL}^\prime =0$. In this
language the Yukawa couplings $h_{ij}$ are $2N\times 2N$ matrices of
the type $\left(\ba{cc} 0&\olh\\\olh&0\ea\right)$ or
$\left(\ba{cc}0&\olh^\prime\\\olh^\prime&0\ea\right)$.}
\be
 \Lc_{\rm Yuk} =
 \olpsi_R \olh\vp\psi_L -\olpsi_L \olh^* \vp^\dagger \psi_R =
 \olpsi\left(
 \ba{cc} \olh\vp & 0 \\
         0       & -\olh^* \vp^\dagger \ea\right)\psi \; .
 \label{Yuk}
\ee
In general, the Yukawa couplings may contain form factors. In a
momentum space representation one has
\be
 S_{\rm Yuk} = \Om \sum_{q,q^\prime} \left\{
 \olh (-q,q^\prime)\olpsi_R(q)\vp(q-q^\prime)\psi_L(q^\prime)-
 \olh^* (-q,q^\prime)\olpsi_L(q)\vp^\dagger (q^\prime -q)
 \psi_R(q^\prime) \right\}
\ee
where we work here on a torus with volume $\Om$ such that momentum
integrals are discrete sums. In particular, this yields the couplings
to a constant scalar fields ($\vp(p)=\vp\dt(p,0)$)
\be
 S_{\rm Yuk} = \Om \sum_q \olpsi (q)
 \left(\ba{cc}
 \olh(q)\vp & 0 \\
 0          & -\olh^*(q)\vp^\dagger \ea\right)\psi (q)
 \label{SYuk}
\ee
with $\olh (q)\equiv \olh (-q,q)$. We adopt conventions with real Yukawa
coupling $\olh$. By a suitable chiral transformation of fermionic phases
this is equivalent to the convention with purely imaginary $\olh$ which
would be closer to the most commonly used phase convention in a
Minkowski space formulation.

By virtue of the $SU_L(N)\times SU_R(N)$ symmetry we can express the
scalar potential $U_k(\vp)$ as a function of a small number of
invariants. The number of independent invariants depends on $N$ and
we want to classify them. For this purpose it is useful to consider
in addition the axial $U_A(1)$ transformation
\be\ba{rclcrcl}
 \psi_L &\ra& \exp\left( +i\vth_A\right) \psi_L
 &,&
 \olpsi_L &\ra& \exp\left(-i\vth_A\right)\olpsi_L \nnn
 \psi_R &\ra& \exp\left( -i\vth_A\right) \psi_R
 &,&
 \olpsi_R &\ra& \exp\left( +i\vth_A\right)\olpsi_R \nnn
 \vp &\ra& \exp\left( -2i\vth_A\right)\vp
 &,&
 \vp^\dagger &\ra& \exp\left( +2i\vth_A\right)\vp^\dagger \; .
\ea\ee
(We observe that the Yukawa coupling (\ref{Yuk}) is invariant under
$U_A(1)$ transformations.) With the help of $SU_L(N)\times
SU_R(N)\times U_A(1)$ transformations we can bring $\vp$ into a
standard diagonal form with real nonnegative eigenvalues
$\widehat{m}_a$. This implies that the $SU_L(N)\times SU_R(N)\times
U_A(1)$ invariants can always be expressed in terms of the $N$
eigenvalues $\widehat{m}_a^2$. There are therefore exactly $N$
independent invariants. Without the $U_A(1)$ symmetry the standard
form of $\vp$ is $\exp\left( i\vth\right){\rm
diag}\left(\widehat{m}_a\right)$ and the additional phase corresponds to
an additional invariant with respect to $SU_L(N)\times SU_R(N)$ which
is not invariant under $U_A(1)$ transformations. If the
symmetry $\vp\ra\vp^\dagger$ is preserved we may choose for this
invariant
\be
 \xi =\det\vp +\det\vp^\dagger \; .
\ee
Consider now the $N$ $SU_L(N)\times SU_R(N)\times U_A(1)$ invariants
corresponding to the $N$ eigenvalues $\widehat{m}_a^2$. The first of
these invariants is simply the trace
\be
 \rho =\tr\left(\vp^\dagger\vp\right) =
 \sum_a \widehat{m}_a^2 \; .
\ee

For $N>2$ this is the only invariant quadratic in the field $\vp$ ---
there is only one singlet in the direct product $(\ol{\bf N},{\bf
N})\otimes ({\bf N},\ol{\bf N})$. Only for $N=2$ the invariant
$\xi $ is also quadratic in $\vp$. It corresponds to the
singlet in $(\ol{\bf N},{\bf N})\otimes (\ol{\bf N},{\bf N})$. The
traceless matrix
\be
 \phi =\vp^\dagger\vp -\frac{1}{N}\rho
\ee
can be used to construct higher invariants, i.e.
\be
 \ttau_i = \tr\phi^i \; .
 \label{HigherInv}
\ee
The invariants $\ttau_i$ can easily be expressed in terms of
the $\widehat{m}_a^2$ using identities of the type
\be
 \tr\phi^2 = \ds{
 \tr\left(\vp^\dagger\vp\right)^2 -
 \frac{1}{N}\left(\tr\vp^\dagger\vp\right)^2 =
 \sum_a \widehat{m}_a^4 -\frac{\rho^2}{N} }
\ee
\be
 \tr\phi^3 = \ds{
 \sum_a \widehat{m}_a^6 -
 \frac{3}{N}\rho\tr\phi^2 -
 \frac{1}{N^2}\rho^3 } \; .
\ee
For $N=2$ the invariant $\ttau_3$ can be expressed as a
function of $\rho$ and $\ttau_2$ and similarly for all higher
$\ttau_i$. More generally, we only need to consider the $N$
invariants $\rho ,\ttau_2,\ldots ,\ttau_N$ and find the
most general $SU_L(N)\times SU_R(N)$ symmetric scalar potential
\be
 U_k (\vp) = U_k (\rho,\ttau_2,\ldots,\ttau_N,
 \xi ) \; .
\ee

\appendix{Scalar mass spectrum}
\label{ScalarMassSpectrum}

In this appendix we compute the scalar mass spectrum in dependence on
the background field for a general form of the potential. We start
with an $SU_L(N)\times SU_R(N)\times U_A(1)$ invariant potential.
Without loss of generality we can then consider configurations of
the type
\be
 \vp_{ab}=\vp_a \dt_{ab} = \widehat{m}_a \dt_{ab} \; .
 \label{Config}
\ee
We parameterize the potential $U_k(\vp,\vp^\dagger)$ in terms of the
invariants $\rho$, $\tau_i$:
\begin{eqnarray}
 U_k (\vp,\vp^\dagger) &=& U_k (\rho,\tau_i) \\[2mm]
 \rho &=& \tr\left(\vp^\dagger\vp\right) \nnn
 \tau_2 &=& \ds{\frac{N}{N-1}\ttau_2
 = \frac{N}{N-1}\tr\left(\vp^\dagger\vp
 -\frac{1}{N}\rho\right)^2 }\nnn
 \tau_3 &=& \ds{
 \frac{N^2}{(N-1)(N-2)}\ttau_3 + \tau_2^{\frac{3}{2}} }
 \label{Invariants}
\end{eqnarray}
whose relation to the eigenvalues $\widehat{m}_a^2$ can be found in
appendix \ref{LinearSigmaModel}. (If needed, suitable definitions of
$\tau_i$ in terms of $\ttau_i$ have to be chosen for $i\geq 4$.)
One obtains for the second derivatives of the potential
\begin{eqnarray}
 \ds{ \frac{\dt^2 U_k}{\dt\vp_R^{ab} \dt\vp_{Rcd}^{ }} }
 &=& \ds{
 U_k^\prime \dt_a^c \dt_b^d +
 2 U_k^{\prpr}\vp_a \vp_c \dt_{ab} \dt^{cd} +
 \sum_i \frac{\prl U_k}{\prl\tau_i}
 \frac{\dt^2 \tau_i}
 {\dt\vp_R^{ab} \dt\vp_{Rcd}^{ }} }\nnn
 &+& \ds{
 \sqrt{2} \sum_i \frac{\prl U_k^\prime}{\prl\tau_i}
 \left(\vp_a \dt_{ab}
 \frac{\dt\tau_i}{\dt\vp_{Rcd}^{ }} +
 \vp_c \dt^{cd}
 \frac{\dt\tau_i}{\dt\vp_R^{ab}} \right) +
 \sum_{i,j} \frac{\prl^2 U_k}{\prl\tau_i\prl\tau_j}
 \frac{\dt\tau_j}{\dt\vp_R^{ab}}
 \frac{\dt\tau_i}{\dt\vp_{Rcd}^{ }} }
 \label{Mass1}\\[2mm]
 \ds{ \frac{\dt^2 U_k}{\dt\vp_I^{ab} \dt\vp_{Icd}^{ }} }
 &=& \ds{
 U_k^\prime \dt_a^c \dt_b^d +
 \sum_i \frac{\prl U_k}{\prl\tau_i}
 \frac{\dt^2 \tau_i}
 {\dt\vp_I^{ab} \dt\vp_{Icd}} }
 \label{Mass2}\\[2mm]
 \ds{ \frac{\dt^2 U_k}{\dt\vp_I^{ab} \dt\vp_{Rcd}^{ }} }
 &=& 0 \; .
 \label{Mass3}
\end{eqnarray}
Here $\vp_{ab}=\frac{1}{\sqrt{2}}\left(\vp_{Rab}+i\vp_{Iab}\right)$
and primes denote partial derivatives with respect to $\rho$. We
observe that $\dt\tau_i/\dt\vp_I^{ab}$ vanishes for the configuration
(\ref{Config}). In order to gain a better understanding of
the mass matrix (\ref{Mass1})---(\ref{Mass3}) we briefly discuss a few
special cases:

For the origin at $\vp_a=0$ we are in the symmetric regime and
the mass matrix has $2N^2$ real eigenvalues $U_k^\prime (0,\ldots,0)$.
Here we use the fact that the invariants $\tau_i$ are at least
quartic in $\vp$.

If the potential is independent of $\tau_i$ it exhibits an
enhanced symmetry $SO(2N^2)$ instead of $SU_L(N)\times SU_R(N)\times
U_A(1)$. In case of spontaneous symmetry breaking the minimum of the
potential occurs at $U_k^\prime (\rho_0)=0$. The $N^2 \times N^2$
matrix $\vp_a \vp_c \dt_{ab}\dt^{cd}$ has exactly one eigenvalue
$\rho$ whereas all other eigenvalues vanish. Together with the
massless fields $\vp_I$ one therefore finds for $\rho =\rho_0$ the
expected $2N^2 -1$ massless Goldstone bosons. The radial excitation
has mass squared $2U_k^{\prpr}(\rho_0)\rho_0$.

We next include the dependence on the invariant $\tau_2
=\frac{N}{N-1}\ttau_2 =\frac{N}{N-1}\tr\left(\vp^\dagger\vp\right)^2
-\frac{1}{N-1}\rho^2$. For the configuration (\ref{Config}) one has
\begin{eqnarray}
 \ds{ \frac{\dt\tau_2}{\dt\vp_{Rcd}^{ }} }
 &=& \ds{
 2\sqrt{2}\dt^{cd}\left(
 \frac{N}{N-1}\vp_c^3 -\frac{1}{N-1}\rho\vp_c
 \right) }\\[2mm]
 \ds{ \frac{\dt^2\tau_2}{\dt\vp_R^{ab}\dt\vp_{Rcd}^{ }} }
 &=& \ds{
 2\dt_a^c \dt_b^d \left(
 \frac{N}{N-1}(\vp_a^2 +\vp_b^2) -\frac{1}{N-1}\rho\right) }\nnn
 &+& \ds{
 \frac{2N}{N-1}\vp_a \vp_b \dt_a^d \dt_b^c -
 \frac{4}{N-1}\vp_a \vp_c \dt_{ab} \dt^{cd} }\\[2mm]
 \ds{ \frac{\dt^2\tau_2}{\dt\vp_I^{ab}\dt\vp_{Icd}^{ }} }
 &=& \ds{
 2\dt_a^c \dt_b^d \left(
 \frac{N}{N-1}(\vp_a^2 +\vp_b^2) -\frac{1}{N-1}\rho\right) -
 \frac{2N}{N-1}\vp_a \vp_b \dt_a^d \dt_b^c }\; .
\end{eqnarray}
The mass matrix for the fields $\vp_I$ is given by (\ref{Mass2}) and
easily evaluated if possible contributions from $\tau_i$, $i>2$ are
neglected. The $N$ fields $\vp_{Iaa}$ do not mix with $\vp_{Icd}$,
$c\neq d$. The corresponding $N$ eigenvalues of the mass matrix are
\be
 M_{Ia}^2 =\frac{2}{N-1}\frac{\prl U_k}{\prl\tau_2}
 \left(N\vp_a^2 -\rho\right) +U_k^\prime \; .
 \label{MIa}
\ee
For $a\neq b$ the fields $\vp_{Iab}$ and $\vp_{Iba}$ mix, but
decouple from $\vp_{Icd}$  if $c\neq a$ or $b$ or if $d\neq a$ or
$b$. There are $N(N-1)/2$ eigenvalues
\be
 \left(M_{Iab}^-\right)^2 =
 \frac{2}{N-1}\frac{\prl U_k}{\prl\tau_2}
 \left[ N\left(\vp_a^2 +\vp_b^2 +\vp_a\vp_b\right) -
 \rho\right] +U_k^\prime
\ee
and $N(N-1)/2$ eigenvalues
\be
 \left(M_{Iab}^+\right)^2 =
 \frac{2}{N-1}\frac{\prl U_k}{\prl\tau_2}
 \left[ N\left(\vp_a^2 +\vp_b^2 -\vp_a\vp_b\right) -
 \rho\right] +U_k^\prime \; .
\ee
The discussion for $\vp_R$ is similar. For $a\neq b$ only $\vp_{Rab}$
and $\vp_{Rba}$ mix and one finds $N(N-1)/2$ eigenvalues
\be
 \left(M_{Rab}^+\right)^2 =
 \frac{2}{N-1}\frac{\prl U_k}{\prl\tau_2}
 \left[ N\left(\vp_a^2 +\vp_b^2 +\vp_a\vp_b\right) -
 \rho\right] +U_k^\prime
\ee
and $N(N-1)/2$ eigenvalues
\be
 \left(M_{Rab}^-\right)^2 =
 \frac{2}{N-1}\frac{\prl U_k}{\prl\tau_2}
 \left[ N\left(\vp_a^2 +\vp_b^2 -\vp_a\vp_b\right) -
 \rho\right] +U_k^\prime \; .
\ee
The mass matrix for the $N$ fields $\vp_{Raa}$ reads
\bea
 \tilde{M}_{Rac}^2 &=& \ds{
 \left[ U_k^\prime +\frac{2}{N-1}\frac{\prl U_k}{\prl\tau_2}
 \left( 3N\vp_a^2 -\rho\right)\right] \dt_{ac} }\nnn
 &+& \ds{
 2\vp_a \vp_c \left[
 U_k^{\prpr} +\frac{2}{N-1}
 \frac{\prl U_k^\prime}{\prl\tau_2}
 \left( N(\vp_a^2 +\vp_c^2 )-2\rho\right) \right. }\nnn
 &+& \ds{ \left.
 \frac{4}{(N-1)^2}
 \frac{\prl^2 U_k}{(\prl\tau_2)^2}
 \left( N\vp_a^2 -\rho\right)
 \left( N\vp_c^2 -\rho\right) -
 \frac{2}{N-1}\frac{\prl U_k}{\prl\tau_2}
 \right] \; .}
 \label{MRac}
\eea
Its eigenvalues $M_{Rac}^2$ have, in general, no particularly
simple form.

If we specialize to $N$ equal values $\vp_a^2 =\frac{1}{N}\rho$
the matrix $\tilde{M}_R^2$ simplifies considerably:
\be
 \tilde{M}_{Rac}^2 =
 \left( U_k^\prime +\frac{4\rho}{N-1}\frac{\prl U_k}{\prl\tau_2}
 \right)\dt_{ac} +
 \left( 2U_k^{\prpr}-
 \frac{4}{N-1}\frac{\prl U_k}{\prl\tau_2}
 \right) \vp_a \vp_c \; .
\ee
This matrix has $(N-1)$ eigenvalues
\be
 \left(M_R^0\right)^2 =U_k^\prime +\frac{4\rho}{N-1}
 \frac{\prl U_k}{\prl\tau_2}
\ee
and one eigenvalue
\be
 \left( M_R^R\right)^2 =U_k^\prime +2U_k^{\prpr}\rho \; .
\ee
For this special case one finds
\be
 \left( M_{Iab}^-\right)^2 =
 \left( M_{Rab}^+\right)^2 =
 \left( M_R^0\right)^2
 \label{AdjScalars}
\ee
and
\be
 M_{Ia}^2 =
 \left( M_{Iab}^+\right)^2 =
 \left( M_{Rab}^-\right)^2 =
 U_k^\prime \; .
 \label{Goldstone}
\ee
For $U_k^\prime =0$ the $N^2$ massless Goldstone bosons
(\ref{Goldstone}) correspond to the symmetry breaking $SU_L(N)\times
SU_R(N)\times U_A(1)\ra SU(N)$ where the unbroken $SU(N)$ is the
diagonal subgroup of $SU_L(N)\times SU_R(N)$. In addition we have
$N^2 -1$ massive scalars (\ref{AdjScalars}) in the adjoint
representation of $SU(N)$. For $U_k^\prime (\rho_0)=0$ their mass
terms are positive provided $\prl U_k/\prl\tau_2 \geq 0$. Finally
there is a singlet with mass term
$2U_k^{\prpr}(\rho_0)\rho_0$.

Another interesting special case occurs for $\vp_a^2 =\rho\dt_{a1}$
which corresponds to the symmetry breaking $SU_L(N)\times
SU_R(N)\times U_A(1)\ra SU_L(N-1)\times SU_R(N-1)\times U(1)\times
U(1)$. The eigenvalues $M_{Ia}^2$ contain $N-1$ values $U_k^\prime
-\frac{2\rho}{N-1}\frac{\prl U_k}{\prl\tau_2}$ and one value
$U_k^\prime +2\rho\frac{\prl U_k}{\prl\tau_2}$. The eigenvalues
$\left( M_{Iab}^+\right)^2$, $\left( M_{Iab}^-\right)^2$, $\left(
M_{Rab}^+\right)^2$ and $\left( M_{Rab}^-\right)^2$ decompose each
into $(N-1)(N-2)/2$ values $U_k^\prime -\frac{2\rho}{N-1}\frac{\prl
U_k}{\prl\tau_2}$ and $N-1$ values $U_k^\prime +2\rho\frac{\prl
U_k}{\prl\tau_2}$. Finally $\tilde{M}_{Rac}^2$ has $N-1$ eigenvalues
$U_k^\prime -\frac{2\rho}{N-1}\frac{\prl U_k}{\prl\tau_2}$ and one
eigenvalue $U_k^\prime +2\rho U_k^{\prpr} +6\rho\frac{\prl
U_k}{\prl\tau_2}+8\rho^2\frac{\prl
U_k^\prime}{\prl\tau_2}+8\rho^3\frac{\prl^2 U_k}{(\prl\tau_2)^2}$.
For $U_k^\prime +2\rho\frac{\prl U_k}{\prl\tau_2}=0$ we observe the
expected $4N-3$ massless Goldstone bosons.

Next we should take into account that the axial $U_A(1)$ symmetry is
broken due to anomalies. The most general potential can therefore
also depend on
\be
 \xi = \det\vp +\det\vp^\dagger \; .
\ee
We observe that $\xi$ violates $U_A(1)$ but is invariant under
$SU_L(N)\times SU_R(N)$. The second possible $U_A(1)$ violating
$SU_L(N)\times SU_R(N)$ invariant, $\om =i\left(\det\vp
-\det\vp^\dagger\right)$, violates the discrete symmetry
$\vp\ra\vp^\dagger$ and hence the $CP$ invariance of the model
and will therefore be discarded here\footnote{We note that
the sum $\om^2 +\xi^2$ is proportional to
$\det\vp\det\vp^\dagger$ and can therefore be expressed in terms of
the invariants $\rho$, $\tau_i$.}. The configuration (\ref{Config})
with $N$ real $\vp_a$ is not the most general configuration in this
case. An overall phase for all $\vp_a$ cannot be removed anymore by
$U_A(1)$ transformations. In contrast to the invariants $\rho$ and
$\tau_i$ the properties of $\xi$ depend crucially on $N$: $\xi$
is of order $\vp^N$. In addition there is an important difference
between $N$ even and odd. For $N$ even the discrete symmetry $\vp\ra
-\vp$ is part of $SU_L(N)\times SU_R(N)$ and is therefore respected
by $\xi$. For $N$ odd $\xi$ is not invariant under this
discrete symmetry. We also note that all invariants observe the
discrete symmetry $\vp\ra\vp^T$ and therefore also $\vp\ra\vp^*$.

We restrict the
discussion here to scalar field configurations with real diagonal
eigenvalues $\widehat{m}_a$ which are sufficient for deriving
the flow equations for all derivatives of $U_k$ with respect to $\rho$
and the $\tau_i$ (but not $\xi$). The discrete symmetry
$\vp\ra\vp^*$ of the potential implies invariance under
$\vp_{Iab}\ra -\vp_{Iab}$. For real $\widehat{m}_a$ we can choose without
further loss of generality the configuration (\ref{Config}) which
respects $\vp_I\ra -\vp_I$. As an immediate consequence the mass
matrix does not mix $\vp_R$ and $\vp_I$ and (\ref{Mass3}) remains
true. The mass matrices (\ref{Mass1}) and (\ref{Mass2}) acquire
additional contributions:
\begin{eqnarray}
 \ds{ \Dt\frac{\dt^2 U_k}{\dt\vp_R^{ab}\dt\vp_{Rcd}^{ }} }
 &=& \ds{
 \sqrt{2}\frac{\prl U_k^\prime}{\prl\xi}
 \left(\vp_a\dt_{ab}\frac{\dt\xi}{\dt\vp_{Rcd}^{ }} +
 \vp_c \dt^{cd}\frac{\dt\xi}{\dt\vp_R^{ab}} \right) }\nnn
 &+& \ds{
 \sum_i \frac{\prl^2 U_k}{\prl\xi\prl\tau_i}
 \left(\frac{\dt\xi}{\dt\vp_{Rcd}^{ }}
 \frac{\dt\tau_i}{\dt\vp_R^{ab}}+
 \frac{\dt\xi}{\dt\vp_R^{ab}}
 \frac{\dt\tau_i}{\dt\vp_{Rcd}^{ }} \right) }\\[2mm]
 &+& \ds{
 \frac{\prl^2 U_k}{(\prl\xi )^2}
 \frac{\dt\xi}{\dt\vp_R^{ab}}\frac{\dt\xi}{\dt\vp_{Rcd}^{ }} +
 \frac{\prl U_k}{\prl\xi}
 \frac{\dt^2 \xi}{\dt\vp_R^{ab}\dt\vp_{Rcd}^{ }} }\nnn
 \ds{\Dt \frac{\dt^2 U_k}{\dt\vp_I^{ab}\dt\vp_{Icd}^{ }} }
 &=& \ds{
 \frac{\prl U_k}{\prl\xi}
 \frac{\dt^2 \xi}{\dt\vp_I^{ab}\dt\vp_{Icd}^{ }} } \; .
\end{eqnarray}
Here we have used $\dt\xi /\dt\vp_I =0$ for the configuration
(\ref{Config}). Writing
\be
 \xi =\frac{1}{N!}\eps^{a_1\ldots a_N}\eps^{b_1\ldots b_N}
 \left(\vp_{a_1b_1}\ldots\vp_{a_Nb_N}+
 \vp_{a_1b_1}^*\ldots\vp_{a_Nb_N}^*\right)
\ee
one has
\bea
 \ds{\frac{\dt\xi}{\dt\vp_{Rcd}^{ }} }
 &=& \ds{
 \frac{1}{\sqrt{2}}\left(
 \frac{\dt\xi}{\dt\vp_{cd}^{ }} +
 \frac{\dt\xi}{\dt\vp_{cd}^*}\right) }\nnn
 \ds{\frac{\dt\xi}{\dt\vp_{Icd}^{ }} }
 &=& \ds{
 \frac{i}{\sqrt{2}}\left(
 \frac{\dt\xi}{\dt\vp_{cd}^{ }} -
 \frac{\dt\xi}{\dt\vp_{cd}^*}\right) }
\eea
with
\bea
 \ds{\frac{\dt\xi}{\dt\vp_{cd}^{ }} }
 &=& \ds{
 \frac{1}{(N-1)!}\sum_{a_2\ldots a_N}
 \eps^{ca_2\ldots a_N}\eps^{da_2\ldots a_N}
 \vp_{a_2}\ldots\vp_{a_N} }\nnn
 \ds{\frac{\dt\xi}{\dt\vp_{cd}^*} }
 &=& \ds{
 \frac{1}{(N-1)!}\sum_{a_2\ldots a_N}
 \eps^{ca_2\ldots a_N}\eps^{da_2\ldots a_N}
 \vp_{a_2}^*\ldots\vp_{a_N}^* }
\eea
for a diagonal configuration $\vp$. Taking into account the reality
of $\vp$ in (\ref{Config}) one recovers $\dt\xi /\dt\vp_I =0$ and
\be
 \frac{\dt\xi}{\dt\vp_{Rcd}} =
\sqrt{2}\dt^{cd}\prod_{a_i\neq c} \vp_{a_i} \; .
\ee
For the particular configuration where all $\vp_a^2$ equal $\rho /N$
one finds
\be
 \frac{\dt\xi}{\dt\vp_{Rcd}} =
 \sqrt{2}\rN^\frac{N-1}{2}
 \dt^{cd}
\ee
 and similarly
\be
 \frac{\dt^2\xi}{\dt\vp_R^{ab}\dt\vp^{cd}_{R}} =
 -\frac{\dt^2\xi}{\dt\vp_I^{ab}\dt\vp^{cd}_{I}} =
 \frac{1}{(N-2)!}\sum_{e_3\ldots e_N}
 \eps_{ace_3\ldots e_N}\eps_{bde_3\ldots e_N}
 \vp^{e_3}\ldots \vp^{e_N}
\ee
or
\bea
 \ds{\frac{\dt^2\xi}{\dt\vp_R^{aa}\dt\vp^{cc}_{R}} }
 &=& \ds{
 \left( 1-\dt_{ac}\right)\prod_{e_i\neq a,c} \vp_{e_i} }\nnn
 \ds{\frac{\dt^2\xi}{\dt\vp_R^{ab}\dt\vp^{ba}_{R}} }
 &=& -\ds{
 \left( 1-\dt_{ab}\right)\prod_{e_i\neq a,b} \vp_{e_i} } \; .
\eea
In the following we will specialize our discussion to the particular
configuration\footnote{The
reader should not get confused by our use of the symbol $\rho$ for two
different purposes --- once for the invariant $\rho\equiv\tau_1$ and
also for the field configuration $\rho=N\si^2$. Quantities like
$U_k^\prime$ always denote derivatives with respect to $\tau_1$ at
fixed $\tau_2$ and $\xi$. On the other hand, for the configuration
(\ref{Config2}) the invariant $\xi$ becomes a nonvanishing function
of $\si^2=\rho /N$.}
\be
 \vp_{ab} =\rN^\hal \dt_{ab}
 \label{Config2}
\ee
We restrict the discussion here to a potential linear in $\xi$
such that $\prl U_k^\prime
/\prl\xi=\prl^2U_k/\prl\xi\prl\tau_2=\prl^2U_k/\prl\xi^2=0$. The terms
$\sim\dt\xi /\dt\vp_R$ vanish in this case and the second functional
derivatives read
\be
 \frac{\dt^2\xi}{\dt\vp_R^{aa}\dt\vp_R^{cc}}
 =- \frac{\dt^2\xi}{\dt\vp_I^{aa}\dt\vp_I^{cc}}
 =\rN^{\frac{N-2}{2}}\left(1-\dt_{ac}\right)
\ee
and for $a\neq b$
\be
 \frac{\dt^2\xi}{\dt\vp_R^{ab}\dt\vp_R^{cd}}
 =- \frac{\dt^2\xi}{\dt\vp_I^{ab}\dt\vp_I^{cd}}
 =-\rN^{\frac{N-2}{2}}\dt_{ad}\dt_{bc}\; .
\ee

The mass matrices for the diagonal and
off--diagonal $\vp_{ab}$ as well as for $\vp_R$ and $\vp_I$ remain
decoupled. For the diagonal part one finds
\begin{eqnarray}
 \ds{ \tilde{M}_{Iac}^2 } &=& \ds{
 U_k^\prime \dt_{ac} }\\[2mm]
 \ds{ \tilde{M}_{Rac}^2 } &=& \ds{
 \left( U_k^\prime +\frac{4\rho}{N-1}\frac{\prl U_k}{\prl\tau_2} -
 \rN^\frac{N-2}{2}
 \frac{\prl U_k}{\prl\xi}\right) \dt_{ac} }\nnn
 &+& \ds{
 \left( 2U_k^{\prpr} -\frac{4}{N-1}
 \frac{\prl U_k}{\prl\tau_2}
 +\rN^\frac{N-4}{2}
 \frac{\prl U_k}{\prl\xi}\right)\frac{\rho}{N} }\; .
\end{eqnarray}
The matrix $\tilde{M}_I^2$ has $N-1$ eigenvalues
\be
 \left( M_I^0\right)^2 =U_k^\prime+\rNt\frac{\prl U_k}{\prl\xi}
 \label{MassEV1}
\ee
and one eigenvalue
\be
 \left( M_I^R\right)^2 =U_k^\prime -
 (N-1)\rNt
 \frac{\prl U_k}{\prl\xi}
 \label{MassEV2}
\ee
whereas $\tilde{M}_R^2$ has $N-1$ eigenvalues
\be
 \left( M_R^0\right)^2 =U_k^\prime +
 \frac{4\rho}{N-1}\frac{\prl U_k}{\prl\tau_2} -
 \rN^\frac{N-2}{2}
 \frac{\prl U_k}{\prl\xi}
 \label{MassEV3}
\ee
and one eigenvalue
\be
 \left( M_R^R\right)^2 =U_k^\prime +
 2U_k^{\prpr}\rho
 +(N-1)\rN^{\frac{N-2}{2}}\frac{\prl U_k}{\prl\xi} \; .
 \label{MassEV4}
\ee
The mass eigenvalues for the off--diagonal fields are
\begin{eqnarray}
 \ds{ \left( M_{Iab}^-\right)^2 } &=& \ds{
 U_k^\prime +\frac{4\rho}{N-1}\frac{\prl U_k}{\prl\tau_2} -
 \rN^\frac{N-2}{2}
 \frac{\prl U_k}{\prl\xi} } \label{MassEV5}\\[2mm]
 \ds{ \left( M_{Iab}^+\right)^2 } &=& \ds{
 U_k^\prime+\rNt\frac{\prl U_k}{\prl\xi} }\label{MassEV6}\\[2mm]
 \ds{ \left( M_{Rab}^+\right)^2 } &=& \ds{
 U_k^\prime +\frac{4\rho}{N-1}\frac{\prl U_k}{\prl\tau_2} -
 \rN^\frac{N-2}{2}
 \frac{\prl U_k}{\prl\xi} } \label{MassEV7}\\[2mm]
 \ds{ \left( M_{Rab}^-\right)^2 } &=& \ds{
 U_k^\prime+\rNt\frac{\prl U_k}{\prl\xi}
 \label{MassEV8} }\; .
\end{eqnarray}
At the potential minimum one has
$U_k^\prime+\rnNt\frac{\prl U_k}{\prl\xi}=0$ and we observe $N^2
-1$ Goldstone bosons (``pions'')
(\ref{MassEV1}), (\ref{MassEV6}), (\ref{MassEV8}), one less than for
$\prl U_k /\prl\xi =0$. The scalar which has acquired a mass
due to the $U_A(1)$ violating term in the potential
--- the ``$\eta^\prime$--meson'' --- has positive mass squared (with
$\prl U_k /\prl\xi \leq 0$), (\ref{MassEV2}):
\be
 M_\xi^2 = -N\rnN^{\frac{N-2}{2}}
 \frac{\prl U_k}{\prl\xi}(\rho_0, \tau_i=0,
 \xi=2\rnNt) \; .
\ee
There are also $N^2 -1$ scalars in the adjoint representation of the
unbroken diagonal $SU(N)$ which have mass squared $\left(
M_R^0\right)^2$, (\ref{MassEV3}), (\ref{MassEV5}), (\ref{MassEV7}):
\be
 \left( M_R^0\right)^2 =M_\tau^2 +\frac{2}{N}M_\xi^2 \; ,\;\;\;
 M_\tau^2 =\frac{4\rho_0}{N-1}
 \frac{\prl U_k}{\prl\tau_2}(\rho_0, \tau_i=0,
 \xi=2\rnNt) \; .
\ee
Finally, there is the singlet (``radial mode'' or ``$\si$--field'')
with mass given by (\ref{MassEV4}).

We observe for $\rho =0$ that $M_\tau^2$ vanishes and $M_\xi^2$
vanishes for $N>2$. In
the symmetric regime one therefore has $2N^2$ real scalar fields with
mass squared $U_k^\prime (0)$. This is different for $N=2$
where
$M_\xi^2 (0)=-2\frac{\prl U_k}{\prl\xi}(0)$. In the symmetric regime
the complex $({\bf 2},{\bf 2})$ representation decays in this
particular case into two irreducible (real) representations with mass
squared $U_k^\prime (0)+\frac{\prl U_k}{\prl\xi}(0)$ and $U_k^\prime
(0)-\frac{\prl U_k}{\prl\xi}(0)$. In an obvious notation with Pauli matrices
$\tau_k$ we can write the complex $({\bf 2},{\bf 2})$ representation
$\vp$ as
\be
 \vp =\hal\left(\si -i\eta^\prime\right)+
 \hal\left(a^k +i\pi^k\right)\tau_k \; .
\ee
The fields $(\si ,\pi^k)$ form an irreducible representation --- they
represent the standard linear $\si$--model \cite{GML60-1} with scalars
in a real four--component vector representation. In the standard
linear $\si$--model the fields $(\eta^\prime ,a^k)$ are
omitted\footnote{This
is a self--consistent truncation for $N=2$,
since $(\eta^\prime ,a^k)$ form a
separate irreducible representation.} and the scalar potential can be
expressed with the help of only one invariant $\rho =\hal (\si^2
+\pi^k\pi_k)$. In terms of $(\si ,\pi^k)$ the Yukawa coupling
(\ref{Yuk}) with imaginary $\olh$ reads
\bea
 \Lc_{\rm Yuk} &=& \ds{i\olh_\si \olpsi\left(\si+i\pi^k\tau_k\olgm
 \right)\psi}\nnn
 && \olh =2i\olh_\si \; ,\;\;\;\olh_\si =\olh_\si^*
\eea
with $\olgm$ being the Euclidean analog of
$\gm^5$.

If we decide to include the fields $(\eta^\prime ,a^k)$ we obtain for the
invariants $\rho$, $\tau_2$ and $\xi$
\bea
 \rho &=& \ds{\hal
 \left(\pi^a\pi_a +a^a a_a\right) }\nnn
 \tau_2 &=& \ds{\left(a_aa^a\right)
 \left(\pi_b\pi^b\right) -\left(a_a\pi^a\right)
 \left(a_b\pi^b\right)}\nnn
 \xi &=& \ds{\hal
 \left(\pi^a\pi_a -a^a a_a\right) } \; .
 \label{LinSiModInv}
\eea
Here we have defined four--vectors $\pi_a =(\pi_k,\si )$ and $a_a
=(a_k,\eta^\prime )$. We observe that all invariants in (\ref{LinSiModInv})
are also invariant under the discrete symmetry $a\ra -a$ which is
equivalent to $\vp\ra\vp^\dagger$. The invariant violating this
symmetry is $\om =i(\det\vp -\det\vp^\dagger )\sim a^a\pi_a$. The
quartic invariant $\tau_2$ can be constructed from $\rho$, $\xi$ and
$\om$.

\appendix{Evolution equation for $\olla_2$}
\label{EvolEquLa2}

In this appendix we discuss in more detail the $\beta$ function for
$\olla_2$ in the SSB regime. The evolution equation
(\ref{BosEvolLambda2SSB}) can formally be
written as
\bea
 \ds{\prlt\olla_2} &=& \ds{
 -\hal\iddq\widehat{\prlt}\left\{
 \frac{N^2}{4}\frac{\olla_2^2}{(Z_\vp P)^2} \right. }\nnn
 &+& \ds{
 \frac{9(N^2-4)}{4}\frac{\olla_2^2}
 {(Z_\vp P+\olla_2 \rho_0)^2}
 + \frac{N^2\olla_2}{2\rho_0}\left[
 \frac{1}{Z_\vp P+\olla_2 \rho_0}-
 \frac{1}{Z_\vp P}\right] }\nnn
 &+& \ds{\left.
 \frac{6\olla_2 (\frac{1}{4}\olla_2+\olla_1)}
      {\rho_0 (\hal\olla_2 -\olla_1)}\left[
 \frac{1}{Z_\vp P+2\olla_1 \rho_0}-
 \frac{1}{Z_\vp P+\olla_2 \rho_0}\right]\right\} }
\eea
where $\widehat{\prlt}$ acts only on the infrared cutoff in $P$, i.e.
$\widehat{\prlt}=\frac{1}{Z_\vp}\frac{\prl R_k}{\prl t}\frac{\prl}{\prl
P}$. Alternatively, we may use the form
\bea
 \ds{\prlt\olla_2} &=& \ds{
 -\hal\iddq\widehat{\prlt}\left\{
 \frac{N^2}{4}\frac{\olla_2^2}{(Z_\vp P)^2}
 +\frac{9(N^2-4)}{4}\frac{\olla_2^2}
 {(Z_\vp P+\olla_2 \rho_0)^2} \right. }\nnn
 &-& \ds{ \left.
 \frac{N^2}{2}\frac{\olla_2^2}{(Z_\vp P)(Z_\vp P+\olla_2\rho_0)} +
 \frac{3\olla_2^2 +12\olla_1 \olla_2}
 {(Z_\vp P+2\olla_1 \rho_0)(Z_\vp P+\olla_2\rho_0)}
 \right\} }
 \label{AltEvolLa2}
\eea
which is close to the $t$--derivative $\widehat{\prlt}$ of the
contribution from the perturbative one--loop graphs, with
$\rho_0$--dependent masses and the infrared cutoff in the propagator
taken into account. For $d=4$, $\rho_0\ra 0$, $P=q^2$ and omitting
the $t$--derivatives one recognizes the one--loop correction for
$\olla_2$ \cite{CGS93-1,BHJ94-1}.

One may ask to what extent the $\beta$ function for $\olla_2$
depends on the choice of the configuration (\ref{configuration}). For
this purpose we have also evaluated $\prl U_k /\prl t$ for a
configuration
\be
 \vp_1^2 =\frac{\rho_0}{N}+\eps \; , \;\;\;
 \vp_2^2 =\frac{\rho_0}{N}-\eps \; , \;\;\;
 \vp_a^2=\frac{\rho_0}{N}\;\; {\rm for}\;\; a\geq 3
 \label{Config3}
\ee
with infinitesimal $\eps$ and
\be
 \tau_2 =\frac{2N}{N-1}\eps^2 \; .
\ee
{}From the contribution proportional to $\eps^2$ in $\prl U_k /\prl t$
one obtains the flow equation
\bea
 \ds{\prlt\widehat{\la}_2} &=& \ds{
 -\hal\iddq\widehat{\prlt}\left\{
 \frac{N(2N-3)}{2}\frac{\olla_2^2}{(Z_\vp P)^2}
 +\frac{(2N^2 +5N-18)}{2}\frac{\olla_2^2}
 {(Z_\vp P+\olla_2 \rho_0)^2} \right. }\nnn
 &-& \ds{ \left.
 \frac{N\olla_2^2}{(Z_\vp P)(Z_\vp P+\olla_2\rho_0)} +
 \frac{3\olla_2^2 +12\olla_1 \olla_2}
 {(Z_\vp P+2\olla_1 \rho_0)(Z_\vp P+\olla_2\rho_0)}
 \right\} }\; .
 \label{AltEvolHatLa2}
\eea
We observe that for $N=2$ the configuration (\ref{Config3}) is
equivalent to (\ref{configuration}) and the evolution equations for
$\olla_2$ and $\widehat{\la}_2$,
 (\ref{AltEvolLa2}) and (\ref{AltEvolHatLa2}), respectively,
agree. For $N\geq 3$, however, the two
equations are not identical and this is the reason why we have chosen
the symbol $\widehat{\la}_2$ in (\ref{AltEvolHatLa2}). The difference is
related to the truncated higher invariants $\tau_i$, $i\geq 3$. The
validity of
the evolution equation (\ref{AltEvolLa2}) corresponds to a
definition of the higher invariants where all $\tau_i$ vanish for
$i\geq 3$ for the configuration (\ref{configuration}). Inserting
(\ref{configuration}) into (\ref{HigherInv}) we find
$\ttau_3=-\frac{(N-1)(N-2)}{N^2}(\tau_2)^{\frac{3}{2}}$ and therefore
(\ref{Invariants}) $\tau_3=0$. On the other hand, we find for the
configuration (\ref{Config3}) that $\ttau_3$ vanishes (but not
$\ttau_4$) which leads to
\bdm
 \tau_3 =\left(\tau_2\right)^{3/2} \; .
\edm
A truncation is well defined only once we specify the exact definition
of the higher invariants which we omit. This gives then a unique
definition of the running of coupling constants. For $N\geq 3$ the
difference in the results for $k\ra 0$ between the use of
(\ref{AltEvolLa2}) or (\ref{AltEvolHatLa2}) can be taken as a rough
estimate of the uncertainty due to the truncation of the higher
invariants.

\appendix{The infrared cutoff for fermions}
\label{InfraresCutoffFermions}

The choice of the infrared cutoff for fermions is not completely
obvious for various reasons. First of all, chiral fermions do not
allow a mass term. Since we want to remain consistent with chiral
symmetries (a necessity for neutrinos, for example), the infrared
cutoff must have the same Lorentz structure as the kinetic term,
i.e. $R_{kF}\sim\gm^\mu q_\mu$ \cite{Wet90-1}. On the other hand, for
$q^2\ra 0$ the infrared cutoff should be
$\sim k$, e.g. $R_{kF}\sim k\slash{q}/\sqrt{q^2}$. The
nonanalyticity of $\sqrt{q^2}$ may then be a cause of problems. We
will develop in this appendix a few criteria for a reasonable infrared
cutoff and finally propose one which seems suitable for practical
calculations.

First of all, the fermionic infrared cutoff term $\Dt_kS_F$ should be
quadratic in the fermion fields.
We next require that $\Dt_kS_F$
should respect all symmetries of the kinetic term for free fermions.
This includes chiral symmetries and Lorentz invariance. (Gauge
symmetries may be implemented by covariant derivatives in a
background gauge field \cite{RW93-1} but this if of no concern in
the present paper.) The symmetry requirement implies in a Fourier
representation
\be
 \Dt_kS_F = \Om\sum_q\ol{\eta}_{\dot{\al}}(q)
 Z_\psi (k)\left(\slash{q}\right)^{\dot{\al}}_{\;\;\al}
 r_F\left(\frac{q^2}{k^2}\right)\eta^\al (q) \equiv
 \Om\ol{\eta}Z_\psi\slash{q}r_F \eta
\ee
where $\ol{\eta}$, $\eta$ are Weyl spinors, $\al,\dot{\al}$ denote
spinor indices,
$\slash{q}=q_\mu\ol{\si}^\mu$ and $\ol{\si}^\mu$ is the
restriction of $\gm^\mu$ for left--handed Weyl spinors (with a
suitable choice for right--handed Weyl spinors). We also have omitted
possible internal indices labeling different Weyl spinors (for
conventions see \cite{Wet90-1}). The wave function renormalization
$Z_\psi$ is chosen for convenience such that it matches with an
approximation for the fermion kinetic term $\Om\ol{\eta}Z_\psi
\slash{q}\eta$ in $\Gm_k$. (Typically $Z_\psi$ is diagonal in the
internal indices $i,j$, but different fermion species may have
different wave function renormalization constants.) The third
condition requires that $\Dt_kS_F$ acts effectively as an infrared
cutoff. This means that for $k\ra\infty$ the combination $Z_\psi
r_F(\frac{q^2}{k^2})$ should diverge for all values of $q^2$. This
divergence should also occur for finite $k$ and $q^2/k^2 \ra 0$ and
be at least as strong as $\left( k^2/q^2\right)^{1/2}$. As a fourth
point we remark that $\Gm_k$ becomes the effective action in the limit
$k\ra 0$ only if $\ds{\lim_{k\ra 0}\Dt_kS_F=0}$. This should hold for all
Fourier modes separately, i.e. for $\ds{\lim_{k\ra 0} Z_\psi (k)={\rm
const.}}$ one requires
\be
 \lim_{k\ra 0}r_F \left(\frac{q^2}{k^2}\right)\slash{q}=0 \; .
 \label{Cond3}
\ee
Even though on a torus with longest circumference $L$ the minimum
value $q_{\rm min}^2=\pi^2/L^2$ does not vanish, we want a smooth limit
to infinite volume and request that (\ref{Cond3}) also holds in
the limit $q^2\ra 0$. Together with the third condition this implies
exactly
\be
 \lim_{q^2/k^2\ra 0} r_F(\frac{q^2}{k^2}) \sim
 \left(\frac{q^2}{k^2}\right)^{-\hal} \; .
 \label{Cond4}
\ee

The requirement (\ref{Cond4}) implies a smooth behavior of $R_{kF}$
for $q^2/k^2\ra 0$. On the other hand, the nonanalyticity of $r_F$
at $q^2=0$ may be a source of worry for practical computations. A
careful choice of $r_F$ is
necessary in order to circumvent this problem. First we note that
$r_F$ appears in connection with the fermion propagator from
$\Gm_k$ and we combine
\bea
 Z_\psi \slash{q} +Z_\psi \slash{q} r_F &=& Z_\psi \slash{q} F \nnn
 P_F =q^2 F^2 &=& q^2 \left( 1+r_F\right)^2 \; .
\eea
Up to the wave function renormalization, $P_F$ corresponds to the
squared inverse propagator of a free massless fermion in the presence
of the infrared cutoff. We will require that $P_F$ and therefore
$F^2$ is analytic in $q^2$ for all $q^2\geq 0$. A reasonable choice
which we will employ in the present paper is
\be
 P_F =P=\frac{q^2}{1-\exp\left\{
 -\frac{q^2}{k^2}\right\}} \; .
 \label{PF}
\ee

\appendix{Scalar wave function renormalization}
\label{ScalarWaveFunctionRenormalization}

In this appendix we provide some details for the calculation of the
flow equation of the scalar wave function
renormalization $Z_{\vp,k}(k)$. We start with the scalar contribution
to (\ref{Zvpfull}). It
is convenient to decompose the complex fields $\vp_{ab}$ and
$\vp_{ab}^\dagger$ into their real and imaginary parts:
\be
 \vp_{ab}(q)=\frac{1}{\sqrt{2}}
 \left[\vp_{Rab}(q)+i\vp_{Iab}(q)\right]\; ,\;\;\;
 \vp_{ab}^\dagger (q)=\frac{1}{\sqrt{2}}
 \left[\vp_{Rba}(q)-i\vp_{Iba}(q)\right]\; .
\ee
For the configuration (\ref{ConfAnDi}) and the ansatz
(\ref{EffActAnsatz}) it is then easy to see that $\Gm_{Sk}^{(2)}$ is
block--diagonal in the indices $R$ and $I$, i.e. it decomposes
into a block $\Gm_{Rk}^{(2)}$ which contains only functional
derivatives with respect to the $\vp_{Rab}$ and an analog block
$\Gm_{Ik}^{(2)}$. We may therefore use the expansion (\ref{TraceExp})
separately for $\Gm_{Rk}^{(2)}$ and $\Gm_{Ik}^{(2)}$. The first term
on the right hand side of (\ref{TraceExp})
is independent of $\dt\vp$, $\dt\vp^*$ and thus
can not contribute to the anomalous dimension. The part of the second
term proportional to $\dt\vp\dt\vp^*$ is independent of $Q$ and hence
does not contribute either to the evolution of $Z_{\vp ,k}(Q)$.
Hence, we are left with the $Q$--dependent part of the third term
whose contribution proportional to $\dt\vp\dt\vp^*$ is obtained by
keeping all terms of $\Dt\Gm_{R/I\, k}^{(2)}$ which are linear in
$\Dt(q,Q)=\dt\vp\dt(q,Q)+\dt\vp^*\dt(q,-Q)$:
\ben
 \ds{ \left[\Dt\Gm_{Rk}^{(2)}\right]_{ab,cd}(q,q^\prime) }
 &=& \ds{
 \frac{\Dt(q^\prime -q,Q)}{\vp} \left\{
 (B+D)\left[\dt_{ad}\Si_{cb}+\dt_{cb}\Si_{ad}\right]
 \right. }\nnn
 &+& \ds{ \left.
 \frac{2}{N}(C-B-D)
 \left[\dt_{ab}\Si_{cd}+\dt_{cd}\Si_{ab}\right]
 \right\}
 +{\cal O}(\Dt^2) }\nnn
 \ds{ \left[\Dt\Gm_{Ik}^{(2)}\right]_{ab,cd}(q,q^\prime) }
 &=& \ds{
 -\frac{\Dt(q^\prime -q,Q)}{\vp} \left\{
 (B+D)\left[\dt_{ad}\Si_{cb}+\dt_{cb}\Si_{ad}\right]
 \right. }\nnn
 &-& \ds{ \left.
 D\left[\dt_{ab}\Si_{cd}+\dt_{cd}\Si_{ab}\right]
 \right\}
 +{\cal O}(\Dt^2) }\nnn
\een
with
\be
 B\equiv \hal M_\tau^2\; ,\;\;\;
 C\equiv \rho U^{\prpr}_k\; ,\;\;\;
 D\equiv \frac{1}{N}M_\xi^2 \; .
\ee
{}From
\ben
 \ds{ \left[\Gm_{Rk,0}^{(2)}\right]_{ab,cd}(q,q^\prime) }
 &=& \ds{
 \left\{ \left(Z_{\vp ,k}(q)q^2
 + U_k^\prime\right) \dt_{ac}\dt_{bd}
 +B\left(\dt_{ac}\dt_{bd}+\dt_{ad}\dt_{bc}\right)
 \right. } \nnn
 &+& \ds{ \left.
 \frac{2}{N} \left( C-B\right)
 \dt_{ab}\dt_{cd}
 \right\} \dt (q,q^\prime) }\nnn
 \ds{ \left[\Gm_{Ik,0}^{(2)}\right]_{ab,cd}(q,q^\prime) }
 &=& \ds{
 \left\{ \left(Z_{\vp ,k}(q)q^2
 + U_k^\prime\right) \dt_{ac}\dt_{bd}
 +B\left(\dt_{ac}\dt_{bd}-\dt_{ad}\dt_{bc}\right)
 \right\} \dt (q,q^\prime) }
\een
we obtain
\ben
 \ds{\left(\Gm_{Rk,0}^{(2)}+R_k\right)_{ab,cd}^{-1}(q,q^\prime ) }
 &=& \ds{
 \frac{\dt (q,q^\prime )}{A(q)+2B}
 \left\{ \frac{A(q)+B}{A(q)}\dt_{ac}\dt_{bd}
 - \frac{B}{A(q)}\dt_{ad}\dt_{bc} \right. }\nnn
 &-& \ds{ \left.
 \frac{2}{N}\frac{C-B}{[A(q)+2C]}
 \dt_{ab}\dt_{cd}\right\} }\nnn
 \ds{\left(\Gm_{Ik,0}^{(2)}+R_k\right)_{ab,cd}^{-1}(q,q^\prime ) }
 &=& \ds{
 \frac{\dt (q,q^\prime )}{A(q)}
 \left\{\frac{A(q)+B}{A(q)+2B}\dt_{ac}\dt_{bd}
 +\frac{B}{A(q)+2B}\dt_{ad}\dt_{bc} \right\} }
\een
where
\be
 A(q)\equiv Z_{\vp ,k}(q)P(q)+U_k^\prime \; .
\ee
Now the traces can be evaluated and one finds for the scalar
contribution to the scalar wave function renormalization
\be
 \left[\prlt Z_{\vp,k}(Q)\right]_S =
 -\frac{k^2}{Q^2}Z_{\vp,k}^2(Q)
 \left[ f_k^d(Q)-f_k^d(0) \right]
\ee
with
\bea
 \ds{ f_k^d(Q) }
 &=& \ds{
 \frac{1}{2\rho k^2} \iddq }\nnn
 && \ds{\hspace{-.8cm}\times
 \widehat{\frac{\prl}{\prl t}}\left\{
 \frac{\left[2\rho U_k^{\prpr}
 -\frac{N-2}{N}M_\xi^2\right]^2}
 {[Z_{\vp,k}(q)P(q)+U_k^\prime-\frac{1}{N}M_\xi^2]
 [Z_{\vp,k}(q+Q)P(q+Q)+U_k^\prime+2\rho
 U_k^{\prpr}-\frac{N-1}{N}M_\xi^2]}
 \right. }\nnn
 && \ds{\hspace{-.8cm}
 +\frac{N^2-4}{2}
 \frac{\left[M_\tau^2+\frac{2}{N}M_\xi^2\right]^2}
 {[Z_{\vp,k}(q)P(q)+U_k^\prime-\frac{1}{N}M_\xi^2]
 [Z_{\vp,k}(q+Q)P(q+Q)+U_k^\prime+M_\tau^2+\frac{1}{N}M_\xi^2]} }\nnn
 && \ds{\hspace{-.8cm}\left.
 +\frac{\left[M_\tau^2-\frac{N-2}{N}M_\xi^2\right]^2}
 {[Z_{\vp,k}(q)P(q)+U_k^\prime+\frac{N-1}{N}M_\xi^2]
 [Z_{\vp,k}(q+Q)P(q+Q)+U_k^\prime+M_\tau^2+\frac{1}{N}M_\xi^2]}
 \right\} }
\eea
and
\bea
 M_\tau^2 &=& \ds{\frac{4\rho}{N-1}\frac{\prl U_k}{\prl\tau_2}}\nnn
 M_\xi^2 &=& \ds{-N\rNt\frac{\prl U_k}{\prl\xi}} \; .
\eea
Here we used the fact that $P(q)$ and $Z_{\vp,k}(q)$ are actually
functions of the invariant $q^2$.

We turn next to the fermionic contribution to the scalar
anomalous dimension. Analogously to the scalar contribution we may
split $\Gm_{Fk}^{(2)}$ into
$\Gm_{Fk,0}^{(2)}$ and $\Dt\Gm_{Fk}^{(2)}$ which contains all
dependence on the scalar background fluctuation $\dt\vp$. For the
truncation (\ref{EffActAnsatz}) of the
average action we obtain
\bea
 \ds{ \left(\Gm_{Fk,0}^{(2)}+R_{Fk}\right)_{ab}^{-1}(q,q^\prime ) }
 &=& \ds{
 \frac{Z_{\psi ,k}(q)[1+r_F (q)]\slash{q}+\olh_k(q)\vp\olgm}
 {Z_{\psi ,k}^2(q)P_F (q)
 +\olh_k^2(q) \vp^2} \dt_{ab}\dt (q,q^\prime )}\nnn
 \ds{\left(\Dt\Gm_{Fk}^{(2)}\right)_{ab}(q,q^\prime)}
 &=& \ds{
 \olh_k(-q,q^\prime)
 \Dt(q^\prime -q,Q)\Si_{ab} }
\eea
where we used the abbreviation
$\olh_k(q)\equiv\olh_k(-q,q)$.
Using an expansion similar to (\ref{TraceExp}) the
fermionic trace can be evaluated and we obtain its
contribution to the scalar anomalous dimension by collecting all
$Q$--dependent terms of order $\dt\vp\dt\vp^*$. One finds
\be
 \left[\prlt Z_{\vp ,k}(Q)\right]_F =
 -2^{\frac{d}{2}-1}N_c \frac{k^2}{Q^2}
 \left[ f_{Fk}^d(Q)-f_{Fk}^d(0)\right]
\ee
with
\bea
 \ds{f_{Fk}^d(Q)}
 &=& \ds{
 k^{-2}\iddq
 \left(q^2+qQ\right)\olh_k^2(-q,q+Q) }\nnn
 &\times& \ds{
 \widehat{\frac{\prl}{\prl t}}\left\{
 \frac{Z_{\psi,k}(q)[1+r_F(q)]}
 {[Z_{\psi,k}^2(q)P_F(q)+\frac{1}{N}\olh_k^2(q)\rho]}
 \frac{Z_{\psi,k}(q+Q)[1+r_F(q+Q)]}
 {[Z_{\psi,k}^2(q+Q)P_F(q+Q)+\frac{1}{N}\olh_k^2(q+Q)\rho]}
  \right\} }\; .
\eea
and we made use of the identity
$\olh_k(-q,q^\prime)=\olh_k(-q^\prime,q)$.

The fermionic contribution to $\eta_\vp$ obtains from the term linear
in $Q^2$ in $f_{Fk}^d$ in close analogy to \cite{BW93-1}, and similarly
for the scalar contribution, where we recover the results of
\cite{Wet93-1} for $\la_2=\nu=0$.
Taking $\rho$ at the minimum of the potential and neglecting all
momentum dependence of the Yukawa coupling and wave function
renormalizations we arrive at
\bea
 \ds{\eta_\vp} &=& \ds{
 4\frac{v_d}{d}\kappa\left\{\left[
 2\la_1-\frac{N-2}{2}\frac{\nu}{N}\kNf\right]^2
 m_{2,2}^d(0,2\kappa\la_1-\frac{N-2}{2}\nu\kNt;\eta_\vp) \right. }\nnn
 &+& \ds{\left[
 \la_2-\frac{N-2}{2}\frac{\nu}{N}\kNf\right]^2
 m_{2,2}^d(\kappa\la_2+\nu\kNt,\frac{N}{2}\nu\kNt;\eta_\vp) }\nnn
 &+& \ds{\left.
 \frac{N^2-4}{2}\left[\la_2+\frac{\nu}{N}\kNf\right]^2
 m_{2,2}^d(0,\kappa\la_2+\nu\kNt;\eta_\vp) \right\} }\nnn
 &+& \ds{
 2^{\frac{d}{2}+2}\frac{v_d}{d}N_c h^2
 m_4^{(F)d}(\frac{1}{N}\kappa h^2;\eta_\psi) } \; .
\eea
The threshold functions $m_{n_1,n_2}^d$ and $m_4^{(F)d}$ are defined
in (\ref{mn1n2d}) and (\ref{m4Fd}), respectively.

\appendix{Yukawa coupling and fermion wave function re\-nor\-malization}
\label{FermionWaveFunctionRenormalization}

We will give here details on the derivation of the evolution equations
for the fermionic wave function renormalization constant and the
Yukawa coupling as defined in section \ref{EvolutionEquationForH}. We
use here the truncation (\ref{EffActAnsatz}) but neglect for simplicity
the dependence of the Yukawa coupling on the scalar momentum and also
the momentum dependence of $Z_\vp$.
For the scalar fields it proves useful to introduce the following
linear combinations:
\bea
 \ds{\vp_{Ra} }
 &=& \ds{
 \frac{1}{\sqrt{2}}\left(
 \vp_{aa}+\vp_{aa}^* \right) }\nnn
 \ds{\vp_{Ia} }
 &=& \ds{
 \frac{-i}{\sqrt{2}}\left(
 \vp_{aa}-\vp_{aa}^* \right) }
 \label{Scalars1}
\eea
and for $a\neq b$
\bea
 \ds{\vp_{R+ab}}
 &=& \ds{
 \hal\left(
 \vp_{ab}+\vp_{ba}+\vp_{ab}^*+\vp_{ba}^* \right) }\nnn
 \ds{\vp_{R-ab}}
 &=& \ds{
 \hal\left(
 \vp_{ab}-\vp_{ba}+\vp_{ab}^*-\vp_{ba}^* \right) }\nnn
 \ds{\vp_{I+ab}}
 &=& \ds{
 \frac{-i}{2}\left(
 \vp_{ab}+\vp_{ba}-\vp_{ab}^*-\vp_{ba}^* \right) }\nnn
  \ds{\vp_{I-ab}}
 &=& \ds{
 \frac{-i}{2}\left(
 \vp_{ab}-\vp_{ba}-\vp_{ab}^*+\vp_{ba}^* \right) }
 \label{Scalars2}
\eea
Using collective indices
$i,j\in\{Ra,Ia,R+ab,R-ab,I+ab,I-ab\}$
this yields
\be
 \frac{\dt^2\Gm_k}
 {\dt\vp_i^* (q)\dt\vp^j (q^\prime )}
 =\left( Z_{\vp}q^2 \dt_j^i
 +(M^2)^i_j \right)
 (2\pi)^d\dt(q-q^\prime)
 \label{BBB}
\ee
with scalar mass matrix
\be
 (M^2)_j^i =\frac{\prl^2 U_k}{\prl\vp_i^* \prl\vp^j}\; .
\ee
This matrix is discussed in appendix \ref{ScalarMassSpectrum} and
does not mix the various fields (\ref{Scalars1}), (\ref{Scalars2}).
The mixed functional derivatives of $\Gm_k$ with respect to one scalar
and one fermion field are easily seen to be of first order in the
fermion fields. Since we are interested in the term bilinear in
$\psi$, we can take $\psi$ to be infinitesimally small and split
\be
 \Gm_k^{(2)}=\Gm_{k,0}^{(2)}+\Dt\Gm_k^{(2)}
\ee
in such a way that all $\psi$--dependence is entailed in
$\Dt\Gm_k^{(2)}$. Hence $\Gm_{k,0}^{(2)}$ is given by (\ref{FFF}),
(\ref{BBB}) and $\Dt\Gm_k^{(2)}$ is determined by the mixed
scalar--fermionic functional derivatives of $\Gm_k$. Using an
expansion similar to (\ref{TraceExp}) we arrive at the following
evolution equation for the bilinear fermionic part $\Gm_{k,2}^{(\psi)}$ of
the effective average action:
\bea
 \ds{ \frac{\prl}{\prl t}\Gm_{k,2}^{(\psi)} }
 &=& \ds{
 \frac{1}{2N}\int\frac{d^dQ}{(2\pi)^d}\iddq
 \olh_k^2(\frac{q+Q}{2})
 \widehat{\frac{\prl}{\prl t}}\left\{
 \frac{Z_{\psi,k}(q)\olpsi_a(Q) \slash{q}
 [1+r_F(q)]\psi^a(Q)}
 {Z_{\psi,k}^2(q)P_F(q)+\olh_k^2(q)\vp^2} \right. }\nnn
 &\times& \ds{
 \left(
 \frac{N^2-1}{Z_{\vp}P(q-Q)+U_k^\prime-\frac{1}{N}M_\xi^2}
 +\frac{N^2-1}{Z_{\vp}P(q-Q)
 +U_k^\prime +M_\tau^2 +\frac{1}{N}M^2_\xi}
 \right. }\nnn
 &+& \ds{ \left.
 \frac{1}{Z_{\vp}P(q-Q)+U_k^\prime
 +2\rho U_k^{\prpr}-\frac{N-1}{N}M_\xi^2}
 +\frac{1}{Z_{\vp}P(q-Q)+U_k^\prime
 +\frac{N-1}{N}M^2_\xi} \right) }\nnn
 &+& \ds{\frac{\olh_k(q)\vp\olpsi^a(Q)\olgm\psi_a(Q)}
 {Z_{\psi,k}^2(q)P_F(q)+\olh_k^2(q)\vp^2} }\nnn
 &\times& \ds{
 \left(
 \frac{N^2-1}{Z_{\vp}P(q-Q)+U_k^\prime-\frac{1}{N}M_\xi^2}
 - \frac{N^2-1}{Z_{\vp}P(q-Q)
 +U_k^\prime +M_\tau^2+\frac{1}{N}M_\xi^2} \right. }\nnn
 &-& \ds{ \left.\left.
 \frac{1}{Z_{\vp}P(q-Q)+U_k^\prime
 +2\rho U_k^{\prpr}-\frac{N-1}{N}M_\xi^2}
 +\frac{1}{Z_{\vp}P(q-Q)+U_k^\prime
 +\frac{N-1}{N}M_\xi^2}
 \right)\right\} }
 \label{EvolGmPsi}
\eea
with
$M_\tau^2=\frac{4\rho}{N-1}\frac{\prl U_k}{\prl\tau_2}$
and $M_\xi^2=-N\rNt\frac{\prl U_k}{\prl t}$.
It is easy to extract from (\ref{EvolGmPsi}) the evolution equations
for the Yukawa coupling and the fermionic anomalous dimension.
For the Yukawa coupling, renormalized according to (\ref{YukRen}),
one finds for $\rho$ at the minimum of the potential
\bea
 \ds{\frac{\prl}{\prl t}h_k^2(Q) }
 &=& \ds{ \left[
 d-4+2\eta_{\psi,k}(Q)+\eta_{\vp,k}\right] h^2_k(Q) }\nnn
 &+& \ds{
 \frac{k^{4-d}}{N}h_k(Q)\iddq
 h_k(q)h^2_k(\frac{q+Q}{2})
 \frac{Z_{\psi,k}^2(\frac{q+Q}{2})}{Z_{\psi,k}(q)Z_{\psi,k}(Q)}Z_\vp
 }\nnn
 &\times& \ds{
 \widehat{\prlt} \left\{
 \frac{1}{P_F(q)+\frac{k^2}{N}h^2_k(q)\kappa}
 \left[
 \frac{N^2-1}{Z_\vp P(q-Q)+U_k^\prime-\frac{1}{N}M_\xi^2}
 \right.\right. }\nnn
 &+&\ds{
 \frac{1}{Z_\vp P(q-Q)+U_k^\prime+\frac{N-1}{N}M_\xi^2}
 -\frac{N^2-1}{Z_\vp P(q-Q)+U_k^\prime+M_\tau^2+\frac{1}{N}M_\xi^2} }\nnn
 &-& \ds{\left.\left.
 \frac{1}{Z_\vp P(q-Q)+U_k^\prime+2\rho U_k^{\prpr}-\frac{N-1}{N}M_\xi^2}
 \right]\right\}  }\; .
 \label{YukEvolMom}
\eea
Analogously we obtain for the fermion anomalous dimension
\bea
 \ds{\eta_{\psi,k}(Q) }
 &=& \ds{
 -\frac{k^{4-d}}{2N}\iddq\frac{qQ}{Q^2}
 h^2_k(\frac{q+Q}{2})
 \frac{Z_{\psi,k}^2(\frac{q+Q}{2})Z_\vp}
 {Z_{\psi,k}(q)Z_{\psi,k}(Q)}
 \widehat{\frac{\prl}{\prl t}}
 \left\{
 \frac{[1+r_F(q)]}{P_F(q)+\frac{k^2}{N}h^2_k(q)\kappa} \right.
 }\nnn
 &\times& \ds{
 \left(
 \frac{N^2-1}{Z_\vp P(q-Q)+U_k^\prime-\frac{1}{N}M_\xi^2}
 +\frac{1}{Z_\vp P(q-Q)+U_k^\prime+\frac{N-1}{N}M_\xi^2}
 \right. }\nnn
 &+& \ds{
 \frac{N^2-1}
 {Z_\vp P(q-Q)+U_k^\prime+M_\tau^2+\frac{1}{N}M_\xi^2} }\nnn
 &+& \ds{ \left.\left.
 \frac{1}
 {Z_\vp P(q-Q)+U_k^\prime+2\rho U_k^{\prpr}-\frac{N-1}{N}M_\xi^2}
 \right)\right\} } \; .
 \label{EtaFEvolMom}
\eea

In the following we do not discuss further the momentum dependence in
the fer\-mionic sector. We are interested in the evolution of
$\eta_\psi=\eta_{\psi,k}(0)$ and $h^2=h_k^2(0)$. Neglecting the
momentum dependence of $Z_{\psi,k}$ and $h_k$ on the right hand side of
(\ref{YukEvolMom}) and (\ref{EtaFEvolMom}) and taking the limit
$Q^2\ra 0$ we obtain
\bea
 \ds{ \prlt h^2}
 &=& \ds{ \left[ d-4+2\eta_\psi +\eta_\vp\right] h^2
 -\frac{4}{N}v_d h^4 \left\{
 (N^2-1)^2l_{1,1}^{(FB)d} (\frac{1}{N}\kappa h^2,\eps;
 \eta_\psi,\eta_\vp) \right] }\nnn
 &+& \ds{
 l_{1,1}^{(FB)d} (\frac{1}{N}\kappa h^2,\eps+\frac{N}{2}\nu\kNt;
 \eta_\psi,\eta_\vp)
 }\nnn
 &-& \ds{
 (N^2-1)l_{1,1}^{(FB)d} (\frac{1}{N}\kappa h^2,
 \eps+\kappa\la_2+\nu\kNt;
 \eta_\psi,\eta_\vp) }\nnn
 &-& \ds{\left.
 l_{1,1}^{(FB)d} (\frac{1}{N}\kappa h^2,
 \eps+2\kappa\la_1-\frac{N-2}{2}\nu\kNt;
 \eta_\psi,\eta_\vp)
 \right\} }
 \label{RunningOfh2WithNy}
\eea
\bea
 \ds{\eta_\psi }
 &=& \ds{
 \frac{4}{N}\frac{v_d}{d} h^2 \left\{
 (N^2-1)m_{1,2}^{(FB)d}(\frac{1}{N}h^2\kappa,\eps;\eta_\psi,\eta_\vp)
 \right. }\nnn
 &+& \ds{
 m_{1,2}^{(FB)d}(\frac{1}{N}h^2\kappa,\eps+\frac{N}{2}\nu\kNt;
 \eta_\psi,\eta_\vp) }\nnn
 &+& \ds{
 (N^2-1)m_{1,2}^{(FB)d}(\frac{1}{N}h^2\kappa,
 \eps+\kappa\la_2+\nu\kNt;
 \eta_\psi,\eta_\vp) }\nnn
 &+& \ds{\left.
 m_{1,2}^{(FB)d}(\frac{1}{N}h^2\kappa,
 \eps+2\kappa\la_1-\frac{N-2}{2}\nu\kNt;
 \eta_\psi,\eta_\vp)
 \right\} } \; .
 \label{EtaPsiWithNu}
\eea
The threshold functions $l_{n_1,n_2}^{(FB)d}$ and
$m_{n_1,n_2}^{(FB)d}$ are defined in (\ref{ln1n2FBd}) and
(\ref{mn1n2FBd}), respectively.
In the symmetric regime one has to set $\kappa=0$ in both expressions,
whereas in the SSB regime $\epsilon=0$.

\appendix{Threshold integrals}
\label{ThresholdIntegrals}

In this appendix we provide explicit expressions for the threshold
integrals introduced in the previous sections which are suitable for a
direct numerical integration. For this purpose we introduce the
dimensionless integration variable $y=q^2/k^2=x/k^2$.  Using the
abbreviations
\be\ba{lcccr}
 p\equiv p(y) = \ds{k^{-2}P(x) }
 &,& \ds{  \dot{p}\equiv \dot{p}(y)
 = \frac{\prl p(y)}{\prl y} }
 &,& {\rm etc.} \nnn
 p_F\equiv p_F(y) = \ds{k^{-2}P_F(x) }
 &,& \ds{  \dot{p}_F\equiv \dot{p}_F(y)
 = \frac{\prl p_F(y)}{\prl y} }
 &,& {\rm etc.} \nnn
 r_F\equiv r_F(y) = \ds{\sqrt{\frac{ p_F(y)}{y}}-1 }
 &,& \ds{ \dot{r}_F\equiv \dot{r}_F(y)
 = \frac{\prl r_F(y)}{\prl y} }
 &,& {\rm etc.}
\ea\ee
we find
\begin{eqnarray}
 \ds{l_n^d(w)} &=& \ds{
 n\int_0^\infty dy\, y^{\frac{d}{2}-1}
 \frac{ p-y \dot{p}}{[ p+w]^{n+1}} }\\[2mm]
 \ds{\hat{l}_n^d(w)} &=& \ds{
 \frac{n}{2}\int_0^\infty dy\, y^{\frac{d}{2}-1}
 \frac{ p-y}{[ p+w]^{n+1}} }
\end{eqnarray}
\begin{eqnarray}
 \ds{l_n^{(F)d}(w)} &=& \ds{
 n\int_0^\infty dy\, y^{\frac{d}{2}-1}
 \frac{ p_F-y \dot{p}_F}{[p_F+w]^{n+1}} }\\[2mm]
 \ds{\check{l}_n^{(F)d}(w)} &=& \ds{
 n\int_0^\infty dy\, y^{\frac{d}{2}}
 \frac{r_F[r_F+1]}{[p_F+w]^{n+1}} }
\end{eqnarray}
\begin{eqnarray}
 \ds{ l_{n_1,n_2}^d(w_1,w_2)}
 &=& \ds{
 \int_0^\infty dy y^{\frac{d}{2}-1}
 \left( p-y \dot{p}\right) }\nnn
 &\times& \ds{ \left\{
 \frac{n_1}
 {[ p+w_1]^{n_1+1}[ p+w_2]^{n_2}}
 +\frac{n_2}
 {[ p+w_1]^{n_1}[ p+w_2]^{n_2+1}}
 \right\} }\\[2mm]
 \ds{ \hat{l}_{n_1,n_2}^d(w_1,w_2)}
 &=& \ds{
 \hal\int_0^\infty dy y^{\frac{d}{2}-1}
 \left( p-y\right) }\nnn
 &\times& \ds{ \left\{
 \frac{n_1}
 {[ p+w_1]^{n_1+1}[ p+w_2]^{n_2}}
 +\frac{n_2}
 {[ p+w_1]^{n_1}[ p+w_2]^{n_2+1}}
 \right\} }
\end{eqnarray}
\begin{eqnarray}
 \ds{l_{n_1,n_2}^{(FB)d}(w_1,w_2) }
 &=& \ds{
 \int_0^\infty dy y^{\frac{d}{2}-1}
 \frac{1}{[ p_F+w_1]^{n_1}[ p+w_2]^{n_2}} }\nnn
 &\times& \ds{ \left\{
 \frac{n_1[ p_F-y \dot{p}_F]}
 {[ p_F+w_1]}
 +\frac{n_2[ p-y \dot{p}]}
 {[ p+w_2]} \right\} }\\[2mm]
 \ds{ \hat{l}_{n_1,n_2}^{(FB)d}(w_1,w_2) }
 &=& \ds{ \frac{n_2}{2}
 \int_0^\infty dy\, y^{\frac{d}{2}-1}
 \frac{ p-y}
 {[ p_F+w_1]^{n_1}[ p+w_2]^{n_2+1}} } \\[2mm]
 \ds{ \check{l}_{n_1,n_2}^{(FB)d}(w_1,w_2) }
 &=& \ds{ n_1
 \int_0^\infty dy\, y^{\frac{d}{2}}
 \frac{r_F[r_F+1]}
 {[ p_F+w_1]^{n_1+1}[ p+w_2]^{n_2}} }
\end{eqnarray}
\begin{eqnarray}
 \ds{ m_{n_1 n_2}^d(w_1,w_2)}
 &=& \ds{
 \int_0^\infty dy y^{\frac{d}{2}}
 \frac{  \dot{p}}{[ p+w_1]^{n_1}[ p+w_2]^{n_2}} }\nnn
 &\times& \ds{ \left\{
 \frac{n_1 \dot{p} [ p-y \dot{p}]}
 {[ p+w_1]}
 +\frac{n_2 \dot{p} [ p-y \dot{p}]}
 {[ p+w_2]} +2y \ddot{p}\right\} }\\[2mm]
 \ds{ \hat{m}_{n_1 n_2}^d(w_1,w_2)}
 &=& \ds{
 \hal\int_0^\infty dy y^{\frac{d}{2}}
 \frac{  \dot{p}}{[ p+w_1]^{n_1}[ p+w_2]^{n_2}} }\nnn
 &\times& \ds{ \left\{
 \frac{n_1 \dot{p} [ p-y]}
 {[ p+w_1]}
 +\frac{n_2 \dot{p} [ p-y]}
 {[ p+w_2]} +2[1- \dot{p}]\right\} }
\end{eqnarray}
\begin{eqnarray}
 \ds{m_4^{(F)d}(w) } &=& \ds{
 2\int_0^\infty dy y^{\frac{d}{2}+1}
 \frac{1}{[ p_F+w]^3}
 \left[ \frac{ \dot{p}[1+r_F]}{ p_F+w}
 -\dot{r}_F\right] }\nnn
 &\times& \ds{
 \left\{\left[\dot{r}_F+y\ddot{r}_F\right]
 [ p_F-w]-y\dot{r}_F \dot{p}_F
 \frac{ p_F-3w}{ p_F+w}
 \right\} } \\[2mm]
 \ds{ \check{m}_4^{(F)d}(w) } &=& \ds{
 \int_0^\infty dy y^{\frac{d}{2}+1}
 \frac{1}{[ p_F+w]^3}
 \left[ \frac{ \dot{p}[1+r_F]}{ p_F+w}
 -\dot{r}_F\right] }\nnn
 &\times& \ds{
 \left\{\dot{r}_F[ p_F-w]
 -r_F \dot{p}_F
 \frac{ p_F-3w}{ p_F+w}
 \right\} }
\end{eqnarray}
\begin{eqnarray}
 \ds{m_{n_1,n_2}^{(FB)d}(w_1,w_2)} &=& \ds{
 \int_0^\infty dy y^{\frac{d}{2}}
 \frac{1} {[ p_F+w_1]^{n_1}[ p+w_2]^{n_2}} }\nnn
 &\times& \ds{ \left\{
 [1+r_F]\left(
 n_2\frac{ \dot{p}[ p-y \dot{p}]}{ p+w_2}
 +y \ddot{p}\right)
 -y\dot{p}\dot{r}_F
 \left( \frac{2n_1p_F}{p_F+w_1}-1\right) \right\} }\\[2mm]
 \ds{\hat{m}_{n_1,n_2}^{(FB)d}(w_1,w_2)} &=& \ds{
 \hal\int_0^\infty dy y^{\frac{d}{2}}
 \frac{[1+r_F]}
 {[ p_F+w_1]^{n_1}[ p+w_2]^{n_2}} }\nnn
 &\times& \ds{\left\{
 n_2\frac{ \dot{p}[ p-y]}{ p+w_2}
 +1- \dot{p} \right\} }\\[2mm]
 \ds{\check{m}_{n_1,n_2}^{(FB)d}(w_1,w_2)} &=& \ds{
 \frac{n_1}{2}\int_0^\infty dy y^{\frac{d}{2}}
 \frac{r_F \dot{p} }
 {[ p_F+w_1]^{n_1+1}[ p+w_2]^{n_2}}
 \left(\frac{2n_1p_F}{p_F+w_1}-1\right) }
\end{eqnarray}

\end{document}